\documentclass[12pt,a4paper]{report}
\usepackage{graphicx}
\usepackage{makeidx}
\usepackage{hyperref,amsfonts}
\usepackage{color,euscript,float,amssymb,amsmath}
\usepackage{amsthm, enumerate}
\usepackage{fancyhdr}

\begin{document}

\begin{titlepage}
\begin{center}
\includegraphics[scale=0.11]{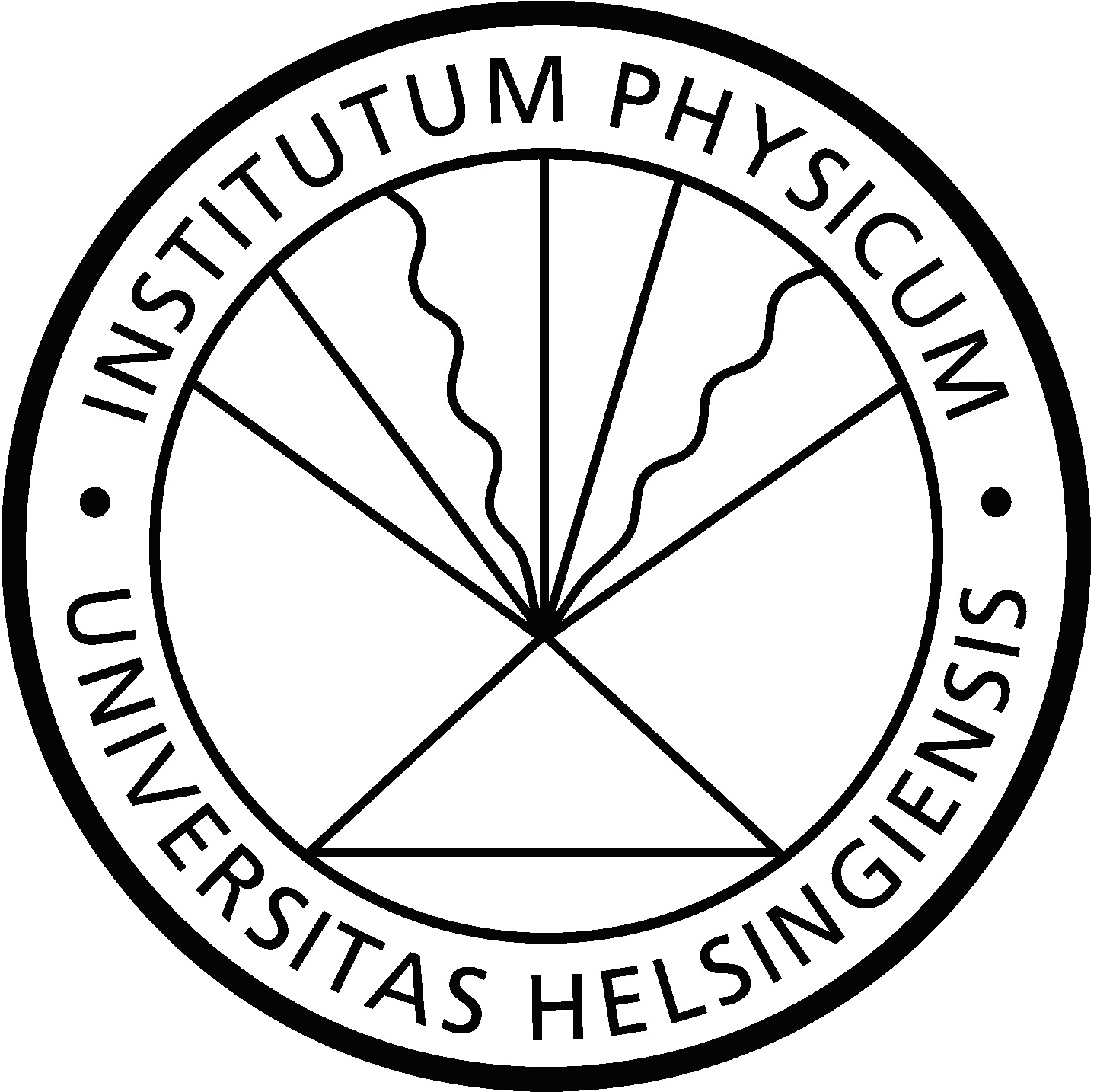}\\[1cm]    
\large \textit{Master's Thesis \\ Degree Programme in Space Sciences}\\ [1.5cm]
\huge \textsc{\textcolor{blue}{Cosmology in the \\ Newtonian limit}}\\[1cm]
\normalsize Ugo Bertello\\
2012  \\[1.5cm]
\normalsize Supervisor: Doc. Syksy R\"{a}s\"{a}nen \\
\normalsize Reviewers: Prof. Kari Enqvist, Doc. Syksy R\"{a}s\"{a}nen\\[2cm]
\normalsize{UNIVERSITY OF HELSINKI} \\
\normalsize DEPARTMENT OF PHYSICS\\ [0.5cm]
\small P.O. BOX 64 (Gustaf H\"{a}llstr\"{o}min katu 2a)\\ 
FI - 00014 University of Helsinki\\
\end{center}
\end{titlepage}

\pagenumbering{roman}


\begin{abstract}
Numerical $N$-body simulations of large scale structure formation in the universe are based on Newtonian gravity. However, according to our current understanding, the most correct theory of gravity is general relativity. It is therefore important to understand which degrees of freedom and which features are lost when the relativistic universe is approximated, or rather replaced, by a Newtonian one. This is the main purpose of our investigation. 
We first define Newtonian cosmology and we give an overview on general relativity, both in its standard and covariant formulations. We show how the two theories deal with inhomogeneous cosmological models and we explain the role that inhomogeneities play in the dynamics of the universe on large scales. We define averaging in cosmology and we introduce the backreaction conjecture. 
Then we review on how Newtonian gravity and general relativity relate to each other in the fully non-linear regime. For this purpose we discuss frame theory, whose aim is to reconcile Newton's and Einstein's theories under the same formal structure. We carry out the same investigation also in the weak-field, small-velocity limit of general relativity, and we derive the Newtonian limit resorting to the framework of post-Newtonian cosmology.
Finally we remark that there are solutions of Newtonian gravity which do not have any relativistic counterpart. This suggests that there are cases in cosmology in which the two theories are irreconcilable and that the reliability of the Newtonian approximation requires further theoretical investigation.
\end{abstract}

\renewcommand{\qedsymbol}{\footnotesize Q.E.D.}

\footnotesize
In this thesis the following notation is adopted:\\
\begin{itemize}
\item The metric tensor $g_{\alpha \beta}$ has Lorentzian signature (-+++).
\item Greek indices run from 0 to 4, Latin indices from 1 to 3. Einstein's summation rules are implied for couples of indices, one being covariant (upstairs) and the other contravariant (downstairs).
\item The statement ``$M_{\alpha}^{\phantom{\alpha}\beta}$ is a tensor" is often used, for simplicity, in the place of the more correct statement ``$M_{\alpha}^{\phantom{\alpha}\beta}$ are the components of the tensor $\mathbf{M} = M_{\alpha}^{\phantom{\alpha}\beta} \, dx^{\alpha} \otimes \partial_{\beta}$".
\item The partial derivative is represented by a comma, or with a nabla symbol with an overline, for instance ${\partial \xi \over \partial x^{\alpha}} = \overline{\nabla}  \xi = \xi_{,\alpha}$.
\item The covariant derivative is represented by a semicolon or with a nabla symbol with an index, for instance $\nabla_{\beta} \xi^{\alpha} = \xi^{\alpha} _{\phantom{\alpha};\beta}$.
\item The Laplace operator is defined as $\overline{\nabla} ^2 = \overline{\nabla} \cdot \overline{\nabla}$.
\item The d'Alembert operator is defined as $\Box = - \frac{\partial ^2}{\partial t^2} + \overline{\nabla} ^2$.
\item Round brackets imply symmetrization on the enclosed indices, while squared ones imply antisymmetrization. For instance $\xi ^{\phantom{\gamma}[\alpha \beta]}_{\gamma} = \frac{1}{2!}(\xi ^{\phantom{\gamma}\alpha \beta}_{\gamma} - \xi ^{\phantom{\gamma}\beta \alpha}_{\gamma}) $.
\item A relativistically geometrized unit system is adopted, i.e.\ the speed of light $c$ is taken to be equal to one, while the gravitational constant $G_N$ is equal to $\frac{1}{8 \pi}$.
\end{itemize}

\newpage \normalsize \tableofcontents \newpage \pagenumbering{arabic}
\pagestyle{fancy}
\renewcommand{\sectionmark}[1]{}

\chapter{Introduction} \label{mot}

\paragraph*{Structure formation in cosmology.}
The inspiration for this thesis comes from the physics of large scale structure (LSS) formation in the universe. Observationally we know that on large scales, luminous matter is distributed in a web-like pattern (see Figure 1.1) known as the \textit{cosmic web}, meaning that galaxies are organised in clusters, and that groups of clusters form superclusters. The clusters are connected to each other by thin filamentary structures, and the regions between clusters, voids, are nearly empty.
\begin{figure} [h!]
\begin{center}
\includegraphics[scale=0.40]{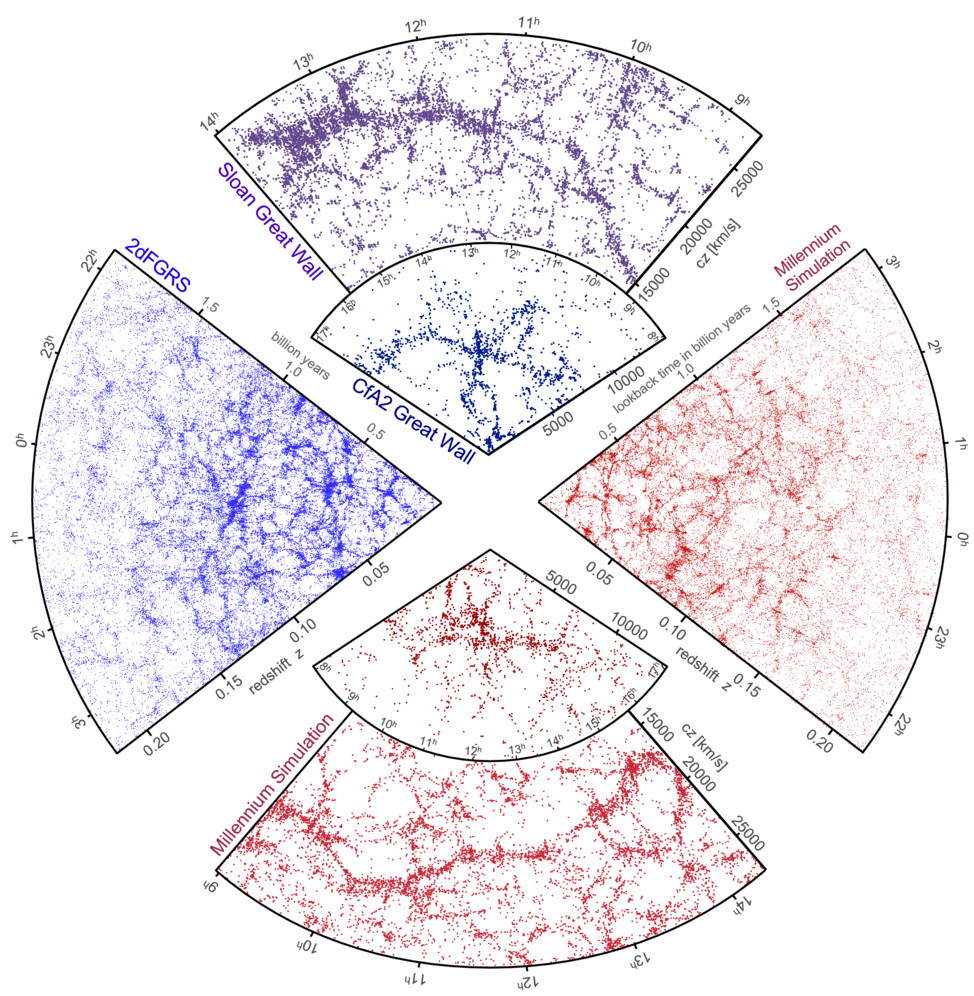}
\caption{\footnotesize LSS from observations (in blue and violet) and numerical simulations (in red and burgundy). One can see features such as filaments, voids and clustering. These are the constituents of the cosmic web. Note that the degree of clustering is roughly constant in space but not in spacetime, i.e.\ structures smooth out with increasing redshift (look at the red and blue panels, where the redshift varies significantly). [Picture credits V. Springel, C. S. Frenk and S. D. M. White \cite{LSSpic}]}
\end{center}
\end{figure}

Studies of LSS formation aim to determine how these structures (inhomogeneities in the density field) that we observe at late times have emerged, through \textit{hierarchical accretion}, from the smooth physics at early times. Usually the initial conditions are given by the \textit{cosmic microwave background} (CMB) (see Figure 1.2).
\begin{figure} [h!]
\begin{center}
\includegraphics[scale=0.08]{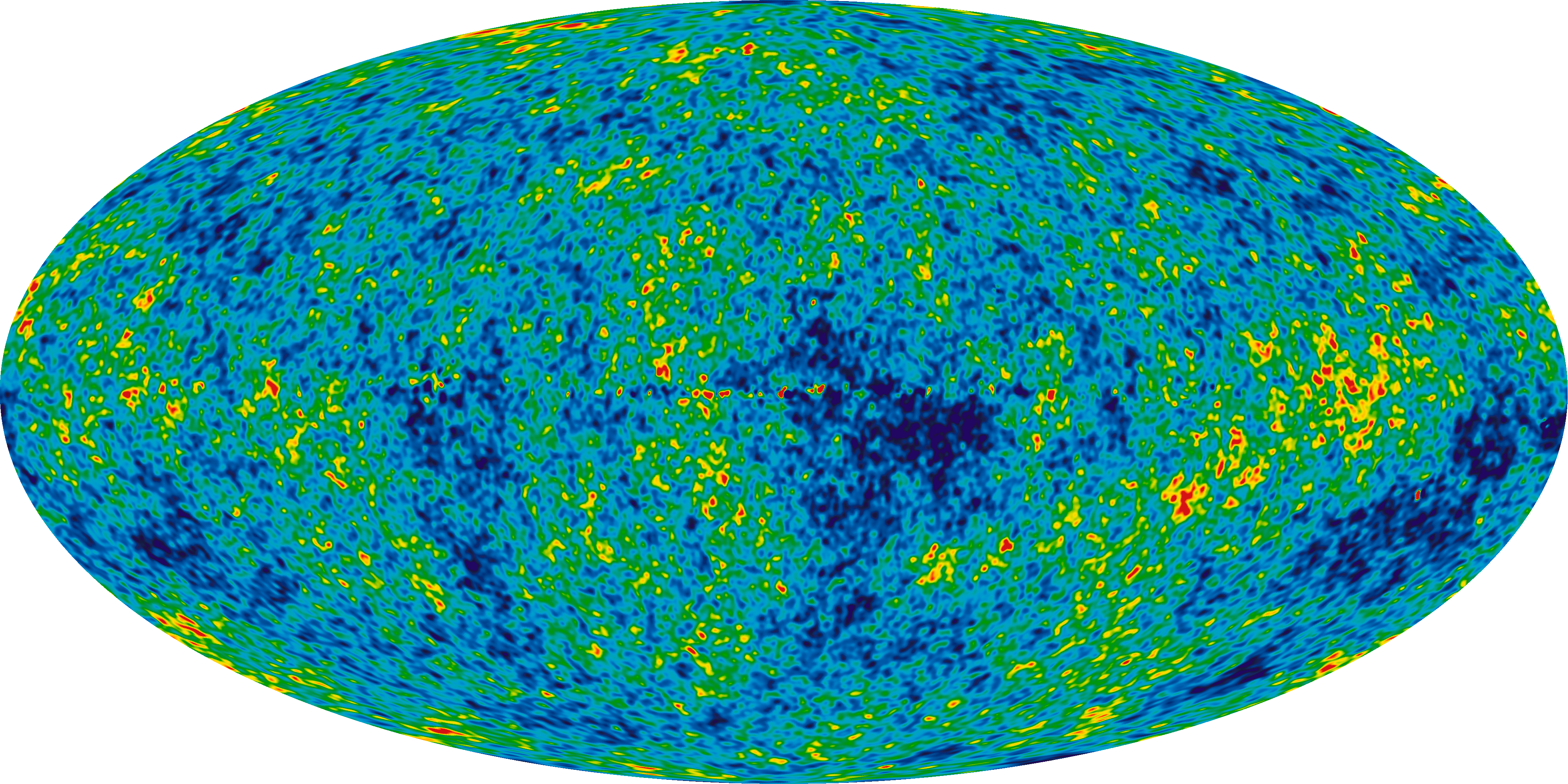}
\caption{\footnotesize WMAP 7-year CMB map. This is a snapshot of the temperature perturbations in the universe after decoupling. It gives the initial conditions required in LSS studies. [Picture credits NASA/GSFC]}
\end{center}
\end{figure}
This is a snapshot of the universe at redshift $z \approx 1090$, when atomic nuclei and electrons recombined to form neutral atoms, photons ceased to be trapped by Compton scattering, and the universe became transparent. Before recombination, photons and baryons were coupled, therefore the CMB gives initial conditions on the matter density field distribution $\rho$ at this time, which is measured to be of the order ${\delta \rho}/{\overline \rho} = \mathcal{O}(10^{-5})$ for baryons, where the overline denotes the mean value, and and $\delta \rho$ is a perturbation.

Another important aim in LSS studies is to understand the global dynamics of structures, meaning the law governing their overall expansion. Up to the time of the CMB, this can be done analytically using the theory of \textit{general relativity} (GR), linearised around a flat and expanding background. However, after that, the perturbations in the density field become locally large and structures enter the non-linear regime. What we would like to do is to follow the accretion process numerically using GR, however this is impossible to put into practice due to computational limitations. 

One way to overcome this limitation is to study LSS formation using codes based on \textit{Newtonian gravity} (NG), which is numerically much lighter than GR. The replacement of the GR ``real" universe with a Newtonian ``mock" universe may not be completely harmless and the purpose of the thesis is to better understand this issue.

\paragraph{The GR universe.} The theory of general relativity describes a generic four-dimensional spacetime which is curved by the presence of matter fields.
Unfortunately, GR is a theory involving a set of ten coupled second-order partial differential equations which cannot be solved numerically in cosmology due to computational limitations. They cannot be solved analytically either, unless the system studied encodes a high degree of symmetry.  Fortunately, the early universe offers this possibility, as it is well described by a model which is \textit{exactly homogeneous and isotropic} up to perturbations.

\paragraph*{Homogeneity and isotropy.}
An intuitive explanation of exact homogeneity and isotropy requires the knowledge of  the concept of \textit{statistical homogeneity and isotropy}.

A universe filled with a medium of density $\rho(x,t)$, which has spatial average \begin{equation} 
\langle \rho(x,t) \rangle \doteq \frac{\int_{V(x)} \rho(x,t) \, d^3x}{V(x)} \end{equation} 
over the domain $V(x)$, is defined to be statistically homogeneous above scale $\lambda _0$  if \cite{statcosmo}
\begin{equation}
| \langle \rho(x,t) \rangle  - \rho_0 | < \rho_0 \; \; \, \forall \, x > \lambda_0 \ ,
\end{equation}
where $\rho_0$ is the ensemble average of the field 
\begin{equation} 
\rho_0 = \lim_{x \to +\infty} \frac{\int_{V(x)} \rho(x,t) \, d^3x}{V(x)} \ .
\end{equation}
The intuitive explanation of statistical homogeneity proceeds as follows: take a box of volume $V$ and size larger than the homogeneity scale $\lambda _ 0$, place it somewhere in the universe at point $r$, measure the particle density there, then  move it to point  $r + r_0$ and repeat the measurement. If the universe is statistically homogeneous, then the density of particles observed within the box when it is centred at either of the two points depends not on the value of $r_0$, but solely on the size of the box and its shape.
Explaining statistical isotropy works in the same way as explaining statistical homogeneity, but in this case, instead of moving the box from one point to another we must keep its centre fixed, rotate it and verify that the density of particles does not depend on the angle of rotation.
The universe is said to be exactly homogeneous and isotropic if the homogeneity scale is infinitesimal. We give a more rigorous definition of exact homogeneity and isotropy in Chapter \ref{FRW}, which is dedicated to the FRW model.

There have been claims, relying on statistical studies of galaxy distribution, that the scale of homogeneity is small, around $100 \, {h}^{-1} {\mathrm{Mpc}}$ \cite{eisenstein}\cite{homoscale}\cite{homoscale2}. However it has also been claimed that no such homogeneity scale has yet been detected \cite{sylos}\cite{inhomo}, because the samples used in the analysis are too small to yield reliable results.

\paragraph*{The FRW model and the $\Lambda$CDM model.}
The \textit{Friedmann-Robertson-Walker model} (FRW) describes an exactly homogeneous and isotropic universe whose space expands or contracts uniformly with a \textit{scale factor}, and it is discussed in detail in Chapter \ref{FRW}. The geometry of the space in the FRW model may be spherical, flat or hyperbolic. 

The flat case agrees with all observations about the expansion history of the universe and its thermodynamics at early times (CMB, big bang nucleosynthesis and baryon acoustic oscillations up to the surface of last scattering).

However the FRW model is not as successful at fitting observations of the late time universe.  
For instance it fails to fit the supernova observations\footnote{The discovery of cosmic acceleration from supernova observations was awarded with the Nobel prize in physics in 2011.} of cosmic acceleration, which began at $z \sim 0.5$ \cite{acc}\cite{acc2}, and it underpredicts distances to the surface of last scattering by a factor of $1.5-2.0$ at late times \cite{vonlanthen}. The unknown entity responsible for the acceleration is called \textit{dark energy}.
The afore-sketched model can be improved upon adding to it a cosmological constant. The model obtained is knows as $\Lambda$CDM, where $\Lambda$ stands for the cosmological constant and CDM stands for cold (meaning slowly moving) dark matter and it contains four kinds of matter sources:  \textit{baryonic matter}, \textit{dark matter}, \textit{photons} and \textit{neutrinos}. The $\Lambda$CDM model is the cornerstone of physical cosmology and agrees with all observations to date.

The the cosmological constant is mathematically equivalent to vacuum energy with negative pressure (thus a fifth matter source in the model), though conceptually they are different. The former introduces a modification in the law of gravity, while the latter represents a new matter component. Vacuum energy and the cosmological constant are some of the possible candidates for dark energy. The energy density of vacuum energy, inferred by cosmological observations is $10^{-120}$ times smaller than the value expected from quantum field theory, which is of the order of $M_{Pl}^4 \sim \mathcal{O}(10^{77}) \, \mathrm{GeV}^4$. This discrepancy indicates that there are some problems to reconcile the dark energy paradigm with standard quantum field theory.

In addition, the $\Lambda$CDM model suffers from a \textit{coincidence problem}: the contribution of the cosmological constant term to the expansion law becomes prominent after structures enter the non-linear regime. This seems to suggest that structure formation plays a role in the overall dynamics at late times. 


\paragraph*{Dark energy candidates.}
There are several ways one can attempt to explain the data on cosmic acceleration, which are different correspond to different ways of understanding dark energy. 

One possibility is  to consider that GR may not hold on large scales, where observations are made. This context requires one to write down a modified theory of gravity, that agrees locally with the prediction of GR (GR has been shown to hold on scales of the solar system), but which on large scales deviates from Einstein's theory enough to reproduce cosmic acceleration. These theories are known as \textit{modified gravity} and $\Lambda$CDM is the simplest and most successful example.

Another possible form of dark energy is \textit{quintessence}, which is an hypothetical dynamical scalar field whose potential energy drives the acceleration.

However, for our purposes the most important attempt to explain cosmic acceleration is the \textit{backreaction conjecture} (see \cite{syksybuchert} for an overview). This conjecture states that the failure of homogeneous and isotropic models with no dark energy or modified gravity to determine distances at late times stems for the fact that, as structures develop, homogeneity and isotropy break down, thus requiring the adoption of an inhomogeneous model. In an inhomogeneous universe, the values of observables such as the density field and the expansion parameter are not the same everywhere, but vary from place to place. The concept of \textit{averaging in cosmology} over a volume is introduced to take this into account. As we review in this thesis, the effect of inhomogeneities on the expansion law of the universe can be parametrised by a quantity known as the \textit{backreaction variable}, whose effect on the dynamics of the universe is to lead acceleration. The most important fact about the backreaction variable is that it is a physical quantity, and that it is non-zero (although it might be small) in any non-exactly homogeneous and isotropic model of the universe. Since we observe inhomogeneities in the real universe, then we should understand whether their contribution to the acceleration is sizeable or not.
Backreaction has been quantified in perturbed homogeneous and isotropic models (as we remark in Chapter \ref{cosmo}), however it hasn't been realistically quantified yet beyond perturbation theory. Nevertheless, some attempts have been made in this direction,  for a recent example, see \cite{enqvist}.

\paragraph*{The Newtonian universe and numerics.} What is done numerically to understand the physics of LSS formation is to take some initial conditions in the density field, to evolve them with Newtonian gravity on an expanding background and to check that the matter distribution matches what is observed in the real universe. This gives good results \cite{LSSpic}. 
However the homogeneity scale of Newtonian $\Lambda$CDM simulations is one order of magnitude smaller then in the real universe \cite{sylos}. The astrophysics involved in numerical simulations is not completely understood and this might be the source of error. A notorious problem in this field is that simulations deal with clumps of dark matter, while in the astronomical observation we see only baryonic matter, whose dynamics inside the gravitational wells of cosmic structures is not clear.

In spite of the success of Newtonian simulations of LSS formation we shouldn't forget that NG is a theory which is deeply different from GR, and that it might fail at some point in reproducing the relativistic universe.
Newtonian gravity has an ill-posed initial value problem when applied to cosmology and presumes an absolute three-space and an absolute time. In contrast, GR has a well-posed initial value problem, embeds extra degrees of freedom, and is defined on a four-dimensional spacetime, that can be split into a three-space plus time only under certain conditions. These differences are important not only conceptually but also because they may lead to conspicuous discrepancies between  cosmological models derived in the two theories. This is the main issue addressed in this thesis. 

\paragraph*{Structure of the thesis.} 
A description of Newtonian gravity is given in Chapter \ref{NG}. In Chapter \ref{NC} we present Newtonian cosmology and its pathologies, in addition we explain how to obtain averaged equations for this theory. General relativity is portrayed in Chapter \ref{gr} in the standard fashion, while a covariant formulation is given in Chapter \ref{1+3form}, where averaging is discussed in the relativistic context. In Chapter \ref{FRW} we discuss the Friedmann-Robertson-Walker model. In Chapter \ref{frame} we review frame theory, which is a theory that reconciles NG and GR on the same geometrical ground. In Chapter \ref{GRC} GR we investigate the linear regime of GR using the machinery of post-Newtonian cosmology and in Chapter \ref{cosmo} we remark that it does not coincide with NG, and we give an example of a solution of NG which does not have any linearised GR counterpart. In the same chapter we calculate backreaction in post-Newtonian cosmology. Chapter \ref{summary} is dedicated to conclusions.

\chapter{Newtonian gravity} \label{NG}

\subsubsection*{Why Newtonian gravity?}

In this chapter we give a review on Newtonian gravity. This theory has been superseded by the more correct theory of general relativity, nevertheless it hasn't been put aside. Thanks to its simplicity it's still the cornerstone of numerical cosmology. 

In what follows we present the dynamics of NG given in terms of forces and in terms of the gravitational potential, considering both discrete and continuous matter models. In addition we discuss the geometry and the symmetries of the theory.

\section{The force picture}
\paragraph*{Discrete case.}
\textit{Newton's inverse-square law of gravitation} states that the force exerted on a test particle with mass $m_j$ located at the point $\mathbf{x}_j$ by a gas of $N$ point particles with masses $m_i$ and position vectors $\mathbf{x}_i$ is
\begin{equation} \label{newtonforcenum}
\mathbf{F}_j = m_j \frac{\partial ^2 \mathbf{x}_j}{\partial t^2} =  \frac{m_j}{8 \pi} \sum \limits _{i=1}^{N}  \frac{\mathbf{x}_i-\mathbf{x}_j}{\left | \mathbf{x}_i-\mathbf{x}_j  \right |^3} m _{i} \ .
\end{equation} 
This is a second order ordinary differential equation and has a well-posed Cauchy problem, i.e.\ it can be solved by giving the initial positions and velocities of the particles at the initial time. This is the equation used in $N$-body simulations of structure formation\footnote{In the simulations however Newton's force is not solved for each particle but clumping techniques are adopted in order to reduce the computing time.}. In these simulations an initial nearly homogeneous distribution of dark matter particles is evolved in time according to (\ref{newtonforcenum}). The overdensities and underdensities give rise, through gravitational accretion, to the objects of the same kind as those that we observe today on large scales in the universe, such as clusters, filaments and voids.

In section \ref{woe1} we study the convergence of the force (\ref{newtonforcenum}) in the cosmological limit of an infinite universe filled with a homogeneous ($\rho(\mathbf{x},t) = const$) medium, and we discuss its range of applicability to the physics of structure formation.

\paragraph*{Continuous case.}
Given a continuous mass distribution $\rho(\mathbf{x'},t)$ and a test particle with mass $m_j$ located at the point $\mathbf{x}_j$, the infinitesimal force d$\mathbf{F}_j$ acting on the test particle, exerted by the mass contained in the volume element $d^3\mathbf{x'}$ located at $\mathbf{x'}$, is 
\begin{equation}
d\mathbf{F}_j = \frac{m_j}{8 \pi} \frac{\mathbf{x'}-\mathbf{x}_j}{\left | \mathbf{x'}-\mathbf{x}_j \right |^3} \rho(\mathbf{x'},t) \, d^3\mathbf{x'} \ .
\end{equation}
This infinitesimal force is well defined.
Summing over the volume domain $\mathcal{D}$ the total force becomes
\begin{equation} \label{newtonforce}
\mathbf{F}_j = \frac{m_j}{8 \pi} \int \limits_{\mathcal{D}} \frac{\mathbf{x'}-\mathbf{x}_j}{\left | \mathbf{x'}-\mathbf{x}_j \right |^3} \rho(\mathbf{x'},t) \, d^3\mathbf{x'} \ .
\end{equation}
In section \ref{woe1} we study the convergence of this integral in cosmology.

\section{The potential picture}
An alternative way of thinking about gravity is not through a force but in terms of a \textit{gravitational potential} $\phi(\mathbf{x},t)$. The two concepts are not completely  interchangeable and some qualitative differences are outlined in the next chapter of this thesis. 

The force and the potential are related to each other by the equation
\begin{equation} \label{forcepot}
\frac{\mathbf{F}}{m} \doteq - \overline{\nabla} \phi(\mathbf{x},t) \ .
\end{equation}
Taking the divergence of (\ref{newtonforce}) and combining it with (\ref{forcepot}) gives the Poisson equation for the gravitational field
\begin{equation} \label{rhophi}
\nabla^2 \phi(\mathbf{x},t) = \frac{1}{2}\rho(\mathbf{x},t) \ .
\end{equation}
Exploiting Gauss' theorem and Green's function theory to solve explicitly (\ref{rhophi}), yields
\begin{equation} \label{phi}
\begin{split}
&\phi(\mathbf{x},t) = \frac{1}{2} \int \limits_{\mathcal{D}} G_f(\mathbf{x},\mathbf{x'}) \rho(\mathbf{x'},t)\, d^3\mathbf{x'} + \\+ \int \limits_{\partial {\mathcal{D}}} [\phi(\mathbf{x'},t) & \nabla ' G_f(\mathbf{x},\mathbf{x'}) - G_f(\mathbf{x},\mathbf{x'})' \nabla ' \phi(\mathbf{x'},t)] \cdot d\mathbf{S}' \ ,
\end{split}
\end{equation}
where d$\mathbf{S}$ is the surface element and $G_f(\mathbf{x},\mathbf{x'})$ is a \textit{Green function}, which in the case of Newtonian gravity is
\begin{equation}
G_f(\mathbf{x},\mathbf{x'}) = - \frac{1}{4 \pi} \frac{1}{\left | \mathbf{x'}-\mathbf{x}  \right |} \ .
\end{equation}
Applying equation (\ref{phi}) to cosmology introduces the difficulty of choosing the boundary term on the right hand side. The reason are explained in detail in section \ref{woe2}.

An interesting feature of Poisson equation (\ref{rhophi}) is that it contains only spatial derivatives, which implies that gravity does not have a finite propagation speed in Newton's theory. Hence a density perturbation affects instantaneously the dynamics at a distance, giving rise to non-local interactions. This problem holds also in the force picture.

On the contrary, this problem doesn't hold in general relativity where the speed of the interaction is finite. In Chapter \ref{1+3form} we remark the consequences that the finiteness of the speed of gravity has on the dynamics of systems governed by GR.


In this thesis  we adopt most of times the potential picture of NG, as this presents several advantages over the force picture.

The first advantage is that NG in the potential picture proves to be well tailored for seeking analogies with general relativity, as the latter theory can be as well defined in terms of fluid quantities and a potential (in the case of weak fields). 

The second advantage is that the potential picture gives more theoretical insights on the nature of NG. For instance equation (\ref{phi}) clearly shows that the potential is made up of a volume term plus a boundary term, and it gives a way of calculation these two quantities separately. On the other hand, looking at the force equation (\ref{newtonforcenum}) it's not clear what happens on the boundary and the choice of this term is much less transparent.

\section{The geometry} \label{NGgeom}
Newtonian gravity lives in an Euclidean space and its infinitesimal element of length, the \textit{line element}, is given in terms of one-forms $dx^i$ by
\begin{equation}
ds^2 = g_{ij} dx^i \otimes dx^j = dx^2 + dy^2 +dz^2 \ ,
\end{equation}
where in the last expression we have suppressed the tensor product symbol for brevity (this holds also for what follows) and where the covariant metric is $g_{ij} = \delta _{ij} = \mathrm{diag}(1,1,1)$. The Euclidean space is flat, meaning that if we take a vector, displace it and bring it back to the origin then its direction is left unchanged\footnote{The definition of curvature is presented rigorously in section \ref{geomgr}.}. This simple fact is very important for our discussion as it implies that in a flat space the order of derivation along different spatial directions does not matter.

After introducing the infinitesimal element of length we define the infinitesimal \textit{volume element} of NG, which is constructed from the metric and the one-form $dx^i$ as
\begin{equation}
\tilde{\epsilon} \doteq \tilde{\epsilon} _{ijk} dx^i  dx^j  dx^k = \sqrt{g} \, d^{3} x = d^{3} x \ .
\end{equation}
Here $g = \mathrm{det}(g_{ij})$ and $\tilde{\epsilon} _{ijk}$ is the Levi-Civita symbol. This is an object that can be generalised to $n$ dimensions $\tilde{\epsilon} _{\alpha \beta \gamma \delta...}$, being equal to $1$ for even permutations of the indices (e.g.\ $0123...$), $-1$ for odd permutations (e.g.\ $1023...$), and $0$ otherwise. In the Newtonian case the indices of the Levi-Civita symbol run from one to three.

\section{The symmetries} \label{NGsymm}
The symmetries of Newtonian gravity are those of the ten-dimensional Galilean group, according to which the laws of physics are invariant under the change of coordinates
\begin{equation}
\begin{cases}
x^i &\rightarrow  \; \; \; x'^{i} = R^i_{\phantom{i}j} x^j + x^i_0 + v^i t \\ \:
t &\rightarrow \; \; \; \; t' = t + t_0
\end{cases}
\end{equation}
Here $x^i$ are the spatial coordinates, $t$ is the absolute time, $R^i_{\phantom{i}j}$ is a rotation matrix (three parameters), $x^i_0$ is a translation vector (three parameters), $v^i$ is a Galilean boost vector (three parameters) and $t_0$ is a time translation scalar (one parameter). Note that the components of the rotation matrix, the translation vector, the boost vector and the scalar are constants in NG.

\chapter{Newtonian cosmology and its sorrows} \label{NC}
\subsubsection*{Brief historical digression} 
The definition of Newtonian cosmology adopted in this thesis was provided in 1959 by Heckmann and Sch\"{u}cking \cite{heck-sch}. It is curious to notice that, albeit Newtonian gravity was formulated around two centuries before general relativity, it has been cosmologically quite sterile. In fact Newtonian cosmology was defined around fifty years later than relativistic cosmology. The latter can be dated back  to the dawn of general relativity, in 1917. 
\subsubsection*{Introduction}
In the first part of this chapter we present the theory of Newtonian cosmology in its Eulerian and Lagrangian formulations. This knowledge is important when in section \ref{1+3cosmo} we study how this theory differs from its relativistic counterpart. In the second part we introduce the concept of averaging in cosmology, which is central to our discussion. In the last part we highlight some problems arising in Newtonian cosmology.

\section{Definition of Newtonian cosmology}
\subsection{Eulerian picture}
Newtonian cosmology consists of an absolute space and an absolute time, defined on a manifold $M = \mathbb{E} \times \mathbb{R}$ where $\mathbb{E}$ is the flat three-dimensional Euclidean space, $\times$ is a Cartesian product and $\mathbb{R}$ is the universal time dimension. 

In the \textit{Eulerian picture} of NG we define three scalar fields living on the manifold $M$: the gravitational potential $\phi(\mathbf{x},t)$, the matter density field $\rho(\mathbf{x},t)$ and the pressure field $p(\mathbf{x},t)$, where the coordinates of the vector $\mathbf{x}(t)$ are given with respect to a fixed coordinate system, called the Eulerian frame. In addition, a Eulerian velocity vector field $\mathbf{v}(\mathbf{x},t)$ is  introduced. These five quantities are related to each other by the laws of fluid dynamics:\\
\\
the \textit{continuity equation}
\begin{equation} \label{continuity}
{\partial \rho \over \partial t} + \mathbf{v} \cdot \overline{\nabla} \rho + \rho \overline{\nabla} \cdot \mathbf{v} = 0 \ ,
\end{equation}
the \textit{Euler equation}
\begin{equation} \label{euler}
{\partial \mathbf{v} \over \partial t} + \mathbf{v} \cdot \overline{\nabla} \mathbf{v} = - \overline{\nabla} \phi - \frac{\overline{\nabla} p}{\rho} =  - \overline{\nabla} \phi + \mathbf{a} \ ,
\end{equation}
the \textit{Poisson equation for gravity}
\begin{equation} \label{poisson}
\nabla^2 \phi = \frac{1}{2} \rho - \Lambda  \ ,
\end{equation}
and the \textit{barotropic equation of state}
\begin{equation} \label{barotropic}
p = p(\rho) \ .
\end{equation}
A cosmological constant is included in the Poisson equation for the sake of generality. In the Euler equation (which corresponds to the viscosity-free Navier-Stokes equation) we gave the gradient of the pressure in terms of a generic acceleration vector $\mathbf{a}$, whose components are 
\begin{equation} \label{ai}
a^i = - \frac{1}{\rho} \frac{\partial p}{\partial x_i}  \ .
\end{equation}
From equation (\ref{euler}) we see that the pressure does enter in the dynamics of NG. From equation (\ref{poisson}) however we see that this quantity does not gravitate. For this reason in NG is not meaningful to associate the cosmological constant with vacuum energy, which is described as a negative pressure of the vacuum.

\subsection{Lagrangian picture} \label{lagr}
\paragraph*{Definition.}
In the Eulerian specification of the fluid flow, the flow quantities are depicted as a function of position with respect to a fixed reference frame or grid. The Eulerian position vector is $\mathbf{x}(t)$ and the velocity is $\mathbf{v}(\mathbf{x},t)$. 

In the \textit{Lagrangian picture}, on the other hand, the velocity field $\mathbf{V}(\mathbf{q},t)$ of the particles is given with respect to a time-independent fiduciary coordinate $\mathbf{q}$. This is often taken to be the position of the center of mass of the fluid at the initial time. 

The Eulerian coordinates are given in terms of the Lagrangian coordinates plus a displacement vector $\mathbf{s}(\mathbf{q},t)$ as
\begin{equation}
\mathbf{x}(t) = \mathbf{q} + \mathbf{s}(\mathbf{q},t) \ .
\end{equation}
The velocities and coordinates in the two pictures are related by
\begin{equation}
\mathbf{V}(\mathbf{q},t) = {\partial \mathbf{s}(\mathbf{q},t) \over \partial t}  = \mathbf{v}(\mathbf{x}(\mathbf{q}),t) \ .
\end{equation}
The Jacobian determinant of the Lagrangian mapping between the initial (Lagrangian) position $\mathbf{q}$ of the fluid elements at time $t_0$ and the final (Eulerian) position $\mathbf{x}$ at time $t$ is
\begin{equation} \label{jacobian}
J \doteq \mathrm{det}(\frac{\partial x_i}{\partial q_j}) = \mathrm{det}(\delta_{ij} + \frac{\partial s_i}{\partial q_j}) \ .
\end{equation}
The volume element in the Eulerian picture at the initial time $t_0$ is related to the volume element in the Lagrangian picture at some later time $t$ as
$d^3x(t_0) \, J(t_0) = d^3q(t) \, J(t)$.
We can choose the two pictures to coincide initially by setting $\mathbf{s}(\mathbf{q},t_0) = 0$, which implies that the Lagrangian volume element evolves in time as
\begin{equation} \label{volumechange}
d^3q(t) = d^3x(t_0) \, J^{-1}(t)\ .
\end{equation}
In Figure 3.1.\ we sketch the situation and we explain the physical meaning of equation (\ref{volumechange}).
\begin{figure} [h!]
\begin{center}
\includegraphics[scale=0.19]{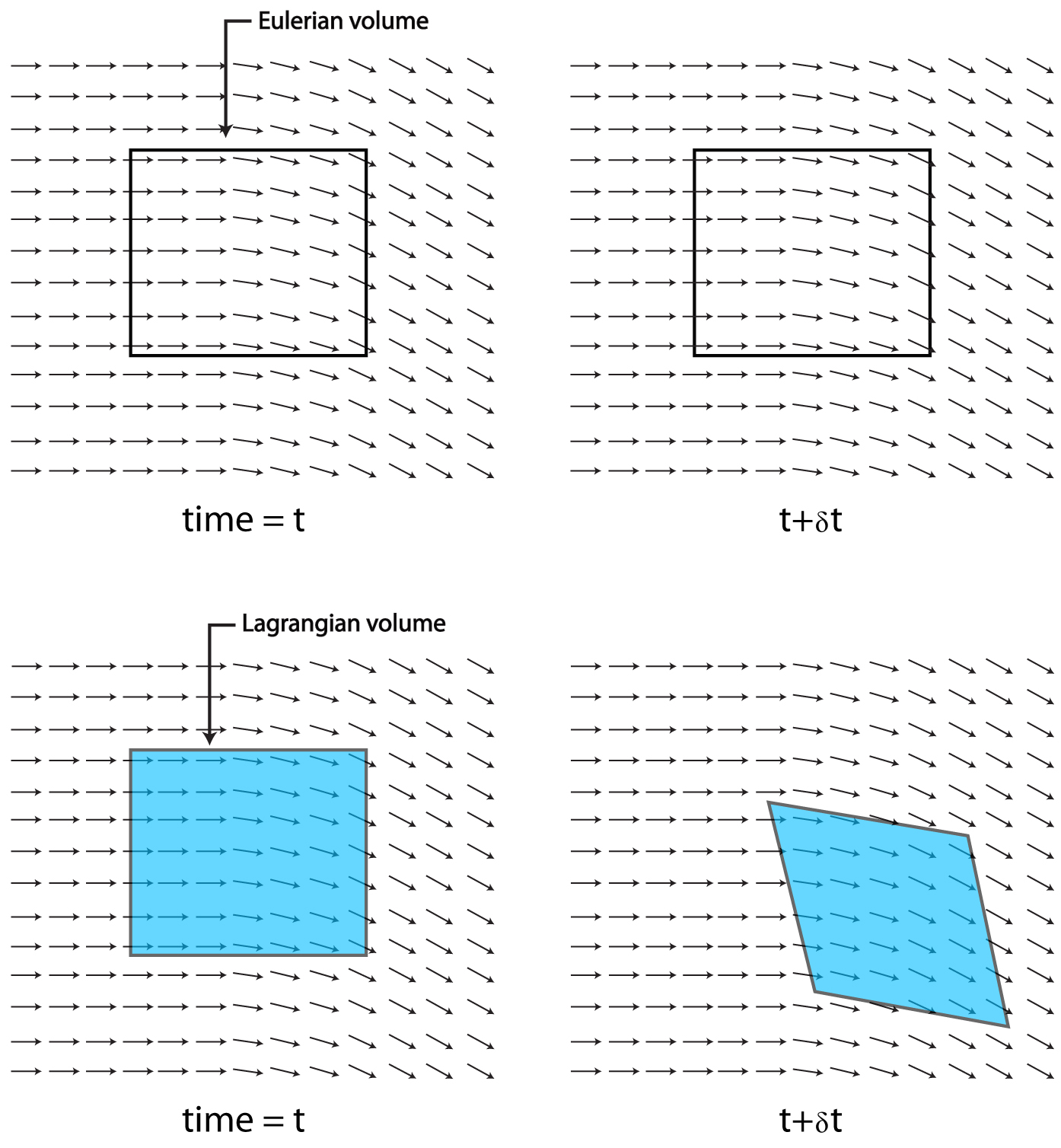}
\caption{\footnotesize Upper panel: the Eulerian volume stays fixed in time. Lower panel: the Lagrangian volume is dragged by the vector field, thus it changes in time according to (\ref{volumechange}).}
\end{center}
\end{figure}
The \textit{Lagrangian time derivative} ${d}/{dt}$, which takes into account the fact that the coordinate system is dragged by the flow, is related to the Eulerian time derivative ${\partial / \partial t}$ and the velocity field $\mathbf{v}$ as
\begin{equation} \label{NGtot}
\frac{d}{dt} \doteq {\partial \over \partial t} + \mathbf{v} \cdot \overline{\nabla} = {\partial \over \partial t} + v^i {\partial \over \partial x^{i}} \ .
\end{equation}
In the thesis we adopt the shortened notation ${d}/{dt} \doteq ()\dot{•}$.
The Lagrangian time derivative does not commute with the partial derivative with respect to Eulerian coordinates, but the two are related by the identity
\begin{equation} \label{identitycomm}
\left[{\partial \over \partial {x^i}},\frac{d}{dt}\right] = v_i^{\phantom{i},k} {\partial \over \partial {x^k}} \ .
\end{equation}
On the other hand the Lagrangian coordinates are independent of time, therefore
\begin{equation} \label{commq}
\left[{\partial \over \partial {q^i}},\frac{d}{dt}\right] = 0 \ .
\end{equation}
If we apply the Lagrangian time derivative to the definition of the Jacobian determinant (\ref{jacobian}), using the identity 
\begin{equation}
\frac{d}{dt}(\det A) = \det(A) \, \mathrm{tr} (A^{-1} \frac{d}{dt}A)
\end{equation}
(where $A$ is a matrix) and the commutator (\ref{commq}), we obtain the evolution equation for $J$
\begin{equation} \label{Jevol}
\dot{J} = v^k_{\phantom{k},k} J \ ,
\end{equation}
which shows that $J$ is stretched in time by a factor which corresponds to the trace of the velocity's gradient.

Equations (\ref{continuity})-(\ref{poisson}), written down using this new concept of Lagrangian time derivative become
\begin{equation} \label{continuity2}
\dot{\rho} + \rho v^k_{\phantom{k},k} = 0
\end{equation}
\begin{equation} \label{euler2}
\dot{v}^i = - \phi ^{,i} + a ^i
\end{equation}
\begin{equation} \label{poisson2}
\phi _{,k}^{\phantom{,k},k} = \frac{1}{2} \rho - \Lambda \ .
\end{equation}
The Lagrangian picture proves to be very useful when we compare the equations of Newtonian cosmology with those of general relativity, which we introduce in section \ref{1+3cosmo}. The reason is that both in GR and in Lagrangian NG observers move with the fluid flow. In NG observers ``\textit{convect}" with the flow, for this reason the Lagrangian derivative is also know as \textit{convective derivative}. Similarly in GR observers can be chosen in a way that they ``\textit{comove}", for instance in an exactly homogeneous and isotropic universe this means that their coordinate system expands at the same rate as the geometry. 

A consequence of this is that in the two theories the volume element in general is not conserved in time, but rather it evolves according to the dynamics of the system. This is of crucial importance in section \ref{averageNG}, where averaging in Newtonian cosmology is discussed.

\paragraph*{Irreducible representation of the fluid equations.} 
The meaning of the equations (\ref{continuity2}) and (\ref{euler2}) can be made clear if we decompose the gradient of the velocity field $v ^i_{\phantom{i},j}$ into its irreducible parts with respect to the Galilean group, i.e.\ if we define:\\
\\
the trace part 
\begin{equation} \label{thetaNG}
\theta \doteq v ^k_{\phantom{k},k} \ ,
\end{equation}
the traceless symmetric part
\begin{equation}
\sigma _{ij} \doteq v _{(i,j)} - \frac{1}{3} \theta \delta _{ij} \ ,
\end{equation}
and the antisymmetric part 
\begin{equation} \label{omegaNG}
\omega _{ij} \doteq v _{[i,j]} \ .
\end{equation}
We can then write $v ^i_{\phantom{i},j}$ as
\begin{equation} \label{gradv}
v ^i_{\phantom{i},j} = \frac{1}{3} \delta ^i_{\phantom{i}j} \theta + \sigma ^i_{\phantom{i}j} + \omega ^i_{\phantom{i}j} \ .
\end{equation}
The tensor $\omega _{ij}$ has three degrees of freedom, since $\omega _{ij} = \omega _{[ij]}$, and it can be re-expressed in vector form $\omega ^i \doteq \frac{1}{2} \tilde{\epsilon} ^{ijk} \omega _{jk}$. On the other hand $\sigma ^i_j$ has five degrees of freedom, since $\sigma ^i_{\phantom{i}i} = 0$ and $\sigma _{ij} = \sigma _{(ij)}$. We can now difine ${\sigma}^2 = \frac{1}{2} \sigma _{ij} \sigma ^{ij}$ and ${\omega}^2 = \frac{1}{2} \omega _{ij} \omega ^{ij}$. The three quantities $\theta$, $\sigma _{ij}$ and $\omega _{ij}$ can be physically interpreted respectively as the rate of volume expansion, the rate of shear and the rate of vorticity (see Figure 3.2).
\begin{figure} [H]
\begin{center}
\includegraphics[scale=0.52]{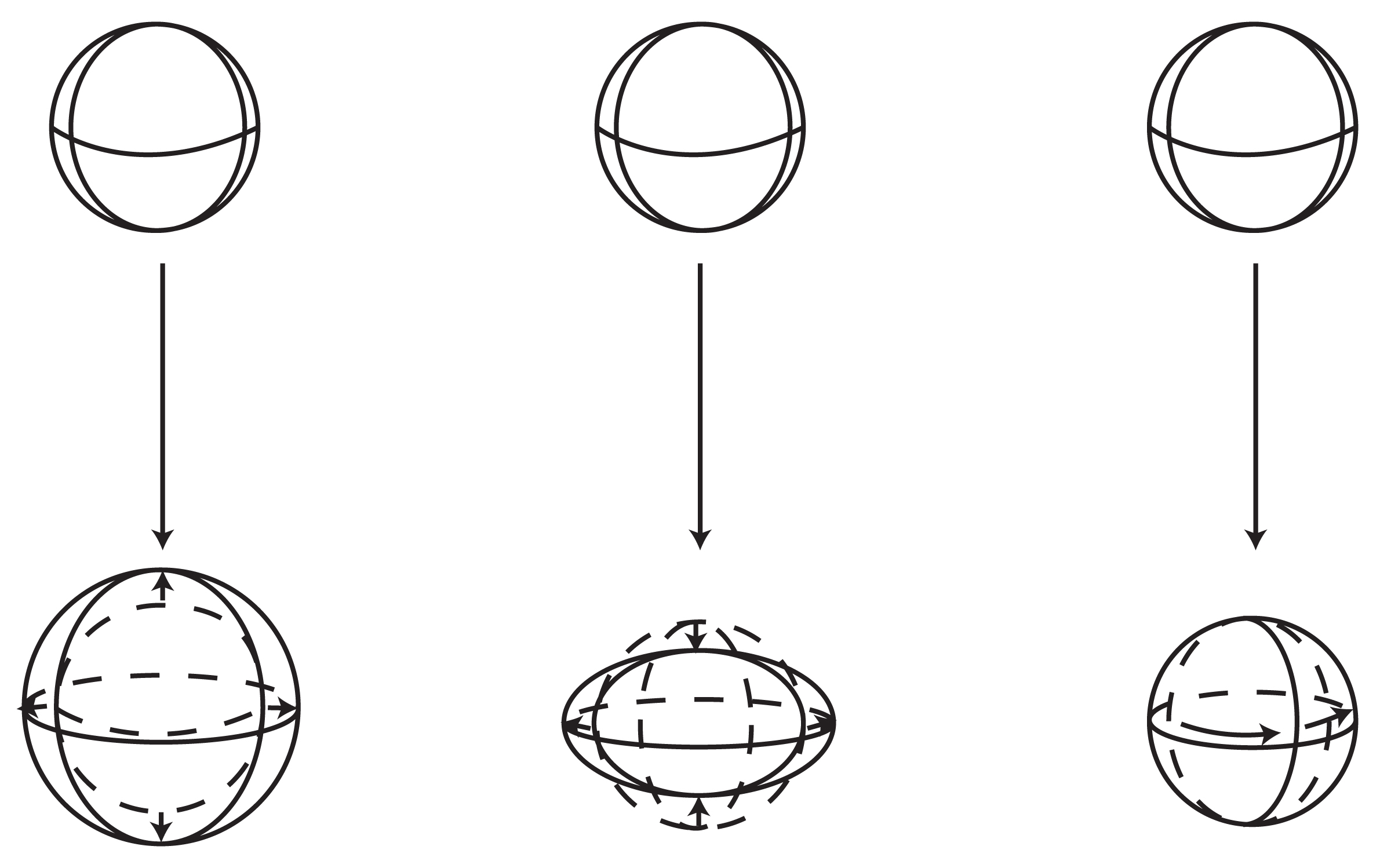}
\caption{\footnotesize If we consider a spherical fluid element, the effect of the expansion is to reduce or increase its radius (on the left). The shear squashes the sphere, for example in the simplest case into an ellipsoid (in the center), leaving its volume constant. The vorticity rotates it (on the right). [Picture credits G. F. R. Ellis.]}
\end{center}
\end{figure}

\paragraph*{Scalar invariants of the velocity's gradient.}
We can denote the second order tensor defined in the previous section as $\mathcal{V} \doteq v^i_{\phantom{i},j}$, and build from it some scalars which are coordinate independent. In the literature on tensor calculus \cite{NGlagrangian} three \textit{principal invariants} are defined, which in the case of the tensor $\mathcal{V}$, living on a flat space, read:
\begin{equation} \label{I}
I \doteq \mathrm{tr}(\mathcal{V}) = v^i_{\phantom{i},i} = \theta \ ,
\end{equation}

\begin{equation} \label{II}
\begin{split}
II \doteq \frac{1}{2}[(\mathrm{tr}(\mathcal{V})^2-\mathrm{tr}(\mathcal{V}^2)] &= \frac{1}{2}[({v^i_{\phantom{i},i}})^2 - v^i_{\phantom{i},j}v^j_{\phantom{j},i}] = \frac{1}{2} \overline{\nabla} \cdot (\mathbf{v} \overline{\nabla} \cdot \mathbf{v} - \mathbf{v} \cdot \overline{\nabla} \mathbf{v}) = \\ &= \omega^2 - \sigma ^2 + \frac{1}{3} \theta^2 \ ,
\end{split}
\end{equation}

\begin{equation} \label{III}
\begin{split}
III \doteq  \mathrm{det}(\mathcal{V}) &= \frac{1}{6}(v^i_{\phantom{i},i})^3 - \frac{1}{2}v^i_{\phantom{i},i}v^i_{\phantom{i},j}v^j_{\phantom{j},i} + \frac{1}{3}v^i_{\phantom{i},j}v^j_{\phantom{j},k}v^k_{\phantom{k},i} = \\ &= \frac{1}{9} \theta^3 + 2 \theta \left(\sigma^2 + \frac{1}{3} \omega^2 \right) + \sigma ^i_{\phantom{i}j} \sigma^j_{\phantom{j}k} \sigma^{k}_{\phantom{k}i} -\sigma^j_{\phantom{j}k} \omega_j \omega^k  \ .
\end{split}
\end{equation}
The fact that the second invariant (\ref{II}) can be written in terms of a divergence solely on a flat space is fundamental for our discussion. In fact in section \ref{woe2} we use (\ref{II}) to highlight an important property of the backreaction variable in NG.

When we move to curved spaces partial derivatives must be replaced by covariant derivatives (defined in Chapter \ref{gr}), which do not commute with each other. Thus equation (\ref{II}) does not have a simple form given in terms of a divergence in GR. We analyse the consequences that this has on the dynamics of the theory in section \ref{buchert}.

\paragraph*{Evolution equations for a Newtonian perfect fluid.}
The purpose of this section is to derive the time evolution equations for the expansion rate, the shear and the vorticity. We follow the references \cite{ellis}\cite{ellis2}. We rewrite the continuity equation (\ref{continuity2}) and the Euler equation (\ref{euler2}) in terms of $\theta$, $\sigma_{ij}$ and $\omega_{ij}$. This is useful since in GR it's possible to derive analogous equations, which makes the comparison between the theories  more transparent.


We begin by substituting (\ref{gradv}) into the continuity equation (\ref{continuity2}) and into the gradient of Euler equations (\ref{euler2}), which yields
\begin{equation} \label{contall}
\dot{\rho} + \rho \theta = 0 \ ,
\end{equation}
\begin{equation} \label{eulerall}
\frac{1}{3} \dot{\theta} \delta ^i_{\phantom{i}j} + \dot{\sigma} ^i_{\phantom{i}j} + \dot{\omega} ^i _{\phantom{i}j}  + \left(\frac{1}{3} \theta \delta ^k_{\phantom{k}j} + \omega ^k_{\phantom{k}j} + \sigma ^k_{\phantom{k}j}\right) \left(\frac{1}{3} \theta \delta ^i_{\phantom{i}k} + \sigma ^i_{\phantom{i}k} + \omega ^i_{\phantom{i}k}\right) = -  \phi ^{,i}_{\phantom{,i},j} +  a ^i_{\phantom{i},j} \ .
\end{equation}

The time evolution of $\theta$ is expressed by the \textit{Raychaudhuri equation} and can be obtained by applying the trace operator $\delta _i^{\phantom{i}j}$ to equation (\ref{eulerall}). It reads
\begin{equation} \label{thetadot}
\dot{\theta} + \frac{1}{3} {\theta}^2 = - \frac{1}{2} \rho  + 2 ({\omega}^2 + {\sigma}^2) + \Lambda    + a ^k_{\phantom{k},k} \ .
\end{equation}


The time evolution of $\omega^i$ is expressed by the so called \textit{vorticity equation} and can be obtained by applying the antisymmetrization operator $\tilde{\epsilon} _i^{\phantom{i}jk}$ to equation (\ref{eulerall}). It reads
\begin{equation} \label{omegadot}
\dot{\omega} ^i + \frac{2}{3} \theta \omega ^i - \sigma ^i_{\phantom{i}k} \omega ^k = - \frac{1}{2} \tilde{\epsilon} ^{ijk} a _{j,k} \ .
\end{equation}

Finally the time evolution of $\sigma_{ij}$ is expressed by the so called \textit{shear equation} and can be obtained lowering the $i$ index of (\ref{eulerall}) and then applying the trace removing symmetric operator $(\delta ^{\phantom{k}(i}_k \delta ^{\phantom{l}j)}_l - \frac{1}{3} \delta ^{ij} \delta _{kl})$. It reads
\begin{equation} \label{sigmadot}
\dot{\sigma} _{ij} + \frac{2}{3} \theta \sigma _{ij} + \sigma _i^{\phantom{i}k} \sigma _{kj} - \delta _{ij} \left(\frac{2}{3}\sigma ^2 \right)   =\delta _{ij}\left(\frac{1}{3}\omega ^2-\frac{1}{3} a ^k_{\phantom{k},k}\right) - \omega _i \omega _j  -E _{ij} +a _{(i,j)} \ ,
\end{equation}
where $E _{ij}$ is identified as the Newtonian \textit{electric part of the Weyl tensor} (this nomenclature is justified in (\ref{GRelectric}))
\begin{equation} \label{Enewton}
E _{ij} \doteq \phi _{,ij} - \frac{1}{3}  \phi _{,k}^{\phantom{,k},k} \delta _{ij} \ .
\end{equation}

Equations (\ref{contall}) and (\ref{thetadot})-(\ref{sigmadot}), together with the Poisson equation of gravity and an equation of state, mathematically define Newtonian cosmology.

It's worth remarking that, since NG does not yield to an evolution equation for the potential (there is no $\dot{\phi}(\mathbf{x},t)$ equation), it also leaves undetermined the behaviour in time of $E _{ij}$. As we outline in Chapter \ref{1+3form}, this represents an important difference with respect to GR, where an evolution equation for the electric part of the Weyl tensor exists and it's coupled to another quantity, absent in NG, called the \textit{magnetic part of the Weyl tensor}.

\paragraph*{Constraint equations for a Newtonian perfect fluid.}
We conclude this section by giving the constraint equations of Newtonian cosmology (without derivation, see \cite{ellis}\cite{ellis2} for references).  These equations are said to be constraints as they do not involve time derivatives of the dynamical quantities. They prove to be useful in Chapter \ref{cosmo}, when we write down the corresponding GR equations and we perform a comparison.

From the first Newtonian Ricci-like identity\footnote{This terminology is justified in Chapter \ref{1+3form}, where we show that these identities are related to the Ricci identities of GR.}
\begin{equation} \label{riccilike1}
{\partial \over \partial t} (v _{i,j}) = \left({\partial v _i \over \partial t}\right) _{,j} \ ,
\end{equation}
we obtain
\begin{equation} \label{constrE1}
E ^{ik}_{\phantom{ik},k} = \frac{1}{3} \rho ^{,i} \ ,
\end{equation}
\begin{equation} \label{constrE2}
E ^{(i}_{\phantom{(i}k,h} \tilde{\epsilon} ^{j)kh} = 0 \ .
\end{equation}
Equations (\ref{constrE1})-(\ref{constrE2} are important as they show that the dependency on space coordinates of the electric part of the Weyl tensor in NG is constrained. Equation (\ref{constrE1}) moreover suggests that choosing a matter model with constant density, i.e.\ a model which is exactly homogeneous and isotropic, implies that the electric part of the Weyl tensor has the trivial behaviour \begin{equation} \label{constrE3}
E ^{ik}_{\phantom{ik},k} = 0 \ .
\end{equation}

From the second Newtonian Ricci-like identity
\begin{equation} \label{riccilike2}
v _{i,jk} = v _{i,kj} \ ,
\end{equation}
we obtain
\begin{equation} \label{NGconstr1}
\omega ^{ik}_{\phantom{ik},k} - \sigma ^{ik}_{\phantom{ik},k} +\frac{2}{3} \theta ^{,i} = 0 \ ,
\end{equation}
\begin{equation} \label{NGconstr2}
\omega ^k_{\phantom{k},k} = 0 \ ,
\end{equation} 
\begin{equation} \label{NGconstr3}
(\omega _{(i}^{\phantom{(i}k,h} + \sigma _{(i}^{\phantom{(i}k,h}) \tilde{\epsilon} _{j)kh} = 0 \ ,
\end{equation}
\begin{equation} \label{constr3}
\sigma _{[h \phantom{[j},k]}^{\phantom{[h}[j \phantom{,k},i]} + \frac{2}{3} \delta _{\phantom{[j}[h}^{[j} \theta _{\phantom{,i},k]}^{,i]} = 0 \ .
\end{equation}

\section{Averaging in Newtonian cosmology and backreaction} \label{averageNG}

\subsection*{Why averaging?}
In this section we discuss averaging in Newtonian cosmology. The reason why we need averaged equations is that the observable universe encodes a large number of degrees of freedom\footnote{The effective number of degrees of freedom in cosmology changes according to the desired level of approximation: it can span from $\mathcal{O}(10^{11})$, i.e. number of galaxies and related dark matter halos in the observable universe at late times, to the number of quantum degrees of freedom if we go down to atomic scale.}, while typically the degrees of freedom in cosmological models are turned down to $\mathcal{O}(10)$ parameters. Cosmological models supply therefore an averaged (or statistical) description for the behaviour of the fluid filling the universe. 
If this fluid is exactly homogeneous and isotropic then the value of physical quantities (take the matter's density field for instance) will be the same from point to point. On the contrary, if the fluid is clumpy, then there are local fluctuations in the density field, and a coarse grained description is needed, which is obtained via averaging. In this section we follow the reference \cite{NGboundary}.

\subsection*{Definition of averaging}
Given a tensor field $\mathcal{A}(x^i,t)$, its spatial average over the domain $\mathcal{D}(t)$ is defined, in the Lagrangian picture, as
\begin{equation}
\langle \mathcal{A} \rangle \doteq \frac{\int _{\mathcal{D}} \mathcal{A} \, \tilde{\epsilon}}{\int _{\mathcal{D}} \, \tilde{\epsilon}}  \ ,
\end{equation}
where
\begin{equation} \label{NGvol}
\int \limits_{\mathcal{D}} \tilde{\epsilon} = \int \limits_{\mathcal{D}} d^3x = V(t)
\end{equation}
is the volume of $\mathcal{D}$.
We define the \textit{scale factor} in NG to be the cubic root of the volume $V(t)$, normalised with the initial time volume $V(t_0)$. It reads
\begin{equation} \label{NGa}
a_{\mathcal{D}} \doteq \left(\frac{\int _{\mathcal{D}(t)} \tilde{\epsilon}}{\int _{\mathcal{D}(t_0)} \tilde{\epsilon}} \right)^{\frac{1}{3}} \ .
\end{equation}
An important property of the Lagrangian picture is that the volume $V$ of the domain is a function of time (as illustrated in Figure 3.1). Its total variation in time is
\begin{equation} \label{dotv}
\dot{V} = \frac{d}{dt} \int \limits_{\mathcal{D}} \, d^3x = \int \limits_{\mathcal{D}(t_0)} \dot{J} \, d^3q = \int \limits_{\mathcal{D}} \theta \, d^3x \ ,
\end{equation}
where we have used $\dot{J}= \theta J$, which was given in equation (\ref{Jevol}).

In the Eulerian picture $\dot{V}$ is zero, since the domain is fixed at the initial time
\begin{equation}
V(t) = V(t_0) =  \int \limits_{\mathcal{D}(t_0)} d^3x \ .
\end{equation}
The fact that $\dot{V}$ is in general not zero in the Lagrangian picture is useful to our purpose of comparing NG to GR, where the situation is similar. This is outlined in Chapter \ref{1+3form}.
From the above discussion and from the definition (\ref{dotv}) we derive the commutation rule
\begin{equation} \label{commute}
{\langle \mathcal{A} \rangle} \dot{} - {\langle \dot{\mathcal{A}} \rangle } = {\langle \mathcal{A} \theta \rangle } - {\langle \mathcal{A} \rangle } {\langle \theta \rangle} \ ,
\end{equation}
which states that the evolution of the average and the average of the evolved field do not commute in general.
From (\ref{NGa}) and (\ref{dotv}) we also deduce that the average of the expansion $\theta (\mathbf{x},t)$ is given, in terms of the scale factor, by
\begin{equation}
\langle \theta  \rangle = \frac{\dot{V}}{V} = 3 \frac{\dot{a}_{\mathcal{D}}}{a_{\mathcal{D}}} \ .
\end{equation}
If we average Raychaudhuri equation (\ref{thetadot}) and then we identify $\langle \frac{\theta}{3} \rangle$ as the \textit{Hubble parameter}
\begin{equation} \label{Ha}
H  \doteq \frac{\dot{a}_{\mathcal{D}}}{a_{\mathcal{D}}} = \frac{1}{3} \langle\theta \rangle
\end{equation}
we obtain the \textit{averaged Raychaudhuri equation} for NG \cite{NGlagrangian}
\begin{equation} \label{secondfriedmann}
3\frac{\ddot{a}_\mathcal{D}}{a_\mathcal{D}} = - \frac{1}{2}  \langle \rho \rangle +  \frac{2}{3} (\langle \theta ^2 \rangle - {\langle \theta \rangle}^2) + 2 \langle {\omega}^2 - {\sigma}^2 \rangle + \Lambda + \frac{1}{3} \langle a ^k_{\phantom{k},k} \rangle \ ,
\end{equation}
where the average of the density field is $\langle \rho \rangle = M V_0^{-1} a_\mathcal{D}^{-3}$, and $V_0$ is the Eulerian volume of the integration domain.
We lump together the terms responsible for inhomogeneity and anisotropy and define the new quantity as the \textit{backreaction variable} 
\begin{equation} \label{backreaction}
\mathcal{Q} \doteq \frac{2}{3} (\langle \theta ^2 \rangle - {\langle \theta \rangle}^2) + 2 \langle {\omega}^2 - {\sigma}^2 \rangle \ .
\end{equation}
In addition, averaging (\ref{continuity2}) gives the \textit{averaged continuity equation}
\begin{equation} \label{contave}
\langle \dot{\rho} \rangle + \langle \rho \theta \rangle = \langle {\rho} \rangle \dot{} + \langle \rho \rangle \langle \theta \rangle = 0 \ .
\end{equation}
From equation (\ref{secondfriedmann}) we see that $\mathcal{Q}$, like $\Lambda$, contributes to the cosmic acceleration. We delve into this aspect in section \ref{woe2}, explaining what's the contribution that $\mathcal{Q}$ brings to the expansion law in Newtonian inhomogeneous cosmologies. Before going into it we discuss two issues which arise when NG is applied to extended systems.

\section{The sorrows of Newtonian cosmology} \label{woes}

\subsubsection*{A matter of context}
In this section we describe the pathologies breaking in when Newtonian gravity is applied to cosmology, i.e.\ to a system described by an extended matter model. For a review on this see \cite{norton}. The fact that NG is unsuitable for studying unbound systems is not surprising since NG is a theory of isolated systems only. The sorrows we are about to describe do not represent intrinsic weaknesses of the theory, they rise indeed just because NG is not applied in the right context.

In the first part of this section we comment on the convergence of Newton's law of gravity when this is applied to an extended system. 

In the second part we discuss the form of the boundary term appearing in the equation for the Newtonian gravitational potential (\ref{phi}) in cosmology, and we comment on the consequences that this has for the physics of structure formation studied with numerical $N$-body simulations.

\subsection{Sorrow 1: Convergence of the inverse square law} \label{woe1}
\paragraph*{Introduction.}
In the first part of this section, 
following the reference \cite{norton}, we discuss the convergence of Newton's law in cosmology. In the second part, 
following \cite{convergence}, we present some results on how a system of particles, which occupies an infinite volume in $\mathbb{R}^n$ and has constant particle density, can be classified according to the convergence properties of the probability distribution function $P_{N}(\mathbf{F})$ of a pair force $|\mathbf{F}(r)| \sim r^{-\gamma +1}$ acting between $N$ particles.

\paragraph*{Force summation.}
Take an infinite universe filled with a medium of constant density. Take an observer in some point $\overline{O}$ which measures the gravitational force exerted by the mass distribution along some axis $\overline{O}-\overline{O}'$. Split the mass distribution around the observer in spherical shells of very small thickness $\Delta l$ and divide each shell in two hemi-shells defined with respect to the plane orthogonal to the axis $\overline{O}-\overline{O}'$. The infinitesimal force per unit mass in the direction $\overline{O}-\overline{O}'$, exerted by each hemi-shell with mass $M$ located at the distance $r$, is
\begin{equation}
dF = \frac{1}{8 \pi} \cos \theta \frac{dM}{r^2} \ ,
\end{equation}
where $\cos \theta$ projects along the line of sight and the angle $\theta$ is defined by identifying the $\overline{O}-\overline{O}'$ axis to be  the  $\hat{z}$ axis in usual polar coordinates. In a similar way we can define the azimuthal angle $\phi$.
The total force exerted by the hemi-shell is then
\begin{equation}
F=\frac{\rho}{8 \pi r^2} \int^{2 \pi}_0 \int^{\frac{\pi}{2}}_0 \int^{r + \Delta l}_{r} r'^2  \sin \theta \cos \theta \, \, d{r'} d\theta d\phi = \frac{1}{8} \rho \Delta l \ ,\footnote{We have used ${\Delta r}^2 = {\Delta r}^3 = 0$ .}
\end{equation}
which is independent of the distance $r$. Thus the net force measure by the observer along the line of sight $\overline{O}-\overline{O}'$ is given by an infinite series, where each term represents the force due to one hemispherical shell
\begin{equation} \label{F}
F = \frac{1}{8}(\rho \Delta l - \rho \Delta l + \rho \Delta l - \rho \Delta l + ...) = \frac{1}{8}{\rho \Delta l} \sum \limits _{n=0}^{\infty} (-1)^n  \ .
\end{equation}
The series has alternating sign since shells on opposite sides exert a force in a opposite direction. The series which appears in (\ref{F}) is known as \textit{Grandi's series}, and it diverges, meaning that the sequence of partial sums does not approach any number, although it has two accomulation points in $0$ and $1$. This is clear if we notice that the value of (\ref{F}) changes based on how we order the sum. For instance $(\rho \Delta l - \rho \Delta l) + (\rho \Delta l - \rho \Delta l) + ... = 0$, but $\rho \Delta l + (- \rho \Delta l + \rho \Delta l) + (- \rho \Delta l + \rho \Delta l) + ... =  \rho \Delta l$.

The Newtonian force in cosmology depends on how we approach the limit to infinity, therefore is not well defined and suffers from a boundary issue. This arbitrariness in the summation order reflects the arbitrariness that we have in the potential picture of Newtonian cosmology in choosing the boundary term in (\ref{phi}).

\paragraph*{Force classification.}
The probability distribution function of a force $\mathbf{F}_j$ exerted by a gas of $N$ particles on a  test particle labelled with $j$ is defined as
\begin{equation} \label{Pdistr}
P _{N}(\mathbf{F}_j) = \int \limits_{\mathcal{D}} \left[\prod_{i=1}^{N} d^n x_i\right] \mathcal{P}_{N}(\{\mathbf{x}_i \}) \delta \left[\mathbf{F}_j +\sum_i \mathbf{f}(\mathbf{x}_i) \right] \ .
\end{equation}
Here $\mathcal{P}_{N}(\{\mathbf{x}_i \})$ is the probability density of having $N$ particles in the realisation $\{\mathbf{x}_i \}$, $n$ is the number of spatial dimensions ($n = 3$ in the usual case of NG), $\delta$ is the Dirac delta function\footnote{From a mathematical viewpoint the Dirac delta function it's not strictly a function, but rather a distribution.} and $\mathbf{f}(\mathbf{x}_i)$ is the $ith$ term in the sum (\ref{newtonforcenum}) with $\mathbf{x}_j = 0$.

For a pair interaction potential $V({r})$ with $V(r \rightarrow \infty)  \sim r^{-\gamma}$ it can be shown that $P_{N}(\mathbf{F}_j)$ converges continuously to a well-defined and rapidly decreasing probability distribution function if and only if the gradient of the pair force is absolutely integrable, i.e.\ for $\gamma > n-2$. We consider the integrability of the gradient of (\ref{Pdistr}) and not (\ref{Pdistr}) itself since in an infinite system without any preferred point the physical meaningful quantity is the relative position of the particles. Moreover in cosmology we are interested in studying perturbations in the density field, which can be shown to be proportional to the gradient of the force, which in this sense is the observable.

We refer to the case $\gamma > n-2$ as \textit{dynamically short range}, since the dominant contribution to the force on a test particle in this limit arises from particles in a finite neighbourhood around it. Here the gradient of the force converges.

In the case of $\gamma \leq n-2$, on the other hand, $P_{N}(\mathbf{F})$ does not converge in general. We can refer to this case as \textit{dynamically long range}, since there is a contribution to the force in this limit which arises from particles infinitely far away from a test particle. 

The case of gravity, $\gamma = n-2$, is the borderline case which yields to a divergent force in this setting. This is in agreement with the previous paragraph where it was shown that the Newtonian force takes contribution from particles infinitely far away. In the next section we present a modification of Newton's law which regularises it in the cosmological limit.

\paragraph*{Regularization of the Newtonian force.}
The pair force can be defined in a weaker sense, in the case of NG, by introducing a regularization of the force summation
\begin{equation}
\mathbf{F} = \lim_{\mu \to 0^{+}} \lim_{\mathcal{D} \to \infty} \frac{m}{8 \pi} \int \limits_{\mathcal{D}} \frac{\mathbf{x'}-\mathbf{x}}{\left | \mathbf{x'}-\mathbf{x}  \right |^3} \rho(\mathbf{x'},t) e^{-\mu \left | \mathbf{x'}-\mathbf{x}  \right |} \, d^3\mathbf{x'} \ .
\end{equation}
In other words $\mathbf{F}$ is defined as a screened version of \ref{newtonforce} (in the limit of an infinite domain of integration) with screening coefficient $\mu$ taken to zero. Thanks to this trick, this regularised version of Newtonian gravity gives rise to a convergent force and becomes a useful tool for cosmology. Note that this solution to the divergence problem introduces a modification in the gravitational law, that rather than purely Newtonian it becomes nearly Newtonian. 

However this is not the solution adopted in numerical cosmology. In this context rather then regularizing the gravitational law what is done is to compactify the matter model. We come back to this in the following section.

\subsection{Sorrow 2: Ill-posedness of Cauchy's initial value problem and backreaction} \label{woe2}
\paragraph*{Introduction.}
The aim of this section is to show transparently that Newtonian cosmology, in any setting, is undefined unless we give arbitrary boundary conditions. We show how these are chosen in numerical cosmology and what consequences this has for the underlying physics. In this context we introduce the backreaction conjecture. 

\paragraph*{The problem.}
Starting from the potential picture, we rewrite equation (\ref{phi})
\begin{equation}
\begin{split}
&\phi(\mathbf{x},t) = \frac{1}{2} \int \limits_{\mathcal{D}} G_f(\mathbf{x},\mathbf{x'}) \rho(\mathbf{x'},t)\, d^3\mathbf{x'} + \\+ \int \limits_{\partial {\mathcal{D}}} [\phi(\mathbf{x'},t) & \nabla ' G_f(\mathbf{x},\mathbf{x'}) - G_f(\mathbf{x},\mathbf{x'})' \nabla ' \phi(\mathbf{x'},t)] \cdot d\mathbf{S}' \ ,
\end{split}
\end{equation}
which shows that the gravitational potential is made up of a volume integral plus a surface integral. 
In astrophysical settings, i.e.\ when studying isolated systems such as galaxies and clusters of galaxies, the boundary term is taken to go to zero at infinity. This is a reasonable assumption, since in these systems the source of the gravitational potential, the matter, is concentrated in an isolated and asymptotically flat region, surrounded approximatively by vacuum\footnote{Galaxies have a typical size of the order of $\mathcal{O}(1-100) \, {h}^{-1} \mathrm{kpc}$ and are usually separated by distances of $\mathcal{O}(1) \, {h}^{-1} \mathrm{Mpc}$. For clusters this numbers become respectively $\mathcal{O} (2-10) \, {h}^{-1} \mathrm{Mpc}$ for the size and over $\mathcal{O}(10) \, {h}^{-1} \mathrm{Mpc}$ for the separation. The factor ${h}$ reflects our ignorance about the value of the Hubble parameter today. This is given as $H_0 = 100 \, {h} \mathrm{km s^{-1} Mpc^{-1}}$.}. This is equivalent, in the force picture, to neglecting the contribution from particles at large radii.
\begin{figure} [h!]
\begin{center}
\includegraphics[scale=0.148]{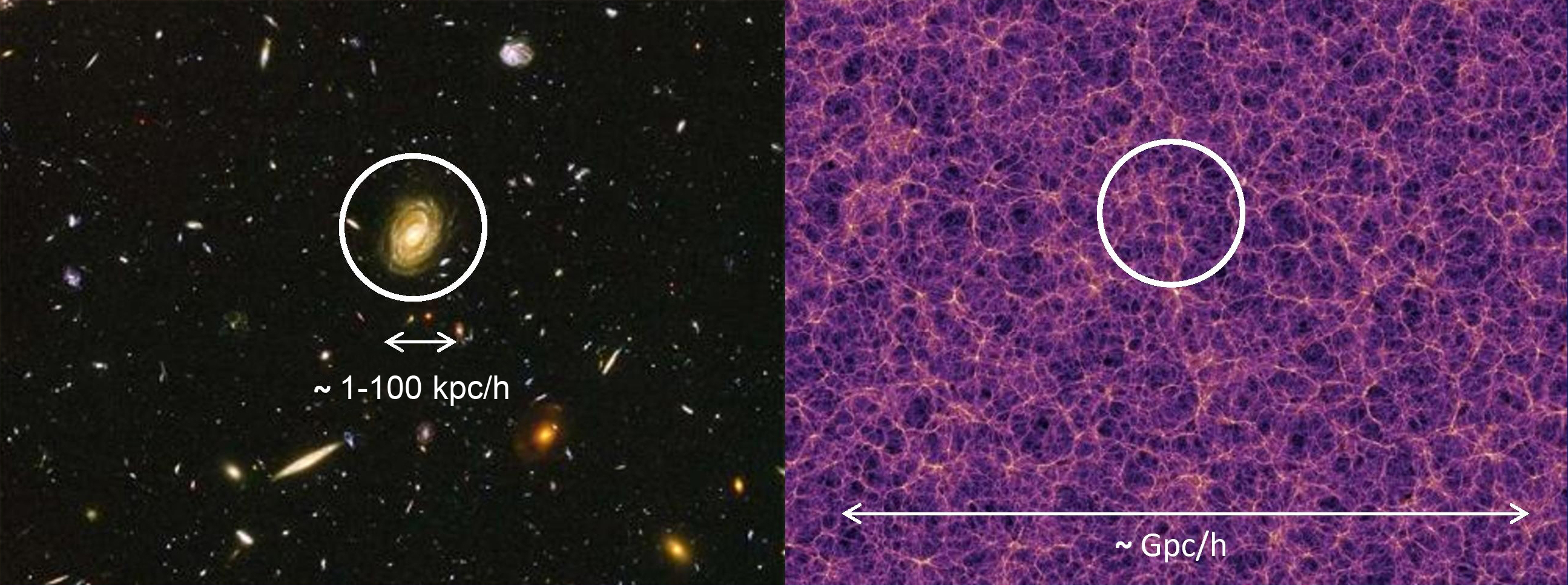}
\caption{\footnotesize On the left, from the Hubble ultra deep field 2003-2004: in astrophysics we are able to isolate single structures. On the right, from the Millennium simulation: in cosmology structures form an extended web and there are no isolated regions. [Picture credits the Hubble deep field project and the Millennium simulation project.]}
\end{center}
\end{figure}

When we are doing cosmology, however, the contribution from the boundary cannot be neglected. The large scale structures of the universe are not organised in compact and isolated island-like systems, but rather they form an extended cosmic web (see Figure 3.3). This is assumed to be \textit{statistically homogeneous} (symmetric under spatial translations) and \textit{isotropic} (symmetric under spatial rotations). An intuitive definition of these two concepts was given in the introduction.

As seen in section \ref{woe1}, statistical homogeneity is enough\footnote{Unless we take into account e.g.\ the possibility that that the matter distribution in the universe is a fractal. For a primer on fractal cosmology see \cite{statcosmo}.}, together with the assumption of an inverse square law of gravity and an extended matter model, to assure the Newtonian force to become long ranged and ill-defined, preventing us from discarding any contribution coming from the boundary. 

The boundary term is  arbitrary in cosmology, and it must be specified for all times: the system of equations (\ref{continuity})-(\ref{poisson}) doesn't provide indeed any information about the time evolution of the potential, therefore a boundary term needs to be given not only for the Poisson equation, but also for all its time derivatives. If we work in the Lagrangian fluid picture of Newtonian gravity then the arbitrariness is shifted from the gravitational potential to the quantity we called the Newtonian electric part of the Weyl tensor, which appears in equation (\ref{sigmadot}). The electric part of the Weyl tensor is however only arbitrary in time, as its spatial derivatives satisfy the constraint equations of NG, given in section \ref{lagr}. This arbitrariness of $E_{ij}$ in NG must be kept in mind when in section (\ref{1+3cosmo}) we discuss the initial value problem of GR.

An explanation for the indeterminacy of the potential picture of Newtonian gravity comes from the theory of partial differential equations \cite{mathmet}: the fact that the equation for the potential (\ref{phi}) was derived from the Poisson equation of gravity (\ref{poisson}), which is an \textit{elliptic equation}. In order to solve its initial value problem, it's necessary to provide an initial condition curve which crosses at least in one point each of the characteristic curves of the differential equation (along which the partial differential equation becomes an ordinary differential equation). In any elliptic equation the curves of constant time are characteristics themselves, thus initial value data cannot be given along such a curve.

\paragraph*{A solution.}
The problem of the indeterminacy of the boundary term in the Newtonian potential has a well established and prosaic solution: in the force picture this corresponds to taking the integral of (\ref{newtonforcenum}) over a compact domain and then mapping together the boundaries of the region. What results from this is a periodic universe, which shares with the real universe the property of being boundary-less\footnote{We don't know whether the universe is infinite or if it is periodic. From observations we can infer a lower bound for the scale of periodicity, which e.g.\ for the topology chosen in \cite{topology} is $24 \, \mathrm{Gpc}$.}. For instance in numerical cosmology this is done running the simulations on a three-torus $\mathbb{T}^3$, meaning that the particles are put in a box and opposite sides of the box are matched together (as shown in Figure 3.4).
\begin{figure} [h!]
\begin{center}
\includegraphics[scale=1.9]{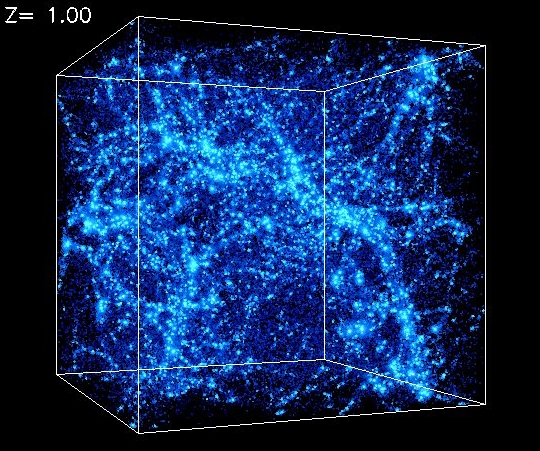}
\caption{\footnotesize A numerical simulation run in a box of size $43 \, {h}^{-1} \mathrm{Mpc}$ with matched sides. This simulation reproduces the history of structure formation between redshift $z \sim 27$ and $z \sim 0$ (today). Here an intermediate snapshot at $z = 1$ is shown. [Picture credits Andrey Kravtsov and Anatoly Klypin]}
\end{center}
\end{figure}
In the potential picture this corresponds to setting the boundary term in equation (\ref{phi}) to be a periodic function. This solution however is unsatisfactory if we want to use Newtonian gravity to study inhomogeneous cosmologies. We explain this aspect in the next section.


\paragraph*{Vanishing of the backreaction on a three-torus.} 
In the introduction of this thesis we described the backreaction variable as a quantity which parametrises the effect that the inhomogeneity in the matter distribution has on the expansion law of the universe. In this section we show which form it takes in the context of NG with periodic boundary conditions. We follow the references \cite{NGlagrangian}\cite{NGboundary}.

We start by taking the gradient of the velocity (\ref{gradv}) and we split it into a sum of a Hubble flow $H(t)$ plus a peculiar velocity gradient $u _{i,j}$, as
\begin{equation} \label{splitting}
v _{i,j} \doteq H(t) \delta _{i j} + u _{i,j} \ .
\end{equation}
This amounts for the fact that the observed recession velocity of an object, take a galaxy, is the sum of the velocity due to the cosmic expansion plus the proper motion of the object due to local gravitational sources.

By averaging (\ref{splitting}) and setting $\langle u_k^{\phantom{k},k} \rangle _{\mathcal{D}} = 0$ we obtain 
\begin{equation} \label{Hu}
H \doteq \frac{\dot{a}_\mathcal{D}}{a_{\mathcal{D}}}=\frac{1}{3} \langle \theta \rangle \ ,
\end{equation}
which is the same as (\ref{Ha}).
We can now average the second invariant of the velocity's gradient (\ref{II}) and work it out as follows: we substitute (\ref{splitting}) in its divergence term, we substitute $\langle \theta^2 \rangle = \langle \theta \rangle\dot{} - \langle \dot{\theta} \rangle + \langle \theta \rangle^2$ which follows from the commutation rule (\ref{commute}), and we insert in this latter equation the evolution equation for the expansion scalar (\ref{thetadot}) and equation (\ref{Hu}). 
What comes out, after exploiting Gauss' theorem to turn the volume integral of the divergence into a surface integral, is 
\begin{equation} \label{comparing}
3\frac{\ddot{a}_\mathcal{D}}{a_\mathcal{D}}= - \frac{1}{2}  \langle \rho  \rangle + a_{\mathcal{D}}^{-3} \int \limits_{\partial \mathcal{D}} (\mathbf{u} \nabla \cdot \mathbf{u} - \mathbf{u} \cdot \nabla \mathbf{u}) \cdot d\mathbf{S} + \Lambda \ .
\end{equation}
Equation (\ref{comparing}) compared to (\ref{secondfriedmann}) in the case $\langle a ^k_{\phantom{k},k} \rangle = 0$ gives the result 
\begin{equation} \label{backreactionint}
\mathcal{Q} = \frac{2}{3} (\langle \theta ^2 \rangle - {\langle \theta \rangle}^2) + 2 \langle {\omega}^2 - {\sigma}^2 \rangle = a_{\mathcal{D}}^{-3} \int \limits_{\partial \mathcal{D}} (\mathbf{u} \nabla \cdot \mathbf{u} - \mathbf{u} \cdot \nabla \mathbf{u}) \cdot d\mathbf{S}  \ .
\end{equation}
The backreaction variable in NG is therefore given in terms of an integral of a function of the peculiar velocity taken on the boundary of the domain $\mathcal{D}$.

If we solve the integral in equation (\ref{backreactionint}) for a spatially compact and periodic universe such as the toroidal one used in numerical simulations, or in another picture we impose periodic boundary conditions, the backreaction variable $\mathcal{Q}$ vanishes identically
This is equivalent with saying that in the periodic Newtonian cosmology the variance term $\frac{2}{3} \langle \theta ^2 \rangle - {\langle \theta \rangle}^2$ cancels identically with the term involving vorticity and shear $2 \langle {\omega}^2 - {\sigma}^2 \rangle$.
This implies that in this periodic topology, inhomogeneities and anisotropies (such as clusters, voids and filaments) do not play any role in the average expansion of the universe, and in particular they cannot contribute to its acceleration through the averaged Raychaudhuri equation (\ref{secondfriedmann}). 

We know observationally that the universe has been undergoing acceleration ever since $z \sim 0.5$ \cite{acc}\cite{acc2}. Since the backreaction is always zero in Newtonian cosmology on a torus it is not possible today to use numerical simulations to quantify the contribution that backreaction brings to the observed acceleration.

It's anyway hard to tell if this lack of Newtonian cosmology is important or not, since it's hard to estimate the magnitude of the backreaction. As we mentioned earlier, the backreaction is expected to become more prominent as the universe becomes more inhomogeneous, and dominated by rapidly expanding and contracting regions, increasing the variance in $\mathcal{Q}$. Note that this is what we observe in the real universe: large voids are expanding faster than the average expansion, and they are taking over the whole volume, enlarging the variance $\langle \theta ^2 \rangle - {\langle \theta \rangle}^2$ as time goes by. 

In the case of general relativity, where space is curved (therefore (\ref{II}) doesn't hold) and where a nearly identical version of the Raychaudhuri equation holds, the backreaction doesn't vanish in a periodic topology. In fact in GR $\mathcal{Q}$ doesn't reduce to a boundary term, but backreaction is a volume phenomenon. Unfortunately GR is numerically intractable in cosmology\footnote{However numerical relativity has been applied successfully to the study of binary systems of black holes and neutron stars.}, therefore a purely numerical GR approach to the backreaction problem must be excluded.
In Chapter \ref{cosmo} we ask whether we can study backreaction in the context of general relativity linearised around a Newtonian background, where relativistic effects are added on top of Newtonian ones.

\chapter{General relativity} \label{gr}
\subsubsection*{Introduction}
In this chapter we briefly portray the theory of general relativity \cite{wald}. First we introduce the geometrical background of the theory, then we give its dynamical equations and classify its symmetries. In the last part we show that, unlike Newtonian gravity, GR has a well-posed initial value problem \cite{weinberg}.

\section{The geometry} \label{geomgr}
\subsection{Geodesic equation and covariant derivative}
The fundamental object for the geometry of GR is the symmetric \textit{metric tensor} $g_{\alpha \beta}$. This determines uniquely the concept of distance on a manifold, given by the line element $ds^2 = g_{\alpha \beta} dx^{\alpha} dx^{\beta}$ It also defines the notion of ``straightest possible path", through the \textit{geodesic equation}
\begin{equation} \label{geodesic}
\frac{d^2 x^{\alpha}}{d s ^2} + \Gamma ^{\alpha}_{\phantom{\alpha} \beta \gamma} \frac{d x^{\beta}}{d s} \frac{d x^{\gamma}}{d s} = \frac{d u^{\alpha}}{d \tau} + \Gamma ^{\alpha}_{\phantom{\alpha} \beta \gamma} u^{\beta} u^{\gamma}  = 0  \ .
\end{equation}
In the first term $s$ is the affine parameter defined as $s =a \tau + b$,  $a$ and $b$ are constants, and $\tau$ is the proper time. In the second term we have chosen $s = \tau$ so that $u^{\alpha} = \frac{d x^{\alpha}}{d\tau}$ is a four-velocity vector. 
Here $\Gamma^{\gamma}_{\phantom{\gamma} \alpha \beta}$ is a \textit{Christoffel symbol}, or connection coefficient for the covariant derivative, which reads
\begin{equation} \label{christoffel}
\Gamma ^{\gamma}_{\phantom{\gamma} \alpha \beta} = \frac{1}{2} g^{\gamma \delta} (g_{\delta \alpha , \beta}+g_{\delta \beta , \alpha}-g_{\alpha \beta , \delta}) \ .
\end{equation}
Equation (\ref{geodesic}) is also known as the parallel-transport equation, and the vector $u^{\alpha}$ is said to be parallel transported if it satisfies it. 

The fundamental derivation operation in GR is called \textit{covariant derivative} and it's defined, for a generic tensor $\mathcal{M}$ with components $M^{\alpha_1 ... \alpha_n}_{\phantom{\alpha_1 ... \alpha_n} \beta_1 ... \beta_m}$, as
\begin{equation}
\begin{split}
&\nabla_{\gamma} M^{\alpha_1 ... \alpha_n}_{\phantom{\alpha_1 ... \alpha_n} \beta_1 ... \beta_m} = M^{\alpha_1 ... \alpha_n}_{\phantom{\alpha_1 ... \alpha_n} \beta_1 ... \beta_m , \gamma}  + \Gamma^{\alpha_1}_{\phantom{\alpha_1} \delta \gamma} M^{\delta ... \alpha_n}_{\phantom{\delta ... \alpha_n} \beta_1 ... \beta_m} + ... \\ & + \Gamma^{\alpha_n}_{\phantom{\alpha_n} \delta \gamma} M^{\alpha_1 ...\alpha_{n-1} \delta}_{\phantom{\alpha_1 ...\alpha_{n-1} \delta} \beta_1 ... \beta_m} - \Gamma^{\delta}_{\phantom{\delta} \beta_1 \gamma} M^{\alpha_1 ...\alpha_n}_{\phantom{\alpha_1 ...\alpha_n} \delta ... \beta_m} - \Gamma^{\delta}_{\phantom{\delta} \beta_m \gamma} M^{\alpha_1 ...\alpha_n}_{\phantom{\alpha_1 ...\alpha_n} \beta_1  ... \beta_{m-1} \delta} \ .
\end{split}
\end{equation}
The formula (\ref{christoffel}) holds only if we assume the connection to be \textit{torsion-free}, i.e\ $\Gamma^{\gamma}_{\phantom{\gamma} \alpha \beta} = \Gamma^{\gamma}_{\phantom{\gamma} (\alpha \beta)}$ and the metric to satisfy  $\nabla_{\gamma} g_{\alpha \beta} = 0$, which is known as the \textit{metric compatibility} condition.

\subsection{The Riemann tensor}
The second important geometric object of GR, the \textit{Riemann tensor}, can be constructed from the first and second derivatives of the metric tensor as
\begin{equation}
R^{\alpha}_{\phantom{\alpha} \beta \gamma \delta} = \Gamma^{\alpha}_{\phantom{\alpha} \delta \beta , \gamma} - \Gamma^{\alpha}_{\phantom{\alpha} \gamma \beta , \delta} + \Gamma^{\alpha}_{\phantom{\alpha} \gamma \epsilon} \Gamma^{\epsilon}_{\phantom{\epsilon} \delta \beta} -\Gamma^{\alpha}_{\phantom{\alpha} \delta \epsilon} \Gamma^{\epsilon}_{\phantom{\epsilon} \gamma \beta} \ .
\end{equation}
The Riemann tensor encodes the information on the curvature of the spacetime. To understand this think of a vector with components $V^{\alpha}$, parallel transport it around a loop, following the parallelogram formed by two vectors with components $A^{\beta}$ and $B^{\gamma}$. Its infinitesimal variation after one loop is given by
\begin{equation}
\delta V^{\alpha} = R^ {\alpha}_{\phantom{\alpha} \epsilon \beta \gamma} V ^{\epsilon} A^{\beta} B^{\gamma} \ .
\end{equation}
The variation depends on our choice of $A^{\beta}$ and $B^{\gamma}$, thus it's path dependent.

The Riemann tensor has twenty degrees of freedom, ten of which describe the curvature of spacetime locally decoupled from matter sources, through the \textit{Weyl tensor} $C_{\alpha \beta \gamma \delta}$, and ten which describe the curvature due to the presence of energy-momentum, through the \textit{Ricci tensor} 
$R_{\alpha \beta}$, defined as 
\begin{equation}
R_{\alpha \beta} \doteq R^{\gamma}_{\phantom{\gamma} \alpha \gamma \beta} \ .
\end{equation}
The Riemann tensor can be decomposed in terms of Weyl and Ricci tensors as
\begin{equation} \label{weyldef}
R_{\alpha \beta \gamma \delta} = g _{\alpha [\gamma} R _{\delta ] \beta} - g _{\beta [\gamma} R _{\delta ] \alpha} + \frac{1}{3} R g _{\alpha [\gamma} g_{ \delta] \beta} + C _{\alpha \beta \gamma \delta} \ .
\end{equation}
Both Weyl and Ricci tensors express the tidal force that a body feels while travelling along a geodesic, but physically the Weyl tensor differs from the Ricci tensor in a way that it does not convey information on how the volume of the body changes (it is indeed the trace-free part of the Riemann tensor), but rather on how the shape of the body is distorted by tidal forces \cite{wald}. 

The Riemann tensor satisfies the Bianchi identities
\begin{equation} \label{bianchi}
\nabla_{\epsilon} R_{\alpha \beta \gamma \delta} + \nabla_{\delta} R_{\alpha \beta \epsilon \gamma} + \nabla_{\gamma} R_{\alpha \beta \delta \epsilon} = 0 \ .
\end{equation}
From the contraction of the Ricci tensor we obtain the \textit{Ricci scalar}
\begin{equation}
R \doteq R^{\alpha}_{\phantom{\alpha} \alpha} \ ,
\end{equation}
which represents the scalar curvature. In the case of a Euclidean space the Riemann tensor is zero, as derivatives of the metric vanish. 

\subsection{The volume element}
To conclude this section about the geometry of GR we give the invariant \textit{volume element} of this theory, which is 
\begin{equation} \label{grvolel}
\epsilon = \epsilon_{\alpha \beta \gamma \delta} dx^{\alpha}  dx^{\beta}  dx^{\gamma}  dx^{\delta} = \sqrt{-g} \, d^{4} x \ ,
\end{equation}
where $g = \mathrm{det}(g_{\alpha \beta})$. The Levi-Civita tensor density  $\epsilon_{\alpha \beta \gamma \delta}$ is related to the Levi-Civita symbol $\tilde{\epsilon}_{\alpha \beta \gamma \delta}$ (defined in section \ref{NGgeom}) as
\begin{equation} \label{epsilonGR}
\epsilon_{\alpha \beta \gamma \delta} = \frac{1}{\sqrt{-g}} \, \tilde{\epsilon}_{\alpha \beta \gamma \delta} \ .
\end{equation}

\section{Einstein field equations} \label{grdef}
\paragraph*{The dynamics.} General relativity is a geometric theory of gravity defined on a generic oriented manifold equipped with a smooth metric tensor, which defines the inner product. The metric signature is not positive-definite, for this reason this kind of manifold is called \textit{pseudo-Riemannian}.

The centerpiece of GR are the \textit{Einstein field equations}, or the \textit{Einstein equation}. These are a set of ten partial differential equations which determine the dynamics of the theory, i.e.\ tell how the curvature of spacetime reacts to the presence of energy-momentum and a cosmological constant $\Lambda$. This $\Lambda$ term was not part of the original definition of GR but it was included in order to obtain a static universe solution from the Einstein equation. We retain it because of its importance in physical cosmology.

The Einstein equation can be derived by applying the principle of least action to the \textit{Einstein-Hilbert action} 
\begin{equation} \label{einhil}
\mathcal{S}= \int d^4x \, \sqrt{-g}\left[\frac{1}{2} R + \mathcal{L}_{M} - \Lambda \right] \ ,
\end{equation} 
where $\mathcal{L}_{M}$ is the Lagrangian which describes the matter fields in the theory, and $R$ is the Ricci scalar.

The simplest alternative theories of gravity in four dimensions, known as \textit{modified gravity} and mentioned in the introduction as candidates for explaining the dark energy, are formulated by replacing the Ricci scalar $R$ in (\ref{einhil}) with some other scalar function $f(R)$\footnote{Studying the Newtonian limit of $f(R)$ theories is an interesting exercise, which helps both in constraining the $f(R)$ class and in understanding how competing theories of dark energy ($f(R)$ gravity itself and backreaction for instance) relate to each other. However in this thesis we focus on GR only.}.

The Einstein field equations obtained from (\ref{einhil}) read
\begin{equation} \label{einstein}
G^{\alpha \beta} = R^{\alpha \beta} - \frac{1}{2}R g^{\alpha \beta} = T^{\alpha \beta} - \Lambda g^{\alpha \beta} \ .
\end{equation}
Here  $T^{\alpha \beta}$ is the symmetric \textit{energy-momentum tensor}, which describes the properties of matter in spacetime. If we consider an ideal fluid it takes the form
\begin{equation} \label{momentumenergy}
T^{\alpha \beta} = (\rho + p) u^{\alpha} u^{\beta} + p g^{\alpha \beta} \ ,
\end{equation}
where $u^{\alpha}$ is the four-velocity of the fluid. In the most general case $T^{\alpha \beta}$ contains terms which take into account anisotropic stress and energy-momentum flux.

\paragraph*{Continuity equations.} From the geometry of the theory it's possible to derive a continuity equation for the energy-momentum tensor. By contracting twice the Bianchi identities (\ref{bianchi}), we find 
\begin{equation} \label{bianchicontr}
\nabla_{\beta}G^{\alpha \beta} = 0 \ ,
\end{equation}
which can be used together with the Einstein equation to derive the \textit{continuity equation}
\begin{equation} \label{grcontinuity}
\nabla_{\beta} T^{\alpha \beta} = T^{\alpha \beta}_{\phantom{\alpha \beta},\beta} + T^{\gamma \beta} \Gamma^{\alpha}_{\phantom{\alpha} \gamma \beta} + T^{\alpha \gamma} \Gamma^{\beta}_{\phantom{\beta} \gamma \beta} = 0 \ .
\end{equation}
Equation (\ref{grcontinuity}) is a energy continuity equation, as from its contraction we can derive a energy-mass continuity equation analogue to (\ref{continuity2}) This is done in Chapter \ref{1+3form}. 

Note that $T^{\alpha \beta}$ is conserved locally but not globally in GR. Let's explain this: ``locally" in this context means that a conservation law can be found only in a differential form, and not in a ``global" integral form. A conservation law in an integral form for $T^{\alpha \beta}$ would state that the rate of change in time of this quantity in a certain volume is equal to its flux through the boundary.
This cannot be formulated in GR since tensors (except for scalars) cannot be added meaningfully in curved spacetimes, and there is not such an equivalent to Gauss' theorem for them. This is easy to understand if we think about two four-vectors living in a curved spacetime. In order to compare them we have first to parallel transport them to the same point, i.e.\ to the same tangent space. On curved manifolds the parallel transport  is a path dependent process, so also the sum is. In a more fundamental way the conservation of energy and momentum comes from the invariance in time and space translations, and since in GR space and time are not absolute but evolve, such an invariance does not hold.

\section{The symmetries} \label{GRsymm}
\paragraph*{Diffeomorphism invariance.} The dynamics of GR is governed by the ten coupled second order Einstein fields equations, which admit analytical solutions only if the metric studied has a high degree of symmetry. This simple fact shows the importance of classifying symmetries in GR.

The coordinate symmetries in Einstein's theory can be identified as invariance under diffeomorphism transformations. This means that physical laws are invariant under arbitrary differentiable coordinate transformations
$x^{\mu} \rightarrow x'^{\mu} (x^{\nu})$.

\paragraph*{Lie derivatives and Killing vectors.} The symmetries of the metric are called \textit{isometries}, and they can be 
classified resorting to the machinery of \textit{Lie derivatives} and \textit{Killing vectors} \cite{carroll}\cite{blau}. Intuitively a Lie derivative evaluates the change of a tensor field along the \textit{congruence} (or the flow) of a given vector field $K^{\alpha}$, which is defined to be the set of curves $\sigma^{\alpha}$ satisfying
\begin{equation}
\begin{cases}
\frac{d\sigma^{\alpha}(t,x_0)}{dt} = K^{\alpha}(\sigma^{\alpha}(t,x_0)) \\ \sigma^{\alpha}(t=0,x_0) = x_0^{\alpha} \ .
\end{cases}
\end{equation}
A rigorous definition of Lie derivative is given in \cite{carroll}\cite{blau}. 
For our purposes it's enough to know that the Lie derivative of the metric tensor along $K^{\alpha}$ can be written in the simple form
\begin{equation}
\mathcal{L}_{K} g_{\alpha \beta} = 2 \nabla_{(\alpha} K_{\beta)} \ .
\end{equation}
The vector $K^{\alpha}$ is defined to be a Killing vector if it satisfies the condition
\begin{equation} \label{killingdef}
\mathcal{L}_{K} g_{\alpha \beta}  = 0 \ ,
\end{equation}
i.e.\ if the Lie derivative of the metric calculated along it vanishes, and one isometry of the metric is generated. A simple way of spotting isometries is the following: take the metric tensor and write it in some coordinate system such that it's independent of one of the coordinates $x_{\sigma}$. Then the vector $K_{\sigma}=\partial_{\sigma}$ is a Killing vector and, together with the other Killing vectors of that metric, it forms the Lie algebra of the isometry group\footnote{In general it's not possible to write the metric in a coordinate system such that it is independent of all the coordinates, and Killing vectors most of time must be spotted one by one.}. For instance it's possible to show that the collection of Killing vectors of the Minkowski metric $g_{\alpha \beta} = \mathrm{diag}(-1,1,1,1)$ forms the Lie algebra of the ten-dimensional Poincar\'{e} group. One isometry comes from time translation, three from space translations, three from rotations in space and three from rotations involving the space dimensions and time (Lorentz boosts). The Galilean group is also ten-dimensional, as mentioned when discussing the symmetries of Newtonian gravity in section \ref{NGsymm}. On the other hand the Euclidean metric has six isometries, associated with space translations and rotations.

The concepts developed in this section will be used in Chapter \ref{FRW} where we define rigorously homogeneity and isotropy in the context of the FRW universe.

\section{Well-posedness of the initial value problem} \label{coordfree}
We can show by applying the Cauchy's problem to the Einstein equation (\ref{einstein}), that this has a well-posed initial value problem. Suppose that initial conditions are given for the metric $g_{\alpha \beta}$ and its first time derivative $\frac{\partial g_{\alpha \beta}}{\partial x^0}$ everywhere on the hypersurface of constant time $x^0 = t$. In order to compute the time evolution of these two quantities we need to extract from the field equations a formula for the ten second derivatives $\frac{\partial^2 g_{\alpha \beta}}{{\partial x^0}^2}$ everywhere at $x^0 = t$. One might think that these ten conditions can be deduced from the ten Einstein  equations, but this is not true. From the $\beta=0$ component of the twice contracted Bianchi identities (\ref{bianchicontr}) we get
\begin{equation}
\frac{\partial G^{\alpha 0}}{\partial x^0} \doteq - \frac{\partial G^{\alpha i}}{\partial x^i} - \Gamma^{\alpha}_{\phantom{\alpha} \beta \lambda} G^{\lambda \beta} - \Gamma^{\beta}_{\phantom{\beta} \beta \lambda} G^{\alpha \lambda} \ .
\end{equation}
The right hand side contains no time derivatives higher than second order, this implies that $G^{\alpha 0}$ on the left hand side cannot contain time derivatives of order higher than one, and therefore it doesn't determine the time evolution of the system. The relation
\begin{equation} \label{metricconstr}
G^{\alpha 0} =  T^{\alpha 0}
\end{equation}
provides solely a constraint on the initial metric. In fact there are only six useful Einstein equations for the purpose of determining the dynamics of the system, they are
\begin{equation} \label{eindyn}
G^{ij} =  T^ {ij} \ .
\end{equation}
This leaves a four-fold ambiguity on the second time derivatives of the metric. This ambiguity can be removed only imposing four coordinate conditions that fix the coordinate system. These conditions are arbitrary and they reflect for \textit{coordinate freedom}, or \textit{gauge freedom} of general relativity. This property corresponds to the freedom in GR of choosing how to slice spacetime in space-like hypersurfaces at a given time, and how this slicing is performed in a different way by two observers whose four-vectors are related by the generic coordinate transformation $x^{\alpha '}  \rightarrow x^{\alpha} + \xi^{\alpha}$. In this thesis we often fix the coordinate system (or the gauge) using harmonic coordinates, i.e.\ we impose
\begin{equation} \label{harm}
\frac{\partial ^2}{\partial x^0 \partial x^{\beta}} (\sqrt{-g} g^{\alpha \beta}) = 0 \ .
\end{equation}
The condition (\ref{harm}) can be equivalently stated using the connection coefficient (\ref{christoffel}), as
\begin{equation} \label{harm2}
g^{\gamma \delta} \Gamma ^{\alpha}_{\phantom{\alpha} \gamma \delta} = 0 \ .
\end{equation}
Given the initial value of $g_{\alpha \beta}$ and $\frac{\partial g_{\alpha \beta}}{\partial x^0}$, the constraint (\ref{metricconstr}) and the ten conditions (\ref{eindyn}),(\ref{harm}), the time evolution of the metric and its first time derivative are completely determined. 
An important reason for the success of this process is that the constraints (\ref{metricconstr}) are not only valid at the initial time, but are carried on by the dynamics of the theory, i.e.\ they hold at all times. In fact the twice contracted Bianchi identities (\ref{bianchicontr}), the continuity equations (\ref{grcontinuity}) and the spatial Einstein equations (\ref{eindyn}) imply the result
\begin{equation}
\frac{\partial}{\partial x^ 0}(G^{\alpha 0} -  T^{\alpha 0}) = 0 \ .
\end{equation}

\chapter{1+3 covariant formalism of GR} \label{1+3form}
\subsubsection*{Introduction}
In this chapter we present the 1+3 covariant formulation of general relativity \cite{ellis}\cite{ellis2}. This 1+3 formalism is said to be covariant since it is not defined with respect to any specific coordinate system. Rather, it's formulated by choosing a fiduciary observer moving along an arbitrary congruence of curves\footnote{A congruence is a set of integral curves of a vector field on a four-dimensional manifold.}, and then by projecting the variables of GR on the one-dimensional space parallel to the four-velocity $u^{\alpha}$ of this observer and on the three-dimensional space orthogonal to it.

When the observer's four-velocity is vorticity-free this corresponds to foliating the manifold into a family of metric hypersurfaces orthogonal to $u^{\alpha}$. Physically this amounts for the fact that spacetime can be split into a time dimension plus three spatial ones. In this sense space and time are derived quantities. The vorticity-free condition is provided by Frobenius' theorem \cite{wald}, which says that three-dimensional spaces orthogonal to $u^{\alpha}$ exist (are given with a global metric) and are integrable if and only if $u^{\alpha}$ is irrotational. Introducing a three-dimensional space is obviously helpful when we are comparing GR and NG.

\section{The 1+3 formalism}
Given a \textit{fundamental observer} with four-velocity $u^{\alpha}$, normalised as $u^{\alpha} u_{\alpha} = -1$, it's possible to define a unique \textit{parallel projection tensor}
\begin{equation}
P ^{\alpha}_{\phantom{\alpha} \beta} = - u^{\alpha} u_{\beta} \ ,
\end{equation}
which satisfies $P ^{\alpha}_{\phantom{\alpha} \gamma} P ^{\gamma}_{\phantom{\gamma} \beta} = P ^{\alpha}_{\phantom{\alpha} \beta}$, $P ^{\alpha}_{\phantom{\alpha} \alpha} = 1$, $P_{\alpha \beta} u^{\beta} = u_{\alpha}$.

We can define also a unique \textit{orthogonal projection tensor}
\begin{equation}
h_{\alpha \beta} \doteq g_{\alpha \beta} +u_{\alpha} u_{\beta} \ ,
\end{equation}
which satisfies $h^{\alpha}_{\phantom{\alpha} \gamma} h^{\gamma}_{\phantom{\gamma} \beta} = h^{\alpha}_{\phantom{\alpha} \beta}$, $h^{\alpha}_{\phantom{\alpha} \alpha} = 3$, $h_{\alpha \beta} u^{\beta} = 0$. In the vorticity-free case this is the metric of the instantaneous three-manifold orthogonal to the fundamental observer. 

Exploiting the orthogonal projection tensor we write the energy-momentum tensor, in the case of an ideal fluid, as
\begin{equation} \label{momen}
T_{\alpha \beta} = \rho u_{\alpha} u_{\beta} + p h _{\alpha \beta} \ .
\end{equation}
Where $\rho$ and $p$ are measured by $u^{\alpha}$.

In addition we can define two new derivatives. The first is the \textit{covariant time derivative along the fundamental worldline}, which for any tensor $U_{\alpha \beta}$ reads
\begin{equation}
\dot{U}_{\alpha \beta} \doteq u^{\gamma} \nabla_{\gamma} U_{\alpha \beta} \ .\footnote{For simplicity we use an overdot notation both for the Lagrangian total time derivative and for the covariant time derivative, however these are two distinct operations.}
\end{equation}
The corresponding operation in NG is the Lagrangian total time derivative (\ref{NGtot}).
The second is the \textit{fully orthogonally projected covariant derivative}, which reads
\begin{equation} \label{tildecov}
\tilde{\nabla}_{\gamma} U^{\alpha}_{\phantom{\alpha} \beta} \doteq h^{\kappa}_{\phantom{\kappa} \gamma} h^{\alpha}_{\phantom{\alpha} \delta}h^{\epsilon}_{\phantom{\epsilon} \beta} \nabla_{\kappa} U^{\delta}_{\phantom{\delta} \epsilon}  \ .
\end{equation}
The corresponding operation in NG is the gradient $U^{i}_{\phantom{i} j,k}$. The derivative (\ref{tildecov}) corresponds  to a proper covariant derivative on a three-space if and only if $u^{\alpha}$ has zero vorticity \cite{ellis2}.
The covariant derivative for $u_{\alpha}$ can be expressed as a sum of a parallel term plus an orthogonal one as
\begin{equation}
\nabla _{\beta} u_{\alpha} = -u_{\beta} \dot{u}_{\alpha} + \tilde{\nabla}_{\beta} u_{\alpha} \ .
\end{equation}
The term $\dot{u}_{\beta}$ is the \textit{relativistic acceleration vector} and it takes into account the forces acting on the energy-mass distribution 
due to the gradient of the pressure. In the vorticity-free case the term $\tilde{\nabla}_{\beta} u_{\alpha}$ is equal, up to a sign, to the \textit{extrinsic curvature} of a three-space $K_{\alpha \beta}$.
In analogy with what was done for Newtonian gravity in section \ref{lagr}, the fully orthogonally projected term $\tilde{\nabla}_{\beta} u_{\alpha}$ can be further decomposed into three parts: \\
\\
the trace part 
\begin{equation} \label{Thetadef}
\Theta \doteq \tilde{\nabla} _{\alpha} u^{\alpha} \ ,
\end{equation}
the traceless symmetric part
\begin{equation} \label{tracelesssymm}
\sigma _{\alpha \beta} \doteq \left(h_{(\alpha}^{\phantom{(\alpha} \gamma} h_{\beta)}^{\phantom{\beta)} \delta} - \frac{1}{3} h_{\alpha \beta} h^{\gamma \delta}\right) \tilde{\nabla}_{\delta} u_{\gamma} \ ,
\end{equation}
and the antisymmetric part 
\begin{equation} \label{omega3+1}
\omega _{\alpha \beta} \doteq \tilde{\nabla}_{[\beta} u_{\alpha]} \ .
\end{equation}
We can then write $\tilde{\nabla}_{\beta} u_{\alpha}$ as
\begin{equation} \label{gradvGR}
\tilde{\nabla}_{\beta} u_{\alpha} = \frac{1}{3} {\Theta} h_{\alpha \beta} + \sigma _{\alpha \beta} + \omega _{\alpha \beta} \ ,
\end{equation}
and $\nabla _{\beta} u_{\alpha}$ as
\begin{equation} \label{parallelcov}
\nabla _{\beta} u_{\alpha} = -u_{\beta} \dot{u}_{\alpha} + \frac{1}{3} {\Theta} h_{\alpha \beta} + \sigma _{\alpha \beta} + \omega _{\alpha \beta} \ .
\end{equation}
The tensor $\omega _{\alpha \beta}$ has the properties $\omega _{\alpha \beta} = \omega _{[\alpha \beta]}$ and $\omega_{\alpha \beta} u^{\beta} = 0$. It can also be written as $\omega^{\alpha} \doteq \frac{1}{2} \epsilon^{\alpha \beta \gamma} \omega _{\beta \gamma}$, where we have defined $\epsilon^{\alpha \beta \gamma} \doteq \epsilon^{\alpha \beta \gamma \delta} u_{\delta}$. For $\sigma_{\alpha \beta}$ we have $\sigma_{\alpha \beta} = \sigma_{(\alpha \beta)}$, $\sigma_{\phantom{\alpha} \alpha}^{\alpha} = 0$ and $\sigma_{\alpha \beta} u^{\beta} = 0$. The physical interpretation of the quantities $\Theta$, $\sigma _{\alpha \beta}$ and $\omega _{\alpha}$ is the same as in the Newtonian case. The acceleration, the vorticity and the shear are all orthogonal to $u^{\alpha}$, and they are in this sense space-like quantities.

At this point we can introduce a new notation and define the traceless symmetric spatially projected part of any rank two tensor as
\begin{equation}
A _{\langle \alpha \beta \rangle} \doteq \left(h_{(\alpha}^{\phantom{(\alpha} \gamma} h_{\beta)}^{\phantom{\beta)} \delta} - \frac{1}{3} h_{\alpha \beta} h^{\gamma \delta}\right) A _{\gamma \delta} \ .
\end{equation}
Similarly for a rank one tensor we have $B_{\langle \alpha \rangle} = h^{\alpha}_{\phantom{\alpha} \beta} B_{\alpha}$. This for instance simplifies (\ref{tracelesssymm}) to $\sigma_{ \alpha \beta} = \tilde{\nabla}_{\langle \gamma} u_{\delta \rangle}$.

\section{Application to cosmology} \label{1+3cosmo}
\subsubsection*{Brief historical digression}
When Einstein formulated the theory of general relativity, in 1915, his understanding of the cosmos was that it was limited to what we acknowledge as the Milky Way. Today we have observational evidence that the universe is immensely wider and more complex, but in spite of it GR is still believed to be the right theory to describe the universe at large scales. This gives an example on how, sometimes, a theory can be even more ambitious than its discoverer.
\subsection{Definition}
In this section we present a model for the evolution of the universe on large scales based on the Einstein equation (\ref{einstein}), decomposed following the 1+3 paradigm, and given in terms of an ideal energy-momentum tensor (\ref{momen}). We follow the references \cite{ellis}\cite{ellis2}. 

The advantage of using the 1+3 approach is that in this framework the degrees of freedom have a simpler physical interpretation compared to the degrees of freedom of traditional GR, namely the components of the metric tensor. Another advantage is that these physical degrees of freedom are closer to those of NG in the Lagrangian picture.

\subsection{Time evolution equations for GR}
The propagation equations of 1+3 cosmology can be obtained from Einstein equation together with the \textit{Ricci identities}\footnote{The Ricci identities are used in differential geometry to define the Riemann tensor.} for the vector field $u^{\alpha}$
\begin{equation} \label{ricci}
2 \nabla _{[\alpha} \nabla _{\beta]} u^{\gamma} = R_{\alpha \beta \phantom{\gamma} \delta}^{\phantom{\alpha \beta} \gamma} u^{\delta} \ .
\end{equation}
Projecting (\ref{ricci}) along $u^{\beta}$, and contracting the $\alpha$ and $\gamma$ indices using the metric, gives the \textit{Raychaudhuri equation} for GR
\begin{equation} \label{thetadotGR}
\dot{\Theta}   + \frac{1}{3} {\Theta}^2 = - \frac{1}{2} (\rho + 3 p) + 2 ({\omega}^2 + {\sigma}^2) +  \Lambda + \tilde{\nabla}_{\alpha} \dot{u}^{\alpha} + \dot{u}^{\alpha} \dot{u}_{\alpha} \ ,
\end{equation}
where ${\sigma}^2 = \frac{1}{2} \sigma _{\alpha \beta} \sigma ^{\alpha \beta}$ and ${\omega}^2 = \frac{1}{2} \omega _{\alpha \beta} \omega ^{\alpha \beta}$.

Projecting (\ref{ricci}) along $u^{\beta}$ and then applying the anti-symmetrization operator $\epsilon^{\alpha \phantom{\gamma} \delta \mu}_{\phantom{\alpha} \gamma}$, defined in (\ref{epsilonGR}), gives the \textit{vorticity equation} for GR
\begin{equation} \label{omegadotGR}
\dot{\omega} ^{\langle \alpha \rangle} + \frac{2}{3} \Theta \omega ^{\alpha} - \sigma ^{\alpha}_{\phantom{\alpha} \beta} \omega ^{\beta} = - \frac{1}{2} \epsilon ^{\alpha \beta \gamma} \tilde{\nabla}_{\beta} \dot{u}_{\gamma} \ .
\end{equation}
As in NG, equation (\ref{omegadotGR}) can be  written in terms of the two-tensor representation of the vorticity.
Projecting (\ref{ricci}) along $u^{\beta}$ and applying the trace removing spatially projecting symmetric operator $h^{(\alpha}_{\phantom{(\alpha} \gamma} h^{\beta)}_{\phantom{\beta)} \delta} - \frac{1}{3} h^{\alpha \beta} h_{\gamma \delta}$ gives the \textit{shear equation} for GR
\begin{equation} \label{sigmadotGR}
\begin{split}
h_{\phantom{\gamma} \alpha}^{\gamma} h_{\phantom{\delta} \beta}^{\delta} \dot{\sigma}_{\gamma \delta} + &\frac{2}{3} \Theta \sigma _{\alpha \beta} + \sigma _{\alpha}^{\phantom{\alpha} \kappa} \sigma _{\kappa \beta} - \frac{2}{3} h _{\alpha \beta} \sigma ^2  = \frac{1}{3} h _{\alpha \beta}  \left(\omega ^2 - \nabla _{\kappa} \dot{u}^{\kappa} \right)- \\ &-\omega _{\alpha} \omega _{\beta} +\dot{u}_{\alpha} \dot{u}_{\beta} -E_{\alpha \beta} +h _{\phantom{\gamma} \alpha}^{ \gamma} h_{\phantom{\delta} \beta}^{ \delta} {\nabla}_{( \delta} \dot{u}_{\gamma )} \ . 
\end{split}
\end{equation}
The three equations (\ref{thetadotGR}),(\ref{omegadotGR}) and (\ref{sigmadotGR}) have a term by term correspondence with their Newtonian counterparts (\ref{thetadot})-(\ref{sigmadot}). What changes is the metric, since here we use $h_{\alpha \beta}$ instead of $\delta_{ij}$, which reflects the fact that in GR space is curved.

Correspondingly the definition of the derivative changes, since we switch from the standard formulation in the flat space to a covariant one in a curved space.
In addition, the role of the \textit{electric part of the Weyl tensor} is different. In GR it is constructed from the Weyl tensor as
\begin{equation} \label{GRelectric}
E_{\alpha \beta} \doteq C_{\alpha \gamma \beta \delta} u^{\gamma} u^{\delta} \ ,
\end{equation}
and it is symmetric, traceless and orthogonal to $u^{\alpha}$. In section \ref{magnele}, while keeping in mind that in Newtonian cosmology $E_{ij}$ is arbitrary up to the constraints (\ref{constrE1})-(\ref{constrE2}), we discuss whether this is the case also in GR or not. A quantity which is closely related to $E_{\alpha \beta}$ is the magnetic part of the Weyl tensor. It's defined as
\begin{equation} \label{GRmagnetic}
H_{\alpha \beta} \doteq \frac{1}{2} \epsilon _{\alpha \gamma}^{\phantom{\alpha \gamma} \tau \kappa} C_{\tau \kappa \beta \delta} u^{\gamma} u^{\delta} \ .
\end{equation}
Like its electric counterpart it's symmetric, traceless and orthogonal to $u^{\alpha}$. The electric and magnetic parts of the Weyl tensor are discussed and put into context in section \ref{magnele}.

\subsection{Constraint equations for GR} \label{constreqGR}
Starting from (\ref{ricci}) it's possible to write down the constraint equations for GR, that we present without derivation (see \cite{ellis}\cite{ellis2} for details). They read
\begin{equation} \label{GRconstr1}
h^{\alpha}_{\phantom{\alpha} \beta}\left(\nabla _{\gamma} \omega^{\beta \gamma} - \nabla _{\gamma} \sigma^{\beta \gamma} +\frac{2}{3} \nabla^{\beta} \theta \right) + (\omega ^{\alpha}_{\phantom{\alpha} \beta} +\sigma ^{\alpha}_{\phantom{\alpha} \beta}) \dot{u}^{\beta} = 0 \ ,
\end{equation}
\begin{equation} \label{GRconstr2}
\nabla _{\alpha} \omega^{\alpha} = 2 \omega ^{\alpha} \dot{u}_{\alpha} \ ,
\end{equation}
\begin{equation} \label{GRconstr3}
H_{\alpha \beta} = 2 \dot{u}_{(\alpha} \omega_{\beta)} - h_{\alpha}^{\phantom{\alpha} \tau} h_{\beta}^{\phantom{\beta} \nu}\left( \nabla ^{\gamma} \omega_{(\tau}^{\phantom{(\tau} \delta} + \nabla ^{\gamma} \sigma_{(\tau}^{\phantom{(\tau} \delta}\right) \epsilon _{\nu) \kappa \delta \gamma} u^{\kappa} \ .
\end{equation}
The constraints (\ref{GRconstr1}) and (\ref{GRconstr2}) correspond in NG respectively to the constraints (\ref{NGconstr1}) and (\ref{NGconstr2}). These equations resemble each other term by term if we impose $\dot{u}_{\alpha} = 0$ (or equivalently, if  the ideal fluid has spatially constant pressure). 

A crucial difference between GR and NG arises from the constraint (\ref{GRconstr3}), which should be compared to the NG constraint (\ref{NGconstr3}). We see that in NG  a $H_{\alpha \beta}$-like term is completely missing. This absence has severe consequences on the dynamics of the two theories, which we remark in the next section. Note that in standard NG the magnetic part of the Weyl tensor $H_{ij}$ is not just equal to zero, rather it isn't defined at all.

In Chapter \ref{frame}, we outline how $H_{\alpha \beta}$ can be defined by introducing a geometrized version of NG. In Chapter \ref{cosmo}, we derive the form that the Weyl tensor assumes in the context of linearised general relativity.
\subsection{Evolution and constraint equations for the Weyl tensor} \label{magnele}
A set of four equations, which governs the behaviour of the magnetic and electric parts of the Weyl tensor, can be obtained from Bianchi identities (\ref{bianchi}), whose gradient satisfies the identity $C^{\alpha \beta \gamma \epsilon}_{\phantom{\alpha \beta \gamma \delta}; \epsilon} = R^{\gamma [\alpha ; \beta]} - \frac{1}{6} g^{\gamma [ \alpha} R^{; \beta ]}$. We expect the equations derive in this section not to have any Newtonian counterpart, since in NG the Bianchi identities are not defined. The evolution and constraint equations for the Weyl tensor in GR are
\begin{equation} \label{divE}
h^{\alpha}_{\phantom{\alpha} \beta} h_{\phantom{\delta} \gamma}^{ \delta} \, \nabla _{\delta} E^{\beta \gamma} -\epsilon ^{\alpha \beta \gamma \delta} u_{\beta} \sigma _{\gamma}^{\phantom{\gamma} \epsilon} \, H _{\delta \epsilon} + 3 H^{\alpha}_{\phantom{\alpha} \beta} \omega ^{\beta} = \frac{1}{3} h ^{\alpha}_{\phantom{\alpha} \beta} \, \nabla ^{\beta} \rho \ ,
\end{equation}
\begin{equation} \label{Hdot}
\begin{split}
 h^{\alpha \gamma} h^{\beta \delta} \dot{H}_{\gamma \delta} - h _{\gamma}^{\phantom{\gamma} (\alpha} \epsilon ^{\beta) \rho \epsilon \delta} u _{\rho} \nabla _{\delta} E^{\gamma}_{\phantom{\alpha} \epsilon} + 2 E _{\epsilon}^{\phantom{\epsilon} (\alpha} \epsilon ^{\beta) \gamma \delta \epsilon} u _{\gamma} \dot{u} _{\delta} + \\  +  h^{\alpha \beta}(\sigma ^{\gamma \delta} H _{\gamma \delta}) + \Theta H ^{\alpha \beta} - 3 H _{\gamma} ^{\phantom{\gamma} (\alpha} \sigma ^{\beta) \gamma} - H _{\gamma} ^{\phantom{\gamma} (\alpha} \omega ^{\beta ) \gamma } = 0 \ ,
\end{split}
\end{equation}
\begin{equation} \label{divH}
h^{\alpha}_{\phantom{\alpha} \beta} h^{\delta}_{\phantom{\delta} \gamma} \, \nabla _{\delta} H^{\beta \gamma} + \epsilon ^{\alpha \beta \gamma \epsilon} u _{\beta} \sigma _{\gamma}^{\phantom{\nu} \delta} E _{\epsilon \delta} - 3 E^{\alpha}_{\phantom{\alpha} \beta} \omega ^{\beta} = (\rho + p) \omega ^{\alpha} \ ,
\end{equation}
\begin{equation} \label{Edot}
\begin{split}
h_{\delta}^{\phantom{\delta} \alpha} h_{\gamma}^{\phantom{\gamma} \beta} \dot{E}^{\delta \gamma} + & h _{\gamma}^{\phantom{\gamma} (\alpha} \epsilon ^{\beta) \rho \epsilon \delta} u _{\rho} \nabla _{\delta} H^{\gamma}_{\phantom{\gamma} \epsilon} - 2 H _{\epsilon}^{\phantom{\epsilon} (\alpha} \epsilon ^{\beta) \gamma \delta \epsilon} u _{\gamma} \dot{u} _{\delta} +  h^{\alpha \beta}(\sigma ^{\gamma \delta} E _{\gamma \delta}) + \\+ &\Theta E ^{\alpha \beta} - 3 E _{\gamma} ^{\phantom{\gamma} (\alpha} \sigma ^{\beta) \gamma} - E _{\gamma} ^{\phantom{\gamma} (\alpha} \omega ^{\beta ) \gamma} = - \frac{1}{2} (\rho + p) \sigma ^{\alpha \beta} \ .
\end{split}
\end{equation}
The structure of (\ref{divE})-(\ref{Edot}) resembles Maxwell's equations for the electric and magnetic field, though in the gravitational case we have some extra source terms. With a close look we see that (\ref{divE}) is an equation which links the four-divergence of $E_{\alpha \beta}$ with the curl of $H_{\alpha \beta}$ (gravitational analogue of the  Gauss' law). Equation (\ref{Hdot}) links the time derivative of $H_{\alpha \beta}$ with the curl of $E_{\alpha \beta}$ (Faraday's law). Equation (\ref{divH}) relates the four-divergence of $H_{\alpha \beta}$ with some source terms (Gauss' law for magnetism, showing that gravitational monopoles do exist). Equation (\ref{Edot}) relates the curl of $H_{\alpha \beta}$ with the time derivative of $E_{\alpha \beta}$ (Amp\`{e}re's circuital law). In analogy with the electromagnetic case, these equations have wave-like solutions. In GR, $E_{\alpha \beta}$ and $H_{\alpha \beta}$ are propagating degrees of freedom, they describe gravitational waves and tidal effects, which convey the gravitational information through spacetime. 

In contrast, in Newtonian gravity $E_{ij}$ and $H_{ij}$ play a different role. A primary reason for this is that in NG there is no counterpart for the Bianchi identities, thus the counterpart of (\ref{divE})-(\ref{Edot}) has to be derived from the Ricci-like identities (\ref{riccilike1})-(\ref{riccilike2}). The constraints (\ref{divE}) and (\ref{Hdot}) correspond roughly to (\ref{constrE1}) and (\ref{constrE2}), while a rough counterpart of (\ref{divH}) and (\ref{Edot}) is given by (\ref{constr3}). If we look at these Newtonian constraints we see that, for instance, there is no term involving $H_{\alpha \beta}$ or $\dot{E}_{\alpha \beta}$. This reflects the fact that, since in NG there is no time evolution equation for the potential (see section \ref{lagr}), a time evolution equation for the $E_{ij}$ is also absent. An analogue to the time evolution equation for $H_{ij}$ is also missing, for the simple reason that the magnetic part of the Weyl tensor is not even defined in NG, or if defined, as in the case of frame theory (developed in Chapter \ref{frame}), it is identically zero. 

A related important difference is that, in the Newtonian case, the evolution of $E_{ij}$ is underdetermined. The dependence of $E_{ij}$ on space coordinates is restricted by the constraints (\ref{constrE1})(\ref{constrE2}), while its time evolution is constrained as equation (\ref{constrE2})  has to be satisfied at all times, and the density may depend on time. However in the case of an exactly homogeneous and isotropic Newtonian universe, where \ref{constrE3} holds, $E_{ij}$ can be set to zero for all times. In GR we can set the initial condition $E_{\alpha \beta}(t_0) = 0$, but because of the evolution equation (\ref{Edot}), the constraint ${E}^{\alpha \beta} = 0$ is propagated in time only if the following conditions hold
\begin{equation} \label{Econd}
H_{\alpha \beta}(t_0) = 0 \; \; \mathrm{and} \; \; \sigma_{\alpha \beta}(t_0) = 0 \ .
\end{equation}
This implies that in GR, given some initial shear or some initial non-zero $H^{\alpha \beta}$, $E^{\alpha \beta}$ grows even if it starts from zero value. 

The quantities $E^{\alpha \beta}$ and $H^{\alpha \beta}$ are coupled to $\Theta$, thus their time evolution in GR affects the expansion rate and is expected to play a role in the backreaction.

\subsection{Continuity equation and four-acceleration equation for GR}
In this section we present the \textit{continuity equation} and the \textit{four-acceleration equation} for GR. These are derived by taking the continuity equation for the energy-momentum tensor (\ref{grcontinuity}), decomposing it as (\ref{momen}) and projecting  parallel and orthogonal to $u ^{\alpha}$.
The two equations obtained  are
\begin{equation} \label{continuityGR}
\dot{\rho} + \Theta (\rho + p) = 0 \ ,
\end{equation}
\begin{equation} \label{adot}
\tilde{\nabla} _{\alpha} p + (\rho + p) \dot{u}_{\alpha} = 0 \ .
\end{equation}
What we can learn from the first one is that in GR the pressure contributes to the inertial mass. The second equation shows that $\dot{u}_{\alpha}$, in analogy with $a_i$, represents a pressure driven acceleration term and vanishes for dust (pressure-less ideal fluid). These equations resemble the Newtonian equations (\ref{contall}) and (\ref{ai}), with the only addition of the pressure term.
\section{Irrotationality assumption} \label{irro}
\subsubsection*{Introduction}
As mentioned in the introduction to this chapter, the assumption that the observer's four-velocity is irrotational plays a crucial role. According to the Frobenius' theorem \cite{wald}, this is the sufficient and necessary condition for the existence of the orthogonal three-space to the fiducial observer, and to take averages on it. In this subsection we point out what this implies for cosmology.
\subsection{The assumption and its consequences}
The irrotationality condition states that that the vorticity $\omega^{\alpha}$ is zero at all times. This can be obtained by requiring that the initial vorticity  vanishes $\omega ^{\alpha}(t_0) = 0$ and that the ideal fluid is barotropic $p = p(\rho)$. In fact in this case it follows from (\ref{omegadotGR}) that $\omega ^{\alpha} = 0 \; \; \forall \, t$.

The irrotationality assumption is sufficient to assure that the fluid flow is hypersurface-orthogonal, and that a \textit{cosmic time function} $t$ exists, such that 
\begin{equation} \label{tdot}
u_{\alpha} = \frac{1}{\dot{t}} \nabla _{\alpha} t \ .
\end{equation}
This can be shown as follows: we multiply (\ref{parallelcov}) by the four-velocity and we take the antisymmetric part of this expression, obtaining
\begin{equation} 
u_{[\gamma}\nabla_{\beta} u_{\alpha]} = u_{[\gamma}\omega_{\alpha \beta]} ,
\end{equation}
from which we deduce that the condition
\begin{equation} \label{frocond}
u_{[\gamma}\nabla_{\beta} u_{\alpha]} = 0
\end{equation} 
holds if and only if the rotation vanishes. Frobenius' theorem asserts that (\ref{frocond}) is the necessary condition for $u_{\alpha}$ to have orthogonal space-like hypersurfaces and to be written as (\ref{tdot}).
So if the irrotationality condition holds then the orthogonal three-space becomes the physical hypersurface at a given time. Its three-metric is $h_{ij}$ and its volume element is
\begin{equation} \label{perpvolel}
\epsilon_{\perp} = \sqrt{h} \, d^3 x \ ,
\end{equation}
where $h = \mathrm{det}(h_{ij})$. 
The total volume of a comoving domain $\mathcal{D}$ of the three-space is
\begin{equation} \label{vgr}
V \doteq \int \limits_{\mathcal{D}} \epsilon_{\perp} = \int \limits_{\mathcal{D}} \sqrt{h} d^3x \ .
\end{equation}

In the irrotational case we also have that \textit{Gauss' formula} \cite{gauss}, which relates the Ricci scalar of the spatial three-surfaces to the four-dimensional Ricci tensor and the extrinsic curvature, holds. It reads
\begin{equation}
^{(3)}R = R + 2 R_{\gamma \delta} u^{\gamma} u^{\delta} - K^2 + K_{ij}K^{ij} \ .
\end{equation}
Combining this expression with the Einstein equation (\ref{einstein}) for a dusty universe gives the \textit{Hamiltonian constraint}, namely
\begin{equation} \label{genfri}
\frac{2}{3} \Theta ^2  = 2( \rho +  \Lambda) +2 \sigma ^2 -  {^{(3)}R} - \nabla_{\alpha}\dot{u}^{\alpha} \ .
\end{equation}
Here $\Theta$ is the spatial part of the expansion rate which, in terms of the three-metric, is taken to be the trace of 
\begin{equation} \label{thetah}
\Theta ^i_j = \frac{1}{2}h^{ik} \frac{d}{dt}\left({h}_{kj}\right) \ .
\end{equation} 

\section{Averaging in the 1+3 formalism and backreaction}
\subsection*{Introduction}
We gave an overview on the problem of averaging in Newtonian cosmology in section \ref{averageNG}. What we said previously however doesn't hold for GR in general, where the situation is more complicated \cite{syksybuchert}\cite{ellis}. 

First, in the GR case we have to make the crucial assumption that the observer's four-velocity is vorticity-free. As anticipated in section \ref{irro}, this assures us that the three-space, where we want to take averages, exists, meaning that it's given with a metric $h_{ij}$ and a volume element $\epsilon_{\perp} = \sqrt{h} \, d^3x$. 

Second, we have to keep in mind, as pointed out in section \ref{grdef}, that integration is meaningful on curved spacetimes only for scalars, because of the path dependency of the parallel transport process, and therefore we perform averages only of scalars\footnote{There are suggestions on how to overcome this difficulty of averaging tensorial objects, however this is still an open issue.}.
\subsection{Definition of averaging}
Given a scalar field $\Psi(x^{\alpha})$, its spatial average over the comoving domain $\mathcal{D}(t)$ is defined, in 1+3 cosmology, as
\begin{equation} \label{whatta}
\langle \Psi \rangle   \doteq \frac{\int _{\mathcal{D}} \Psi \, \epsilon_{\perp}}{\int _{\mathcal{D}} \, \epsilon_{\perp}}  \ .
\end{equation}
The denominator of (\ref{whatta}) is the volume of the hypersurfaces of constant time over the domain $\mathcal{D}$, which was defined in (\ref{vgr}).

We define the \textit{scale factor} in GR to be the cubic root of the volume $V(t)$, normalised with the initial time volume $V(t_0)$. It reads
\begin{equation} \label{GRa}
a_{\mathcal{D}} \doteq \left(\frac{\int _{\mathcal{D}(t)} \epsilon_{\perp}}{\int _{\mathcal{D}(t_0)} \epsilon_{\perp}} \right)^{\frac{1}{3}} \ .
\end{equation}
The variation of $V$ with respect to the cosmic time $t$ is
\begin{equation}
\dot{V} = \int \limits_{\mathcal{D}} \frac{d}{dt}(\sqrt{h}) d^3x \ .
\end{equation}
By using this expression together with the definition of the expansion rate (\ref{thetah}), we find the commutation rule for 1+3 cosmology between the processes of averaging and time derivation
\begin{equation}
{\langle \Psi \rangle}\dot{}   - {\langle \dot{\Psi} \rangle } = {\langle \Psi \Theta \rangle} - {\langle \Psi \rangle } {\langle \Theta \rangle} \ .
\end{equation}
This relation looks the same as in the Lagrangian picture (\ref{commute}) of NG, though the definitions of average and time derivative are different. This proves the advantage given by the fluid dynamic Lagrangian picture over the Euclidean picture, when we are comparing Newtonian gravity to general relativity.
\subsection{The Buchert equations} \label{buchert}
\paragraph*{Derivation of the equations.}
In the previous section we have shown how to take averages in 1+3 cosmology, we can now apply this knowledge to the scalar equations derived earlier on in this chapter. For simplicity in what follows we neglect the pressure, $p = 0$. By averaging (\ref{continuityGR}) we obtain the \textit{averaged continuity equation}
\begin{equation} \label{continuityaveGR}
\langle \rho \rangle \dot{}   + 3 \frac{\dot{a}_\mathcal{D}}{a _\mathcal{D}}  \langle \rho \rangle= 0 \ ,
\end{equation}
where the average of the expansion rate is taken to be $\langle \Theta \rangle = 3 \frac{\dot{a}_\mathcal{D}}{a_\mathcal{D}}$.
By averaging (\ref{thetadotGR}) we obtain the \textit{averaged Raychaudhuri equation} for GR
\begin{equation} \label{rayaveGR}
3\frac{\ddot{a}_\mathcal{D}}{a_\mathcal{D}}    = - \frac{1}{2}  \langle \rho \rangle + \mathcal{Q} + \Lambda   \ .
\end{equation}
Integration in GR can be performed only in the irrotational case, for this reason here the backreaction variable is defined as 
\begin{equation}
\mathcal{Q} \doteq \frac{2}{3} \left(\langle \Theta ^2 \rangle - {\langle \Theta \rangle}^2 \right) - 2 \langle  {\sigma}^2 \rangle \ .
\end{equation}
If we average (\ref{genfri}) we obtain
\begin{equation} \label{hamconstr}
3\left(\frac{\dot{a}_\mathcal{D}}{a _\mathcal{D}}\right)^2   = \langle \rho \rangle  - \frac{1}{2}\mathcal{Q}  - \frac{1}{2} \langle {^{(3)}} R \rangle  + \Lambda \ ,
\end{equation}
which is known as the \textit{averaged Hamiltonian constraint}.
An \textit{integrability condition}, making sure that (\ref{rayaveGR}) is compatible with the first time derivative of (\ref{hamconstr}), is
\begin{equation} \label{intcond}
\dot{\mathcal{Q}} + \langle ^{(3)}\mathcal{R} \rangle \dot{•} = - 6 \frac{\dot{a}_\mathcal{D}}{a _\mathcal{D}} \mathcal{Q} - 2 \frac{\dot{a}_\mathcal{D}}{a _\mathcal{D}} \langle ^{(3)}\mathcal{R} \rangle \ .
\end{equation}
Equations (\ref{continuityaveGR}), (\ref{rayaveGR}) and (\ref{hamconstr}) form the set of the \textit{Buchert equations} \cite{buchert1}. These three equations are all independent, meaning that we cannot choose one of them and derive it from the two others, or if we try to do so a fourth equation (\ref{intcond}) appears. In the Buchert equations four unknowns appear, these are the scale factor, the density field, the backreaction variable and the curvature of the three space. The set is closed by giving $\mathcal{Q}$ or $\langle ^{(3)}\mathcal{R} \rangle$. 

\paragraph*{The independence of the equations in NG.} In the previous paragraph we said that all the Buchert equations are independent in GR.

This is not the case in Newtonian gravity\footnote{Nor in the case of the Friedmann-Robertson-Walker model, as we show in Chapter \ref{FRW}.}. Newtonian gravity provides independently a counterpart for equations (\ref{continuityaveGR}) and (\ref{rayaveGR}), given by the averaged continuity equation for NG (\ref{contave}) and the averaged Raychaudhuri equation for NG (\ref{secondfriedmann}). However the counterpart of the Hamiltonian constraint (\ref{hamconstr}) can be derived from the latter two equations. If we combine the first Lagrangian total time derivative of the averaged Raychauduri equation for NG (\ref{secondfriedmann}) with the continuity equation (\ref{contave}) we obtain the Newtonian equivalent to the Hamiltonian constraint
\begin{equation} \label{hamNG}
3\left(\frac{\dot{a}_\mathcal{D}}{a _\mathcal{D}}\right)^2  = \langle \rho \rangle  + \Lambda + \mathcal{Q} +  \frac{1}{a^2_{\mathcal{D}}} \int a^2 _\mathcal{D} \dot{\mathcal{Q}} \, dt  \ .
\end{equation}
A consequence of the fact that in NG (\ref{hamNG}) is derived from the averaged continuity and the Raychaudhuri equation (\ref{secondfriedmann}) is that in this theory there is no need for an integrability condition, since (\ref{hamNG}) and (\ref{secondfriedmann}) are compatible by construction. In NG an equation corresponding to the integrability condition (\ref{intcond}) doesn't exist, thus the time evolution of ${\mathcal{Q}}$ is not constrained, nor it is coupled to the curvature of the three space. This is related to the fact that $\langle ^{(3)}\mathcal{R} \rangle$ is zero in a flat space.

\paragraph*{The integrability condition.}
The integrability condition (\ref{intcond}) gives us a hint on the behaviour of the backreaction in clumpy cosmological models: if we set $\mathcal{Q} = 0$ in  (\ref{intcond}) we see that the spatial curvature in this case must be inversely proportional to the scale factor squared $\langle{ ^{(3)}\mathcal{R} \rangle} \propto a_{\mathcal{D}}^{-2}$. This behaviour of the spatial curvature is characteristic of exactly homogeneous and isotropic models, so choosing zero backreaction implies choosing a cosmological model where the spatial curvature must have the trivial behaviour mentioned above.
On the other hand if we retain the backreaction then the GR equation (\ref{hamconstr}) allows the curvature scalar to have a non-trivial behaviour depending on the evolution of the backreaction itself. It couples also to the acceleration through (\ref{rayaveGR}), letting the averaged curvature of the three-space to take part in the dynamics of the system.
\paragraph*{Magnitude of the backreaction vs.\ spatial curvature.}
We can get a hint from the averaged Hamiltonian constraint on the relative size of backreaction and averaged spatial curvature in a universe without dark energy. In fact the left hand size of (\ref{hamconstr}) is the Hubble parameter squared, and we know that the matter density is not enough alone to explain the magnitude of the Hubble constant today, $H_{0} = 100 \, {h} \mathrm{km}\mathrm{s}^{-1} \mathrm{Mpc}^{-1}$, therefore we need a positive contribution from the term $- \frac{1}{2}(\mathcal{Q} + \langle {^{(3)} R} \rangle)$ on the right hand side. This implies the spatial curvature to be negative and larger than the backreaction in its absolute value, as we need the backreaction to be positive to get acceleration from (\ref{rayaveGR}).


\subsection{On the relation between the backreaction and the Weyl tensor}
We conclude this chapter by studying the contribution of the Weyl tensor to the backreaction and its consequences on the expansion rate. 

First, we start looking at the averaged ${\sigma}^2$ term in $\mathcal{Q}$ as defined in NG and in GR. This is crucially different: in GR $\dot{\sigma}_{\alpha \beta}$ is coupled to $E_{\alpha \beta}$, which is coupled to $H_{\alpha \beta}$, so the electric part of the Weyl tensor propagates and contributes to $\dot{\sigma}_{\alpha \beta}$, no matter what is its initial value, as long as the condition (\ref{Econd}) doesn't hold. In fact the system of equations (\ref{divE})-(\ref{Edot}) has wavelike solutions, as mentioned in section \ref{1+3cosmo}. On the contrary, in NG $\dot{\sigma}_{ij}$ is coupled to $E_{ij}$, which is not coupled to $H_{ij}$, because $H_{ij}$ is not defined \cite{weyl}. Moreover NG doesn't provide an equation for $\dot{E}_{ij}$, so if this quantity is zero initially it doesn't grow following of the dynamics of the system, but it does only if we impose it arbitrarily (up to the constraints (\ref{constrE1})(\ref{constrE2})). These considerations can be reformulated by saying that in GR the propagating degrees of freedom, i.e.\ the magnetic and electric part of the Weyl tensor, always contribute to the ${\sigma}^2$ term in the backreaction unless the condition (\ref{Econd}) holds, while in NG they contribute unless we choose an homogeneous and isotropic which implies, through the constraint constraint (\ref{constrE2}), that $H_{ij}$ can be set to zero at all times. In this extent the case of the \textit{silent universe} \cite{silent} \cite{silentuni} of GR, where $H_{\alpha \beta} = 0$, $\omega ^{\alpha} = 0$, $p = 0$ seems to be close to NG. 

The effect of $C_{\alpha \beta \gamma \delta}$ on the variance of the backreaction is not as easy to understand explicitly. However, since the Buchert equations and the evolution equations for $E_{\alpha \beta}$ and $H_{\alpha \beta}$ are coupled, we expect the Weyl tensor to play a role in the backreaction and to contribute to the expansion law given by the Raychaudhuri equation. 

It would be interesting to have an analytical relation giving explicitly $\mathcal{Q}$ in terms of $C_{\alpha \beta \gamma \delta}$ in GR, since it would show how the information conveyed by gravitational waves and tidal fields affects the backreaction and how important these non-local effects are in its economy. However this is impossible, as the equations involved are numerous, coupled and non-linear. What we can do is to work with the perturbation theory of GR, and to calculate the linearised backreaction. This approach is presented in section (\ref{PNback}), where we calculate the backreaction variable in post-Newtonian cosmology and we show that it is a boundary term at leading order.

\chapter{Friedmann-Robertson-Walker universe} \label{FRW}
\section*{Introduction}
This chapter is dedicated to the solution of GR due to Friedmann, Lema\^{i}tre Robertson and Walker \cite{carroll}\cite{wald}. This is a solution which corresponds to a spacetime foliated into exactly homogeneous and isotropic space-like hypersurfaces, whose expansion law is determined by a scale factor. The FRW model is a cornerstone of physical cosmology since it has proven so far to be the most successful model in describing the real universe. In our discussion it's important to present this model as we are interested in understanding how it relates to the Newtonian universe and the 1+3 formulation of the inhomogeneous universe.

\section{FRW metric}
The FRW line element is given, in terms of a spherical set of coordinates $(r,\theta,\phi)$, as
\begin{equation} \label{FRWmetric}
ds^2 = -dt^2 + a(t)^2 \left[\frac{dr^2}{1-k r^2} + r^2 d\theta^2 + r^2 \sin ^2\theta d\phi^2 \right] \ ,
\end{equation}
where $t$ is the cosmic time, $a(t)$ is a scale factor which determines the expansion law of the spatial dimensions, and $k$ is a constant which is related to the curvature of the three-space of constant time and to $a$ as
\begin{equation}
^{(3)}R= 6 \frac{k}{a^2} \ .
\end{equation}

\paragraph{Hyperspherical case.} If $k>0$ then the FRW metric is singular at $r=k^{-\frac{1}{2}}$. This is just a coordinate singularity, as it vanishes if we introduce a new coordinate system, defining $r=k^{-\frac{1}{2}} \sin \chi$, with $0 \leq \chi \leq \pi$, and we apply it to (\ref{FRWmetric}), obtaining
\begin{equation}
ds^2 = -dt^2 + a(t)^2 k^{-1} \left[d\chi^2 + \sin^2\chi d\theta^2 + \sin^2\chi \sin ^2\theta d\phi^2 \right] \ .
\end{equation}
This is the line element of a hypersphere with finite volume. This is called a closed model of the universe.

\paragraph{Hyperbolic case.} If $k<0$ there is no coordinate singularity, the radial coordinate $r$ is free to span from zero to infinity, the universe is infinite and the model is said to be open. The coordinate transformation $r=|k|^{-\frac{1}{2}} \sinh \chi$, with $0 \leq \chi < \infty$, if introduced in (\ref{FRWmetric}), shows the hyperbolic nature of this space
\begin{equation}
ds^2 = -dt^2 + a(t)^2 |k|^{-1} \left[d\chi^2 + \sinh^2\chi d\theta^2 + \sinh^2\chi \sin ^2\theta d\phi^2 \right] \ .
\end{equation}

\paragraph{Flat case.} In the case $k=0$, the FRW metric reads
\begin{equation}
ds^2 = -dt^2 + a(t)^2 \left[d\chi^2 + \chi^2 d\theta^2 + \chi^2 \sin ^2\theta d\phi^2 \right] \ ,
\end{equation} 
where $\chi$ is the radial coordinate spanning from zero to infinity. The spatial part corresponds to an Euclidean space which is contracted or expanded by the scale factor, in the $(x,y,z)$ base it reads
\begin{equation} \label{FRWxyz}
ds^2 = -dt^2 + a(t)^2\left[dx^2 + dy^2 +dz^2\right] \ .
\end{equation}

\paragraph{General form for the line element.}
We can rescale for simplicity the curvature $k$ to values $(1,0,-1)$ in the hyperspherical, flat and hyperbolic cases, and we can rewrite the three cases discussed above in a compact form as
\begin{equation} \label{FRWf}
ds^2 = -dt^2 + a(t)^2 \left[d\chi^2 + f_k(\chi)^2 d\theta^2 + f_k(\chi)^2 \sin ^2\theta d\phi^2 \right] \ ,
\end{equation}
where
\begin{equation} \label{fk}
f_k(\chi)=\left\{
  \begin{array}{l l}
    \sin\chi & \quad {\text{if} \; k=1}\\
    \chi & \quad {\text{if} \; k=0}\\ \sinh\chi & \quad {\text{if} \; k=-1} \ .
  \end{array} \right. 
\end{equation}
The line element expressed as (\ref{FRWf}) is useful when we derive all the isometries of the FRW metric in the next section.

\section{Homogeneity and isotropy revisited} In the introduction we said that the $\Lambda$CDM model of physical cosmology relies on the FRW metric (up to perturbations), which has exactly homogeneous and isotropic space-like hypersurfaces. After we have introduced the machinery of Killing vectors in section \ref{GRsymm}, and we have discussed the FRW metric, we are now ready to give a rigorous definition of homogeneity and isotropy, following \cite{blau}.

We deduce from the definition of Killing vector (\ref{killingdef}), combined with the Ricci identities (\ref{ricci}) and with the cyclic identity for the Riemann tensor
\begin{equation}
R^{\alpha}_{\phantom{\alpha} \beta \gamma \delta} + R^{\alpha}_{\phantom{\alpha} \delta \beta \gamma} + R^{\alpha}_{\phantom{\alpha} \delta \gamma \beta} = 0 \ ,
\end{equation}
that Killing vectors and the Riemann tensor are related by
\begin{equation} \label{killingriemann}
\nabla_{\gamma}\nabla_{\beta} K_{\alpha} = R^{\delta}_{\phantom{\delta} \gamma \beta \alpha} K_{\delta} \ .
\end{equation}
Equation (\ref{killingriemann}) shows that a Killing vector $K_{\alpha}(x)$ is determined everywhere by the values of $K_{\alpha}(x_0)$ and $\nabla_{\beta} K_{\alpha}(x_0)$ in a point $x_0$ of the spacetime.

\paragraph{Definition of homogeneity.} A space is defined to be homogeneous if it has infinitesimal isometries that carry each point $x_0$ into any other point in its neighbourhood. Thus homogeneity requires the existence of arbitrary Killing vectors $K_{\alpha}(x_0)$ at any arbitrary point $x_0$. This means that a $n$-dimensional space must admit $n$ translational Killing vectors in order to be homogeneous. 

\paragraph{Definition of isotropy.} A space is defined to be isotropic if it has isometries that leave the given point $x_0$ fixed and  rotate any vector at $x_0$ into any other vector at $x_0$. Therefore the metric must admit Killing vectors such that $K_{\alpha}(x_0)=0$ and such that $\nabla_{\beta} K_{\alpha}(x_0)$ is an arbitrary rotational matrix. This means that a $n$-dimensional space must admit $n(n-1)/2$ rotational Killing vectors in order to be isotropic. This follows from the fact that in a $n$-dimensional space there are $n$ axis of rotation, and each of them can be rotated around $n-1$ other axes. The one half coefficient takes into account that for instance we shouldn't count a rotation of an axis labelled $\hat{y}$ into an axis labelled $\hat{z}$ as separate from a rotation of $\hat{z}$ into $\hat{y}$.

\paragraph*{Maximally symmetric spaces.} A $n$-dimensional space is \textit{maximally symmetric} if is has the maximum number of translational and rotational Killing vectors
\begin{equation} \label{maxsymm}
n_{K} = n +\frac{1}{2}n(n-1) = \frac{1}{2}n(n+1) \ .
\end{equation}
If a space is maximally symmetric then it is also homogeneous and isotropic and the Riemann tensor is given \cite{carroll} as
\begin{equation} \label{Rmax}
R_{\alpha \beta \gamma \delta} = \frac{R}{n(n-1)}(g_{\alpha \gamma}g_{\beta \delta}-g_{\alpha \delta}g_{\beta \gamma}) \ ,
\end{equation}
with $R$ constant over the whole manifold.
\paragraph{Isometries of FRW.} 
The FRW model has six isometries. This can be shown by solving explicitly the Killing equation (\ref{killingdef}), or by using physical intuition. We resort to the second approach. The argument goes as follows: from the flat line element in the $(x,y,z)$ base (\ref{FRWxyz}) it is clear that three of the Killing vectors are $K_{(1)}= \partial_x$, $K_{(2)}= \partial_y$ and $K_{(3)}= \partial_z$. We can generalise these vector to the three different geometries of the FRW model by introducing the coordinate transformation
\begin{equation} \label{basechange}
\begin{cases}
x=f_k(\chi) \sin \theta \cos \phi \\
y=f_k(\chi) \sin \theta \sin \phi \\
z=f_k(\chi) \cos \theta 
\end{cases} \ ,
\end{equation}
where $f_k(\chi)$ was defined in (\ref{fk}).
Going from the $(x,y,z)$ base to the $(\chi,\theta,\phi)$ base using (\ref{basechange}) gives the three translational Killing vectors for the FRW metric as 
\begin{equation}
\mathbf{K}_{1} = \left( \sin \theta \cos \phi \right) \partial_{\chi} + \left( \frac{f_k(\chi)_{,\chi}}{f_k(\chi)} \cos \theta \cos \phi \right) \partial_{\theta} - \left( \frac{f_k(\chi)_{,\chi}}{f_k(\chi)} \csc \theta \sin \phi \right) \partial_{\phi} \ ,
\end{equation}
\begin{equation}
\mathbf{K}_{2} = \left( \sin \theta \sin \phi \right) \partial_{\chi} +  \left( \frac{f_k(\chi)_{,\chi}}{f_k(\chi)} \cos \theta \sin \phi \right) \partial_{\theta} + \left( \frac{f_k(\chi)_{,\chi}}{f_k(\chi)} \csc \theta \cos \phi \right) \partial_{\phi} \ ,
\end{equation}
\begin{equation}
\mathbf{K}_{3} = \left( \cos \theta \right) \partial_{\chi} - \left( \frac{f_k(\chi)_{,\chi}}{f_k(\chi)} \sin \theta \right) \partial_\theta \ .
\end{equation}
The form of the fourth Killing vector is straightforward if we notice that the FRW line element (\ref{FRWf}) does not depend on the azimuthal angle $\phi$. From this it follows that
\begin{equation}
\mathbf{K}_4 = \partial_{\phi} \ .
\end{equation}
In the $(x,y,z)$ base it reads $\mathbf{K}_4 = -y \partial_x + x \partial_y$,
which generates rotations around the $\hat{z}$ axis. From this we deduce that the two last Killing vectors must be associated with rotations around the $\hat{y}$ axis and the $\hat{x}$ axis. Therefore they are $\mathbf{K}_5=z\partial_x  -x\partial_z$ and $\mathbf{K}_6= -z\partial_y + y\partial_z$. In the $(\chi,\theta,\phi)$ base, given with the change of coordinates (\ref{basechange}), the last two rotational Killing vectors for the FRW metric read
\begin{equation}
\mathbf{K}_5=\left(\cos \phi \right) \partial_{\theta} - \left(\cot \theta \sin \phi \right) \partial_{\phi} \ ,
\end{equation}
\begin{equation}
\mathbf{K}_6=\left(\sin \phi \right) \partial_{\theta} - \left(\cot \theta \cos \phi \right) \partial_{\phi} \ .
\end{equation}

We have shown that the spatial part of the metric \ref{FRWh} has six Killing vectors, or six isometries. From (\ref{newtonforce}) it follows that it generates a maximally symmetric space, which is exactly homogeneous and isotropic.
If we write the FRW line element in the form
\begin{equation} \label{FRWh}
ds^2 = -dt^2 + a(t)^2 h_{ij}(x^k) dx^i dx^j \ .
\end{equation}
we can show, using equation (\ref{Rmax}), that the Riemann tensor of the three-space is
\begin{equation} \label{riemann3}
^{(3)}R_{i j k l} = \frac{^{(3)}R}{6} (h_{i k}h_{j l}-h_{i l}h_{j k}) \ .
\end{equation}
Equation (\ref{riemann3}) is useful in section \ref{fromto}, where we study how the FRW model is related to relativistic cosmology in the 1+3 formulation.

In this section for simplicity we started by taking the FRW metric and then we derived the associated Killing vectors. Note however that normally we work the other way around, i.e.\ normally we choose which are the isometries of the space that we are studying by setting its Killing vectors, and from these we deduce the form of the metric.

\section{The dynamics of the FRW universe}
The purpose of this section is to derive the dynamical equations of the FRW universe, given in terms of the time derivative of the scale factor. In order to do so we take Einstein equation (\ref{einstein}), we substitute the FRW metric in the Einstein tensor, we write the energy-momentum tensor in the perfect fluid form (\ref{momentumenergy}) (to preserve the requirement of homogeneity and isotropy) and solve. In this case the given metric embeds a high degree of freedom, therefore the solutions of the Einstein equation are analytical. What we obtain are: \\ \\
the \textit{first Friedmann equation}
\begin{equation} \label{fried1}
3 \left( \frac{\dot{a}}{a}\right)^2 = \rho - 3 \frac{k}{a^2} +  \Lambda \ ,
\end{equation}
and the \textit{second Friedmann equation}
\begin{equation} \label{fried2}
3 \frac{\ddot{a}}{a} = - \frac{1}{2} (\rho + 3p) + \Lambda \ .
\end{equation}
Combining the two Friedmann equations yields
\begin{equation} \label{frwcont}
\dot{\rho} + 3 \frac{\dot{a}}{a}(\rho + p) = 0 \ ,
\end{equation}
which is the continuity equation. This equation can be reformulated as
\begin{equation}
\left(a^3 \rho \right) \dot{} + p\left(a^3\right) \dot{} = 0 \ ,
\end{equation}
which shows intuitively that in a dynamical homogeneous and isotropic universe the change in internal energy (first term) is balanced by the pressure work (second term). This same equation can be derived similarly from the first law of thermodynamics, plus the assumption of the absence of a heat flow, which would violate the isotropy requirement of the FRW model.

What's important here is that the equation (\ref{frwcont}) has been derived from the two Friedmann equations, without resorting to the relativistic continuity equation (\ref{grcontinuity}). This shows how a high degree of symmetry in the metric increases the redundancy of the GR equations. We delve further into this aspect in the conclusions of this chapter.

\section{Newtonian derivation of the Friedmann equations}
In this section we show how the Friedmann equations can be derived from the basic principles of Newtonian gravity. Our derivation is not rigorous, for more details see \cite{rigore}.

Identify a spherical region around an arbitrary observer the universe. The matter density in the region is homogeneous and since the matter outside the sphere cannot play any role in its dynamics (its pull would violate isotropy) the size of the sphere is arbitrary.
The Eulerian coordinate $x(t)$ of a test mass $m$ located on the boundary of the sphere with respect to the center of the mass distribution is mapped from its initial Lagrangian coordinate $q$ as 
\begin{equation} \label{disp}
x(t) = a(t) q \ .
\end{equation}

As time goes on the test mass moves with respect to the observer as described by the $a(t)$ factor.

The equation of motion of a test mass at the edge of the spherical expanding region, located in $x(t)$, is
\begin{equation}
m \ddot{x} = - G\frac{m M}{x^2} \ ,
\end{equation}
which, by using $M=\frac{4}{3} \pi x^3 \rho$ and (\ref{disp}), becomes
\begin{equation}
3 \frac{\ddot{a}}{a} = - \frac{\rho}{2} \ .
\end{equation}
This is the second Friedmann equation without the pressure term or the $\Lambda$ term.

On the other hand the total energy of the test mass is
\begin{equation}
E = m \frac{\dot{x}^2}{2} - \frac{1}{8 \pi} \frac{m M}{x} \ .
\end{equation}
It can be rewritten as
\begin{equation}
3 \left( \frac{\dot{a}}{a}\right)^2 = \rho + \frac{2 E}{m q^2 a^2} \ ,
\end{equation}
which is the second Friedmann equation without the $\Lambda$ term and with the curvature term identified with $k= -2 E m^{-1}q^{-2}$. This identification is however only formal, as in Newtonian gravity the curvature is zero. The Newtonian derivation of the Friedmann equations can only be made a posteriori, after the relativistic result is known.

\section{From the 1+3 formulation of GR to FRW} \label{fromto}
In this section we show how to recover FRW cosmology as a particular case of GR in the 1+3 formalism.
Starting from the symmetries of the Riemann tensor of a maximally symmetric three-space (\ref{riemann3}), we can re-derive the equations of cosmology in the 1+3 formulation as we did in section \ref{1+3cosmo}. 

If we plug equation (\ref{riemann3}) into the Ricci identities (\ref{ricci}) we find out that the only time evolution equation which survives among the Raychaudhuri equation, the vorticity equation and the shear equation is the first. This is equivalent with saying that in the FRW universe we can identify a geodesic observer $u^{\alpha}$ such that $\dot{u}^{\alpha}$, $\omega_{\alpha \beta}$, $\sigma_{\alpha \beta}$, $E_{\alpha \beta}$ and $H_{\alpha \beta}$ are all zero. The vanishing of these quantities also assures that the constraint equation of GR, presented in section \ref{constreqGR}, are zero in the FRW universe. The same consideration holds for the evolution and constraint equation of the Weyl tensor, discussed in section \ref{magnele}.

From the continuity equation for the energy-momentum tensor we re-derive the continuity equation (\ref{continuityGR}), and this completes the set of the Friedmann equations.

\section{Dynamics and redundancy}
In this chapter we pointed out that both in FRW cosmology and in NG one equation among the continuity equation, the first Friedmann equation and the second Friedmann equation is redundant. As we remarked in section \ref{buchert}, this implies that NG doesn't provide an equation for the time evolution of the backreaction in the form of the integrability condition (\ref{intcond}). In FRW cosmology the situation is similar, as an equation corresponding to (\ref{intcond}) exists, but it's always zero. FRW cosmology is therefore close to the case of Newtonian gravity for what concerns the dynamics of $\mathcal{Q}$.

\chapter{Frame theory} \label{frame}
\subsubsection*{Introduction}
In the previous chapters we delved into an analysis of the aspects which differentiate NG from GR.
In this chapter we present a way of assimilating the two theories, defining them under the same formal structure. This purpose is achieved by \textit{frame theory} (FT), which is a geometric theory of gravity designed to include Newtonian gravity and general relativity as special cases. This theory is interesting because it offers an example on how to unify physical theories, and it gives insight on the geometric structure of NG.

In particular frame theory sheds light on the relation between the magnetic part of the Weyl tensor and the Riemann tensor in NG. In addition FT suggests that it's possible to build a Newtonian limit of GR using a perturbative approach when solving the Einstein field equations. This hint leads in Chapter \ref{GRC} to the formulation of post-Newtonian cosmology. In this thesis we do not aim to give an exhaustive presentation of FT, rather we introduce only the aspects which are useful for the discussion.

\section{Definition}
Frame theory (see \cite{frameth} for references) is defined by the following set of objects: a four-manifold $M$, a contravariant symmetric \textit{temporal metric}  $g_{\alpha \beta}$, a covariant \textit{inverse spatial metric} $g^{\alpha \beta}$, a covariant \textit{energy-momentum tensor}  $T^{\alpha \beta}$, a symmetric \textit{connection coefficient} $\Gamma^{\alpha}_{\phantom{\alpha} \beta \gamma}$, a \textit{cosmological constant} $\Lambda$, and a \textit{causality constant} $\lambda$.

An intuitive way of deriving FT equations is the following: first, we write down the Einstein equation (\ref{einstein}) with $R=-T$ (which follows from the contraction of Einstein equation itself) and we temporarily restore the speed of light unit $c$
\begin{equation} \label{einsteinFT}
R_{\alpha \beta} = c^{-4}\left(T_{\alpha \beta} - \frac{1}{2} g_{\alpha \beta}T \right) + c^{-2}\Lambda g_{\alpha \beta} \ .
\end{equation}
Then we make an apparently trivial change of notation: $g_{\alpha \beta} \doteq - \lambda^{-1} t_{\alpha \beta}$, $g^{\alpha \beta} \doteq s^{\alpha \beta}$, we define the rule for lowering indices as $\xi_{\alpha} = - t_{\alpha \beta} \xi^{\beta}$ and we introduce the important axiom
\begin{equation}  \label{axiom}
g_{\alpha \beta} g^{\beta \gamma} = - \lambda \delta_{\alpha}^{\phantom{\alpha}\gamma}\ .
\end{equation}
Also a change of units is introduced 
\begin{equation} \label{units}
c^{-2} \doteq \lambda \ ,
\end{equation}
where $\lambda$ is dimensionless.

After these changes equation (\ref{einsteinFT}) becomes
\begin{equation} \label{einsteinFT2}
R_{\alpha \beta} = \left(t_{\alpha \gamma}t_{\beta \delta} - \frac{1}{2} t_{\alpha \beta} t_{\gamma \delta} \right)T^{\gamma \delta} - \Lambda t_{\alpha \beta} \ ,
\end{equation}
where $T^{\alpha \beta}= (\rho + \lambda p) u^{\alpha} u^{\beta} + p g^{\alpha \beta}$ and $g_{\alpha \beta}u^{\alpha}u^{\beta} = - \lambda^{-1}$.

This is clearly the equation of GR for $\lambda = 1$, and in this sense we say that FT reduces to GR for this value of $\lambda$. 

Maybe not as transparently the case $\lambda = 0$ provides the \textit{Newtonian limit} of GR. This is due to the fact that, for $\lambda = 0$, the temporal metric $g_{\alpha \beta}$ and the spatial metric $g^{\alpha \beta}$ decouple, as the axiom (\ref{axiom}) suggests. 
This can be made more evident if we know a result of FT: in the case $\lambda = 0$ FT admits a scalar field $t$, which gives the time direction, such that $t_{,\alpha} = t_{\alpha}$, $t_{\alpha \beta}=t_{\alpha}t_{\beta}$, $s^ {\alpha \beta} t_{\beta} = 0$, $t_{; \alpha \beta} = 0$. Using these properties of $t$ in equation (\ref{einsteinFT2}) this simplifies to
\begin{equation} \label{einsteinsimple}
R_{\alpha \beta} = \left(\frac{1}{2} \rho - \Lambda \right) t_{, \alpha} t_{, \beta} \ ,
\end{equation}
where $\rho = T = T^{\alpha}_{\phantom{\alpha} \alpha}$.
Newton's equations can be derived explicitly if we make a coordinate transformation such that in the new reference frame $t_{\alpha \beta} = \mathrm{diag}(1,0,0,0)$, $s^{\alpha \beta} = \mathrm{diag}(0,1,1,1)$ and $\Gamma^{\alpha}_{\phantom{\alpha} \beta \gamma} = 0$ except for $\Gamma^{i}_{\phantom{i} 0 0} \doteq - g^i$, $\Gamma^{i}_{\phantom{i} 0 j} \doteq s^{ik} \epsilon_{jkh} \omega ^{h}$.
Plugging these connection coefficients in (\ref{einsteinsimple}) returns the following field equations
\begin{equation} \label{fieldFT}
\omega^i_{\phantom{i},i} = 0 \ ,
\end{equation}
\begin{equation}
\epsilon^ {ijk}g_{k,j} + 2 \frac{\partial {\omega}^i}{\partial t} = 0 \ ,
\end{equation}
\begin{equation}
\epsilon^ {ijk}\omega_{k,j} = 0 \ ,
\end{equation}
\begin{equation}
g^i_{\phantom{i},i} -  \omega_i \omega^i = -\frac{1}{2} \rho + \Lambda \ .
\end{equation}
The geodesic equation (\ref{geodesic}), given in terms of these connection coefficients, reads
\begin{equation}
\frac{\partial^2 {x}^i}{\partial t^2} = g^ i + 2 \epsilon ^{ijk}  \omega_k \frac{\partial{x}_j}{\partial t} \ .
\end{equation}
The covariant conservation law for the energy-momentum tensor (\ref{grcontinuity}) provides the continuity equation (\ref{continuity}) for NG and a modified version of the Euler equation (\ref{euler})
\begin{equation}
{\partial \rho \over \partial t} + v^i \rho_{,i}  + \rho v^i_{\phantom{i},i} = 0 \ ,
\end{equation}
\begin{equation} \label{eulerFT}
{\partial v^i \over \partial t} + v^i  v^k_{\phantom{k},k} = g^i - \frac{1}{\rho} p^{,i} + 2 \epsilon ^{ijk} v_j \omega_k  \ .
\end{equation}
Equations (\ref{fieldFT})-(\ref{eulerFT}) are often referred to as the Newton-Cartan theory (NCT), which we have given in 1+3 notation. They do not correspond to pure NG yet, but rather to what we denote from now on as \textit{geometrized Newtonian gravity} (GNG). Comparing the Euler equation of NG and GNG unveils an important difference between the theories: in GNG there is no general inertial coordinate system relative to which the vorticity field $\omega^i$ vanishes, unlike in NG. This is related to the fact that in GNG the vorticity is changing both in space and in time. Keeping this is mind in the next section we restrict GNG to NG. 

Before going into it we define the Weyl tensor in the notation of FT, as
\begin{equation} \label{weylFTbad}
C _{\alpha \beta \gamma \delta} =  R_{\alpha \beta \gamma \delta} - \delta _{\alpha [\gamma} R _{\delta ] \beta} - \lambda^ {-1} \left( t _{\beta [\gamma} R _{\delta ] \alpha} + \frac{1}{3} R \delta _{\alpha [\gamma} t_ {\delta] \beta} \right) \ .
\end{equation}
This expression is meaningless for $\lambda = 0$. However, if we use the field equation (\ref{einsteinsimple}) to eliminate $R_{\alpha \beta}$, we obtain
\begin{equation} \label{weylFT}
C _{\alpha \beta \gamma \delta} =  R_{\alpha \beta \gamma \delta} -( \delta _{\alpha [\gamma} T _{\delta ] \beta} +t _{\beta [\gamma} T _{\delta ] \alpha} - \frac{2}{3} T \delta _{\alpha [\gamma} t_ {\delta] \beta}) +  \frac{2}{3} \Lambda \delta _{\alpha [\gamma} t_ {\delta] \beta} \ ,
\end{equation}
which makes sense for any value of $\lambda$ and which we take to define the Weyl tensor in FT. In the case $\lambda = 0$ the last equation simplifies to
\begin{equation}
C _{\alpha \beta \gamma \delta} =  R_{\alpha \beta \gamma \delta} - \frac{2}{3}t_{\beta}\delta_{\alpha[\gamma}t_{\delta]} \left(\frac{1}{2}\rho - \Lambda \right) \ .
\end{equation}
From this expression  the magnetic and electric part of the Weyl tensor can be defined in GNG, in analogy with what we did for GR, with respect to a fiduciary four-velocity. They are
\begin{equation}
E_{\alpha \beta} = R_{\alpha \gamma \beta \delta} u^{\gamma}u^{\delta} - \frac{1}{3} \left(\delta^{\alpha}_{\phantom{\alpha} \beta}-u^{\alpha}t_{\beta}\right) \left(\frac{1}{2} \rho - \Lambda \right) \ ,
\end{equation}
\begin{equation} \label{HFT}
H_{\alpha \beta} = \frac{1}{2}\epsilon_{\alpha \gamma \lambda \mu} R^{\lambda \mu}_{\phantom{\lambda \mu} \beta \delta} u^{\gamma} u^{\delta} \ .
\end{equation}
In GNG $H_{\alpha \beta}$ is proportional to the Riemann tensor $R^{\alpha \beta}_{\phantom{\alpha \beta} \gamma \delta}$. When we discussed Newtonian gravity we said that in this theory $H_{ij}$ vanishes identically. This is not the case in GNG since in this theory the Riemann tensor $R^{\alpha \beta}_{\phantom{\alpha \beta} \gamma \delta}$ is completely decoupled from the Riemann tensor $R_{\alpha \beta \gamma \delta}$, and the vanishing of the latter (as in any flat space) does not imply the vanishing of the former.
In the next section we show that in order to restrict GNG to NG we have to impose a condition stating that the Riemann tensor $R^{\alpha \beta}_{\phantom{\alpha \beta} \gamma \delta}$ is zero in all its components. 

\section{Restriction to Newtonian gravity} \label{newlim}
As suggested in the previous section the equation of fluid dynamic Newtonian gravity seems to emerge from FT when $\lambda = 0$ and $\omega^i = \omega^i (t)$. It can be shown that the condition $\omega^i = \omega^i (t)$ is equivalent to \textit{Trautman's condition}
\begin{equation} \label{traut}
R^{\alpha \beta}_{\phantom{\alpha \beta} \gamma \delta} = 0 \ ,
\end{equation}
and implies that we can find a frame where $\omega = 0$. The connection coefficients in this case can be written as $\Gamma^{\alpha}_{\phantom{\alpha} \beta \gamma} = t_{, \beta}t_{, \gamma}h^{\alpha \delta} U_{, \delta}$, where we have redefined $g^i$ in terms of a \textit{connection potential} $U$ as $g^i = -U^{,i}$. The restriction to NG implies that $H_{\alpha \beta}$ is always zero, while $E_{\alpha \beta}$ is given by
\begin{equation}E_{\alpha \beta} = \left(s _{\alpha \gamma} s _{\beta \delta} - \frac{1}{3}s _{\alpha \beta} s _{\gamma \delta} \right) U^ {,\gamma \delta} \ .
\end{equation}
Restricting GNG by imposing Trautman's condition still does not give NG properly. As we remarked in section \ref{woe2}, NG is a theory of isolated systems, so to recover it fully we have to give an isolatedness condition. This condition holds if the two requisites are met: first, $\rho$ has a compact support, i.e. the density field is defined in a finite volume of the rest space of a given observer $u^{\alpha}$, second the scalars $R^{\lambda}_{\phantom{\lambda}\gamma \alpha \delta} R^{\gamma \phantom{\lambda} \delta}_{\phantom{\gamma}\lambda \phantom{\delta} \beta} u^{\alpha}u^{\beta}$ and $R^{\lambda}_{\phantom{\lambda} \alpha \mu \beta} R^{\mu}_{\phantom{\mu}\gamma \lambda \delta} u^{\alpha}u^{\beta}u^{\gamma}u^{\delta}$ tend to zero at spatial infinity. These two conditions imply not only (\ref{traut}),  they also imply that the connection coefficients can be decomposed uniquely and tend to the flat connection at spatial infinity. 

This is sufficient to assure that the connection potential $U$ coincides with the Newtonian potential and that pure Newtonian gravity is recovered.

\section{Towards post-Newtonian cosmology}
We reviewed a way of obtaining a consistent Newtonian limit of GR performing a change in the notation, writing the metric, the energy-momentum tensor and the Einstein equation as a function of a parameter $\lambda = c^{-2}$, and then taking $\lambda$ to go to zero. This suggest that we could go further and assume that these quantities depend on $\lambda$ differentially, expanding them in terms of a power series in this parameter. The lower term in the expansion would account for the Newtonian limit, while the higher terms would give post-Newtonian corrections, accounting for GR effects. This approach is developed in the next chapter.

\chapter{Post-Newtonian Cosmology} \label{GRC}

\section{Introduction} \label{PNC}
In this chapter we address the issue of how to quantify general relativistic effects in cosmology and how Newtonian gravity emerges from the perturbative approach to GR. 

A systematic framework for this is given by post-Newtonian cosmology (PNC), which is a theory derived from the weak-field, small-velocity approximation of GR. The post-Newtonian approximation is suitable for studying systems in which Newtonian gravity has a dominant role, while  relativistic effects are small but non-negligible. Well known applications are the precession of Mercury's perihelion and other solar system tests of the Schwarzschild metric of GR. In recent times the post-Newtonian approach has been used to study the gravitational wave production of binary stars, and it has proven to be surprisingly effective also in describing systems of binary neutron stars and black holes, where one would expect the post-Newtonian approximation to break down because of fast motion and strong gravitational fields \cite{binary}. 

In this chapter we focus on recovering the Newtonian limit of GR and we show that, thanks to the greater richness in equations of PNC, in this theory it's possible to define a well-posed initial value problem for the gravitational potential. This fixes two problems of NG: first, it makes the speed of interaction finite, this is due to the fact that in PNC the gravitational potential is solved from a hyperbolic partial differential equation. Second, it allows to treat LSS formation as an initial  value problem in the potential picture. As we have shown is section \ref{woe2}, this is not the case in pure Newtonian gravity.

The reason why we are interested in PNC is that in principle we can use this theory to study the hydrodynamics of LSS in the universe, and to quantify the contribution that non-Newtonian, or better post-Newtonian effects have in the structure formation budget. This topic is developed further in Chapter \ref{cosmo}.

\section{The post-Newtonian approach}
In this section we present PNC following the reference \cite{weinberg}. The quantity $c$ is restored in all equations.
The basic idea of PNC is to take the GR metric $g_{\alpha \beta}$, the typical four-velocity $u^{\alpha} = \gamma (c, v^i)$, the energy-momentum tensor $T_{\alpha \beta}$ and to expand them in powers of the parameter $\beta = {|v_0|}c^{-1}$, where $|v_0|$ is a typical velocity\footnote{We remind that $|v| = \sqrt{v^i v_i}$ is the absolute value of the three-velocity. Here $|v_0|$ is just a parameter that we use to make $\beta$ dimensionless.} and is taken to be $|v_0| \approx 1 \, \mathrm{m s^{-1}}$, which implies $\beta \approx c^{-1}$, whereas $c \approx 3 \cdot 10^8 \, \mathrm{m s^{-1}}$. 

The condition $\beta \ll 1$ provides both the \textit{small-velocity approximation} and the \textit{weak-field approximation} of GR. The first point is evident, while the second point becomes clear if we consider that, in a nearly virialised Newtonian system, the typical kinetic energy $1/2 M |v| ^2$ is of the same order as the typical potential energy $G M^2 x^{-1}$. From this follows $|v| ^2 \sim {G M}/{x}$, which shows that small velocities imply weak gravitational fields. In a non-virialised system the condition $\beta \ll 1$ doesn't have a clear physical interpretation, but the weak-field assumption is guaranteed by the requirement that the metric is close to the flat Minkowski metric $\eta_{\alpha \beta} = \mathrm{diag}(-1,1,1,1)$. The range of applicability of PNC is very wide, as the Minkowskian metric is believed to describe well the geometry of spacetime from scales much larger than the Schwarzschild radius up to scales where the cosmic expansion becomes important. 

One might point out that, since we are doing cosmology, we'd better to use the FRW metric (\ref{FRWxyz}) as a background. However that would result in an increased number of terms in the equations, and for our purpose of comparing PNC to NG a post-Minkowskian approximation is satisfactory. 

In the next section we derive the equations of PNC and we discuss their Newtonian limit and the associated initial value problem.

\section{Post-Newtonian expansion and the Newtonian limit}

\subsection{The expansion}
The metric is taken to behave asymptotically, for $\beta \ll 1$, as 
\begin{equation}
g_{\alpha \beta} \cong \eta_{\alpha \beta} + \epsilon_{\alpha \beta} \doteq  g_{\alpha \beta}^{(0)} + g_{\alpha \beta}^{(1)} +  g_{\alpha \beta}^{(2)} + \mathcal{O}({\beta^3})  \ .
\end{equation}
Where $\epsilon_{\alpha \beta}$ represents the perturbation in the metric around the background Minkowki metric $g_{\alpha \beta}^{(0)} = \eta_{\alpha \beta}$ and where the superscript $(n)$ indicates that the associated term is of the same order as $\beta ^n$.

We can separate the metric into a scalar, a vector and and a tensor part as
\begin{equation} \label{g00pure}
g_{0 0}  \cong  g_{0 0}^{(0)} + g_{0 0}^{(2)} + g_{0 0}^{(4)} + g_{0 0}^{(6)} + \mathcal{O}(\beta ^8)
\end{equation}
\begin{equation} \label{g01pure}
g_{0 i}  \cong  g_{0 i}^{(3)} + g_{0 i}^{(5)} + \mathcal{O}(\beta ^7) 
\end{equation}
\begin{equation} \label{gijpure}
g_{i j}  \cong  g_{i j}^{(0)} + g_{i j}^{(2)} + g_{i j}^{(4)} + g_{i j}^{(6)} + \mathcal{O}(\beta ^8) \ ,
\end{equation}
where the magnitude of each term in the expansion is an ansatz.
If we take (\ref{g00pure})-(\ref{gijpure}), impose harmonic coordinates (\ref{harm}) and require consistency with Newtonian gravity, then the expansion coefficients can be expressed \cite{weinberg} as 
\begin{equation} \label{g00}
g_{0 0} \cong  -1 - 2 \phi \beta^2 - 2 (\phi ^2 + V) \beta^4 - 2 \alpha ' \beta ^6 + \mathcal{O}(\beta ^8)
\end{equation}
\begin{equation} \label{g0i}
g_{0 i} \cong \zeta_i \beta ^3 + \zeta_i ' \beta^5 + \mathcal{O}(\beta ^7)
\end{equation}
\begin{equation} \label{gij}
g_{i j}  \cong  \delta_{i j} - 2 \psi \delta_{i j} \beta^2 + \alpha_{i j} \beta^4 + \alpha_{i j}' \beta^6 + \mathcal{O}(\beta ^8) \ . \footnote{In this thesis the energy-momentum tensor has the perfect-fluid form, this implies that $\psi$ coincides with $\phi$ \cite{Hannu}.}
\end{equation}
All the coefficients of the expansion are assumed to be of order $\mathcal{O}(\beta^0)$, and dimensionless. If for instance we want to identify $\phi$ with the physical gravitational potential per unit mass then its dimensionality must be fixed by multiplying it by $|v_0|^2$. 

We remark here that the expansion (\ref{g00})-(\ref{gij}) is not arbitrary, rather it comes from taking the most general expansion for the metric and then choosing to work in the harmonic gauge \cite{weinberg}\cite{PNcosmo} with a perfect fluid matter model.  An example of an alternative gauge choice, which is often adopted in PNC, is the synchronous and comoving gauge \cite{PNlagr} \cite{PNlagrback}. 

There are many other possible ways of specifying (\ref{g00pure})-(\ref{gijpure}), and each of them corresponds with a specific gauge choice that needs to be made to fix the coordinate freedom of GR.

The next step is to derive the Newtonian limit of PNC, but before going into it we must introduce the asymptotic form of the energy-momentum tensor for $\beta \rightarrow 0$, needed to solve the Einstein equation. This tensor is defined, for a perfect fluid, by equation (\ref{momentumenergy}), in terms of the pressure, the energy density and the four-velocity.  We can expand these three quantities as
\begin{equation} \label{Ppost}
P \rightarrow P  + P' \beta^2 + P'' \beta^4 + \mathcal{O}(\beta ^6)
\end{equation}
\begin{equation} \label{rhopost}
\rho \rightarrow \rho +   \rho' \beta^2 + \rho'' \beta^4 + \mathcal{O}(\beta ^6)
\end{equation}
\begin{equation}  \label{upost}
u_{\alpha} \rightarrow  \gamma (-c, v_i + {v_i {'}} \beta^2 + \mathcal{O}(\beta ^4)) \ .
\end{equation}
It follows from $g_{\alpha \beta}u^{\alpha}u^{\beta} = - c^2$ that
\begin{equation} \label{PNvel}
\gamma ^2 \cong 1 + (2 \phi + v^2) \beta ^2 + (2 \phi ^2 + 2 V + 6 \phi v^2 + v^4 + 2 \zeta ^i v _i + 2 v^i {'} v_i) \beta^4 + \mathcal{O}(\beta ^6) \ .
\end{equation}
In order to determine the post-Newtonian form of Einstein equation we should similarly expand explicitly in series of $\beta$ the connection coefficients (\ref{christoffel}) and the Ricci tensor. However we don't do it here because these objects are completely defined in terms of the metric, whose perturbative behaviour has been presented\footnote{For details we refer to \cite{weinberg}.}.

\subsection{The Newtonian limit} \label{newtonlimit}
We obtain the Newtonian limit of PNC as follows:  we keep the perturbed metric up to second order, i.e.\
\begin{equation} \label{epsilonphi}
\epsilon_{\alpha \beta} \cong -2 \phi \delta_{\alpha \beta} \beta^2 \ .
\end{equation} 
In order to remove the coordinate freedom of GR we define a new quantity 
\begin{equation} \label{defharm}
h_{\alpha \beta} \doteq \epsilon_{\alpha \beta} - \frac{1}{2} \eta_{\alpha \beta} \epsilon^{\gamma}_{\phantom{\gamma}\gamma}
\end{equation}
and impose harmonic coordinate condition, as in (\ref{harm}), such that 
\begin{equation}
h^{\alpha}_{\phantom{\alpha} \beta ,\alpha}  = 0 \ .
\end{equation}
The Einstein equation (\ref{einstein}) for $\Lambda = 0$ and $c$ is
\begin{equation}
R_{\alpha \beta} - \frac{1}{2} R g_{\alpha \beta} = \frac{1}{c^4} \left(\rho u_{\alpha}u_{\beta} + \frac{p}{c^2}u_{\alpha}u_{\beta} + p g_{\alpha \beta}\right) \ .
\end{equation}
If we plug in this equation the perturbed metric (\ref{epsilonphi}) and the energy-momentum tensor given in terms of (\ref{Ppost})-(\ref{upost}) we obtain, up to second order, the expression
\begin{equation} \label{einsteinlin}
- \frac{1}{2} (h_{\alpha \beta})^{, \gamma}_{\phantom{, \gamma} , \gamma} \cong 2 \phi^{, \gamma}_{\phantom{, \gamma} , \gamma}  \cong \rho u_{\alpha}u_{\beta} \beta^4 \ .
\end{equation}
The $\alpha = \beta = 0$ component gives the Poisson equation of Newtonian gravity
\begin{equation}
\phi ^{,i}_{\phantom{,i},i} = \frac{1}{2} \rho \ .
\end{equation}
Note that in (\ref{einsteinlin}) we assumed the derivative with respect to $x^0$ to be first order in $\beta$, as ${\partial} /{\partial x^0} = \beta {\partial} /{\partial t}$, consequently the $\phi^{, 0}_{\phantom{, 0} , 0}$ term has been discarded.

Repeating this process for the linearised Bianchi identities
\begin{equation}
T^{\alpha \beta}_{\phantom{\alpha \beta},\beta} = G^{\alpha \beta}_{\phantom{\alpha \beta},\beta} = 0
\end{equation}
gives in the $\alpha = 0$ component the equation 
\begin{equation}
\dot{\rho} + \rho v^k _{\phantom{k},k} = 0 \ ,
\end{equation}
while in the $\alpha = i$ component it gives the equation
\begin{equation}
\dot{v}^i = - \frac{p ^{,i} }{\rho} \ .
\end{equation}
The former is the continuity equation (\ref{continuity2}), while the latter is the Euler equation (\ref{euler2}), with the $\phi ^{,i}$ term missing. The absence of this term is important: it forces us to look for the complete Newtonian limit in a higher perturbative order, and it shows that the Newtonian limit as usually understood is in practice a mixture of second order and higher order approximations of GR. The $\phi ^{,i}$ term in fact appears only if we keep perturbations up to fourth order. In this case we recover the complete Newtonian equations, plus a few other equations which  play an auxiliary role in the theory \cite{PNcosmo}. However, in analogy with the second order case, also at fourth order PNC leads to a Newtonian limit given with an ill-posed initial vale problem.

This is surprising since in section \ref{coordfree} we stated that GR has a well-posed initial value problem, hence we would expect any theory derived from it to behave the same way. 

The indeterminacy of the initial value problem in the Newtonian limit is due to the way we derive the equations of post-Newtonian cosmology. What we do is to write the Einstein equation as a sum of perturbations up to a certain order, then we obtain a set of equations reading off order by order successive terms in the Einstein tensor and the energy-momentum tensor. This produces a set of constraint equations which are not carried on in time by the dynamics of the system, and therefore it results in an ill-posed initial value problem.  

This problem can be fixed if, instead of peeling off Einstein equation order by order, we chop it off at successive levels, first at second order, then at fourth order (retaining the second order terms) and finally at sixth order (retaining the second and fourth order terms). The system of equations which comes out from this process in the harmonic gauge is well-posed \cite{PNcosmo} since, in analogy with what happens in GR (see section \ref{coordfree}), the constraint equations are carried on by the dynamics of the system and they hold at any time.  In this case the gravitational potential is not solved  from a Poisson equation, rather from an hyperbolic partial differential equation 
\begin{equation} \label{boxphi}
\Box \phi \doteq -\frac{1}{c^2} \ddot{\phi} + \nabla^2{\phi} =  \frac{1}{2} \rho + \frac{1}{4} \beta^2 (\nabla^2{\theta} - A) \ ,
\end{equation}
where $\theta$ is the coefficient of the fourth order $h_{00}$ term expanded by plugging the metric expansion (\ref{g00})-(\ref{gij}) in its definition (\ref{defharm}), while $A$ is a term which depends on the mass density $\rho$, on the gravitational potential and on its first spatial derivatives. Equation (\ref{boxphi}) has oscillating solutions, corresponding to wave fronts with constant potential, representing gravitational waves\footnote{With this approach we have arrived to the well-known result saying that gravitational waves are present only from $2.5$PN order in the harmonic gauge \cite{2.5PN}, where  $2.5$PN  order means $g_{00} \sim \mathcal{O}(\beta ^7)$, $g_{0i} \sim \mathcal{O}(\beta ^6)$, $g_{ij} \sim \mathcal{O}(\beta ^7)$.}.  

Equation (\ref{boxphi}) proves the advantage of deriving the Newtonian limit in the harmonic gauge. In fact this gauge offers the possibility to verify what happens to the initial value problem of gravity when we are going from the elliptic ill-posed NG equation to the hyperbolic well-posed PNC equation for the potential.

This result derived in the harmonic gauge is interesting, however it's not completely consistent. In fact when we expanded the metric in powers of $\beta$ we supposed its coefficients to be independent from $\beta$, but a solution of (\ref{boxphi}) gives the coefficient $\phi$ as a function of $\beta$.

The differences between PNC and NG do not end to the issue of the initial value problem. In the next chapter we give an example of incompatibility between the solutions of NG and PNC. Moreover in that same chapter we give the post-Newtonian form of the Weyl tensor and the backreaction, and we comment on how these relate to their Newtonian counterparts.

\paragraph*{Brief detour on post-Newtonian cosmology vs.\ cosmological perturbation theory}
Two different approaches are commonly used in the literature to describe perturbative GR: one is post-Newtonian cosmology, while the other is cosmological perturbation theory (CPT) \cite{bardeen}. 

The former is obtained by expanding the metric in powers of ${|v_0|} c^{-1}$ around the Minkowski or FRW metric. It is commonly adopted when the perturbative approach is applied to the study of compact systems (binary objects emitting gravitational radiation, three-body problems in the solar system) or when the focus is explicitly on finding  equations where the purely Newtonian and GR terms can be identified separately, as in the case of this thesis. 
The latter is obtained by expanding the metric as a sum of the FRW metric plus small perturbations and is adopted in LSS studies.

In the literature the two approaches are often treated as qualitatively different from each other \cite{PNlagrback} \cite{PNhydro}. However it seems to us that the two approaches have the same range of applicability, the only important difference being that in CPT theory the perturbed metric component are all first order in $\beta$, while in PNC they can be highest order, as in the case of the perturbed $g_{0i}$ which is $\mathcal{O}(\beta ^3)$.

\chapter{On the relation between NC and PNC} \label{cosmo}
\subsubsection*{Introduction}
The aim of this chapter is to discuss the theory of general relativity, given in the limit of weak fields and small velocities, and to show how it is related to Newtonian gravity. 

In the first part of this chapter we delve into this topic through an example: in GR any shear-free dust universe must be either expanding or rotating, while NG admits both rotating and expanding solutions. This statement is equivalent to saying that in GR the \textit{shear-free theorem} holds, while in NG it doesn't. This theorem is derived in section \ref{shear} following the reference \cite{theorems}. In the same section we show that this theorem holds also in the case of linearised GR, including PNC.

There are other examples of discrepancies arising between PNC and NG that can be found in the literature. In \cite{k0} it's shown that PNC is a better approximation to the flat FRW solution of GR than NG. A reason is that in PNC pressure gravitates, thus this theory is sensitive to the equation of state of the matter model, while in NG pressure doesn't gravitate. In \cite{anisocosmo} the author treat the case of anisotropic cosmologies: in this context PNC outshines NG in reproducing the anisotropic GR universes classified under the nine Bianchi types \cite{CMBaniso}. We do not look deeper into these two cases since the conceptual breakup between PNC and NG is the same as for the shear-free theorem.

In the second part of this chapter we write down the post-Newtonian expansion for the Weyl tensor and the backreaction variable and we comment on the magnitude of the terms which correspond to GR corrections.

\section{Shear-free theorem} \label{shear}
\subsection{The conjecture}
The shear-free theorem hasn't been proven yet for a barotropic fluid with a generic equation of state, therefore in its widest formulation this is not really a theorem, but rather a conjecture.
\newtheorem*{conj}{Shear-free conjecture}
\begin{conj}
In general relativity, every shear-free barotropic perfect fluid must be either expansion-free or rotation-free, i.e.\ \\ hp: $\rho + p \neq 0 \; \; \wedge \; \; p = p(\rho) \; \; \wedge \; \; \sigma = 0$. \\
ts: $\theta = 0 \; \; \lor \; \; \omega =0 $.
\end{conj}
Here the wedge logic operator between two propositions $A  \; \; \wedge \; \; B$ states that \textit{both} $A$ \textit{and} $B$ are true, while the inclusive wedge logic operator $A \; \; \lor \; \; B $ means that $A$ \textit{or} $B$ is true, if not \textit{both}.
In this thesis we treat the geodesic case of the conjecture, which has been proven.
\subsection{The theorem in GR} \label{shearfreeGR}
\newtheorem*{sht}{Geodesic shear-free theorem}
\begin{sht}
In general relativity, every shear-free geodesic perfect fluid must be either expansion-free or rotation-free, i.e.\ \\ hp: $\dot{u}_{\alpha} = 0  \; \;  \wedge \; \;  \sigma = 0 $. \\
ts: $\theta = 0 \; \;  \lor \; \;  \omega =0 $.
\end{sht}
The quantities $p$, $\rho$, $\sigma$, $\omega$, $\theta$ were defined in Chapter \ref{1+3form}.
In this section we use the covariant approach to GR to review a proof of this theorem. Similarly we attempt a proof for Newtonian gravity in the potential hydrodynamical picture, following the same steps as in the GR case, as prescribed by \cite{theorems}. 

Before starting to prove the theorem we need to introduce three lemmas which are useful in the discussion.
\newtheorem*{lemma1}{Lemma 1}
\begin{lemma1}
If there exists a function f such that $\tilde{\nabla}_{\beta} f = 0$ then either $f$ is covariantly constant in time or the rotation vanishes, i.e.\ \\
hp: $\exists f \; \;  / \; \;  \tilde{\nabla}_{\beta} f = 0$. \\
ts: $\dot{f} = 0 \; \;  \lor \; \;  \omega = 0$.
\end{lemma1} 
\begin{proof}
If $\dot{f} = 0$ then $f$ is constant.  \\
If $\dot{f} \neq 0$ then we can write $u_{\alpha} = - {\dot{f}}^{-1} \nabla_{\alpha} f$. It follows from section \ref{irro} (Frobenius' theorem) that $\omega = 0$. \qedhere
\end{proof}
\newtheorem*{lemma2}{Lemma 2}
\begin{lemma2}
If the perfect fluid is geodesic then either the pressure is constant or the rotation vanishes, i.e.\ \\
hp: $\dot{u}_{\alpha} = 0$. \\
ts: $\dot{p} = 0 \; \;  \lor \; \;  \omega = 0$.
\end{lemma2}
\begin{proof}
From equation (\ref{adot}) we know that if the acceleration vanishes then $\tilde{\nabla} _{\alpha} p = 0$. Lemma 1 implies the result.
\end{proof}
\newtheorem*{lemma3}{Lemma 3}
\begin{lemma3}
If the perfect fluid is geodesic, shear-free and if the density can be written as
\begin{equation} \label{rhoparam}
\rho = (c_1 -1) p + c_2 \omega^2 \ ,
\end{equation}
where $c_1$ and $c_2$ are real constants, then either the rotation or the expansion rate vanishes, i.e.\ \\
hp: $\sigma = 0 \; \;  \wedge \; \;  \dot{u}_{\alpha} = 0 \; \;  \wedge \; \;  \rho = (c_1 -1) p + c_2 \omega^2$.   \\
ts: $\dot{f} = 0 \; \;  \lor \; \;  \omega = 0$.
\end{lemma3}
\begin{proof}
Lemma 2 shows that  either the pressure of a geodesic perfect fluid is constant or its rotation vanishes. If the rotation vanishes then lemma 3 is satisfied. To prove this lemma therefore we need to focus only on the case of constant pressure.
The application of the geodesic and the shear-free conditions to the evolution equation for the vorticity (\ref{omegadotGR}) gives
\begin{equation} \label{omegadotlemma}
\dot{\omega}^{\alpha} = - \frac{2}{3} \theta \omega^{\alpha} \ .
\end{equation}
This same equation holds if we have $\omega^{\alpha \beta}$ or $\omega$ in the place of $\omega^{\alpha}$. Applying the operator $u^{\beta} \nabla_{\beta}$ to equation (\ref{rhoparam}), in other words \textit{time evolving} it, and substituting for $\dot{\rho}$ and $\dot{\omega}$ with (\ref{continuityGR}) and (\ref{omegadotlemma}), returns the expression
\begin{equation}
\theta (c_1 p - \frac{1}{3} c_2 \omega ^2) = 0 \ .
\end{equation}
Here if $\theta$ is zero then lemma 3 holds, on the other hand if the expression in brackets equals to zero then $\omega$ is a constant, being proportional to $p$. If $\omega$ is a constant then {\ref{omegadotlemma}} implies $\theta \omega = 0$, proving lemma 3.
\end{proof}
\paragraph*{Proof of the shear-free theorem in GR.}
\begin{proof}
We are now ready to prove the shear-free theorem. The calculations required to develop the proof are lengthy, but this is worthy the effort, as the consequences that the theorem carries are of fundamental importance. 

We start the proof by time evolving the constraint equation (\ref{GRconstr2})\footnote{All the terms involving shear or acceleration in this and the following equations vanish according to the hypothesis of the theorem.}, we use (\ref{omegadotlemma}), and obtain 
\begin{equation}
2 \tilde{\nabla}_{\alpha} \rho - 13 h_{\alpha}^{\phantom{\alpha} \beta} \omega^{\gamma} \nabla_{\beta} \omega_{\gamma} -3 h_{\alpha}^{\phantom{\alpha} \beta} \omega^{\gamma} \nabla_{\gamma} \omega_{\beta} = 0 \ .
\end{equation}
Exploiting the constraint equation (\ref{GRconstr1}) contracted with $\omega_{\alpha \gamma}$ and using the vorticity identity $\omega_{\alpha \beta} \nabla_{\gamma} \omega ^{\beta \gamma} = h_{\alpha}^{\phantom{\alpha} \beta} \omega^{\gamma} \nabla_{\gamma} \omega_{\beta} - h_{\alpha}^{\phantom{\alpha} \beta} \omega^{\gamma} \nabla_{\beta} \omega_{\gamma}$ (derived using the standard identity $\omega_{\alpha}^{\phantom{\alpha} \gamma} \omega_{\gamma}^{\phantom{\gamma} \beta} = \omega_{\alpha} \omega^{\beta} - \omega ^2 h_{\alpha}^{\phantom{\alpha} \beta}$) to transfer between the vector and two-tensor representation of the vorticity gives
\begin{equation} \label{gradients}
\tilde{\nabla}_{\alpha} \rho - 8 \omega \omega_{\alpha}^{\phantom{\alpha} \beta} \nabla_{\beta} \omega + \omega_{\alpha}^{\phantom{\alpha} \gamma} \omega_{\gamma}^{\phantom{\gamma} \beta} \nabla_{\beta} \theta = 0 \ .
\end{equation}
From this relation containing the covariant derivatives of the density, the vorticity and the expansion scalar we want to get to an expression for the expansion scalar only. We do this through a two step iterative process. 

In the first step we time evolve equation (\ref{gradients}), then we get rid of covariant time derivatives of $\rho$, $\theta$ and $\omega$ using the Raychaudhuri equation, the continuity equation and the vorticity equation. We finally get rid of $\nabla_\beta \omega$, substituting it from (\ref{gradients}). 
What comes out is an expression which relates the covariant derivatives of $\theta$ and $\rho$:
\begin{equation} \label{firststep}
\left(\rho + p - \frac{16}{3} \omega^2 \right) \tilde{\nabla}_{\alpha} \theta = \frac{1}{2} \omega_{\alpha}^{\phantom{\alpha} \gamma} \omega_{\gamma}^{\phantom{\gamma} \beta} \nabla_{\beta} \theta + \frac{1}{3} \theta \tilde{\nabla}_{\alpha} \rho \ .
\end{equation}

In the second step we use the same procedure as in the first step, this time applied to equation (\ref{firststep}). This process returns the expression
\begin{equation} \label{dimstratlong}
\begin{split}
\frac{1}{2}  \left(\rho + p - \frac{16}{3} \omega ^2 \right) \theta   \omega_{\alpha}^{\phantom{\alpha} \beta} \nabla_{\beta} \theta - \frac{1}{4} \theta \omega_{\alpha}^{\phantom{\alpha} \gamma} \omega_{\gamma}^{\phantom{\gamma} \delta} \omega_{\delta}^{\phantom{\delta} \beta} \nabla_{\beta} \theta + \left(\frac{1}{3} \theta^2 + \omega ^2  - \frac{\rho + 3p}{4} \right) \cdot \\ \cdot \omega_{\alpha}^{\phantom{\alpha} \gamma} \omega_{\gamma}^{\phantom{\gamma} \beta} \nabla_{\beta} \theta + \left[\frac{32}{9} \theta^2 \omega^2 + \left(\rho + p - \frac{16}{3} \omega^2 \right) \left(\frac{\rho + 3p}{2} -2 \omega ^2 \right) \right] h_{\alpha}^{\phantom{\alpha} \beta} \nabla_{\beta} \theta = 0 \ .
\end{split}
\end{equation}
This can be projected along $\omega ^{\alpha}$, giving the relation
\begin{equation} \label{dimstart}
\left[\frac{32}{9} \theta^2 \omega^2 + \left(\rho + p-\frac{16}{3} \omega^2 \right) \left(\frac{\rho + 3 p}{2} - 2 \omega^2 \right) \right]\omega^{\beta} \nabla_{\beta} \theta = 0 \ .
\end{equation}
The proof of the theorem follows from (\ref{dimstart}) and from the three lemmas derived previously. \\
Equation (\ref{dimstart}) is satisfied in the two cases a) and b): \\ \\
a) \\
\begin{equation}
\frac{32}{9} \theta^2 \omega^2 + (\rho + p-\frac{16}{3} \omega^2)(\frac{\rho + 3 p}{2} - 2 \omega^2) = 0 \ .
\end{equation}
We can time evolve this equation and after getting a rid of the dotted quantities as done earlier on we get to the equation
\begin{equation}
\theta \left[\frac{64}{9} \omega ^4 - 2 \rho  \omega^2 - \frac{38}{3} p \omega ^2 + p (\rho +p) \right] = 0 \ .
\end{equation}
This equation is satisfied in the two cases a') and a''): \\ \\
a') \\
\begin{equation} \theta = 0 \ , \end{equation} which proves the theorem. \\ \\
a'') \\
\begin{equation}
\frac{64}{9} \omega ^4 - 2 \rho  \omega^2 - \frac{38}{3} p \omega ^2 + p (\rho +p) \ .
\end{equation}
This equation can be time evolved again, leading to
\begin{equation}
\left(\rho + \frac{7}{3}p - \frac{40}{9} \omega^2 \right) \omega^2 = 0 \ .
\end{equation}
This equation proves the theorem in the case a) since it's satisfied either if the rotation vanishes, or if $\rho + \frac{7}{3}p - \frac{40}{9} \omega^2 = 0$. In the first case the theorem is trivially satisfied, in the second case, according to lemma 3, we have $\theta \omega = 0$. \\ \\
b) \\
In this case we start from (\ref{dimstratlong}) and we exploit equation (\ref{omegadotlemma}) to derive the expression
\begin{equation} \label{GRequi}
{\theta} \left[29 \omega^2 - 6 (\rho + p) \right] \omega_{\alpha}^{\phantom{\alpha} \beta} \nabla_{\beta} \theta = 0 \ .
\end{equation}
This equation is verified in the three cases b'), b''), b'''): \\
b') \\
\begin{equation}
\theta = 0 \ .
\end{equation}
This is the trivial case in which the theorem is satisfied. \\
b'') \\
\begin{equation}
29 \omega^2 - 6 (\rho + p) = 0 \ .
\end{equation}
In this case the theorem is satisfied via lemma 3. \\
b''') \\
\begin{equation} \label{omeganablatheta}
\omega_{\alpha}^{\phantom{\alpha} \beta} \tilde{\nabla}_{\beta} \theta = 0 \ .
\end{equation}
This is the most subtle case, the reason being that at this point the Newtonian proof of the theorem, which as we see in the next section follows step by step the GR case, breaks up. In this case the qualitative difference between the two theories shows up. 

However here we have to verify equation (\ref{omeganablatheta}), which finally proves that the shear-free theorem holds in GR. This can be done by showing that (\ref{omeganablatheta}) can be rewritten as $\omega_{\alpha \gamma} h^{\gamma \beta}\nabla_{\beta} \theta = 0$, which holds either if $\omega_{\alpha \gamma} = 0$, in which case the theorem is proven, or if $h^{\gamma \beta}\nabla_{\beta} \theta = 0$. In the latter case we are free to lower down an index, getting $\tilde{\nabla}_{\alpha} \theta = 0$. Inserting this equation and (\ref{omeganablatheta}) in (\ref{dimstratlong}) shows that $\tilde{\nabla}_{\alpha} \rho = 0$. Inserting this expression in (\ref{gradients}) gives $\tilde{\nabla}_{\alpha} \omega = 0$. Now if we set $f= \omega$ in lemma 1 we get that the vorticity must be covariantly constant in time. This implies, through equation (\ref{omegadotlemma}), that the vorticity must vanish, proving the theorem.
\end{proof}
In the next section we proceed with an attempt of proving the shear-free theorem for Newtonian gravity.

\subsection{The theorem in NG} \label{shearfreeNG}
As anticipated above, the geodesic shear-free theorem doesn't hold in NG, so we initially downgrade it to the status of conjecture, we enunciate it, and we go on showing that it is wrong.
\newtheorem*{shtN}{Non-accelerating shear-free conjecture}
\begin{shtN}
In Newtonian gravity, every shear-free non-accelerating perfect fluid must be either expansion-free or rotation-free, i.e.\ \\ hp: $\dot{v}_{i} = 0  \; \;  \wedge \; \;  \sigma = 0 $. \\
ts: $\theta = 0 \; \;  \lor \; \;  \omega =0 $.
\end{shtN}
We don't get into details of the Newtonian proof, because it proceeds the same way as the GR proof. One just need to replace 1+3 GR equations with the equations of Newtonian cosmology and work them out as done in the previous section. As we have already mentioned, the Newtonian proof breaks down when we get to the analysis of the relation
\begin{equation} \label{proofNG}
{\theta}(29 \omega^2 - 6 \rho)\omega_{ij} \theta^{j} = 0 \ ,
\end{equation} 
which is the NG equivalent of the GR formula (\ref{GRequi}) in the case b''')\footnote{In order to get to the result (\ref{proofNG}), lemmas 2 and 3 need to be exploited, this is a fair procedure since they both hold also in NG, in the case of lemma 2 trivially, as NG is pressure less by definition.}. Asking (\ref{proofNG}) to be true and working on it as for the GR case, leads to the NG conditions
\begin{equation} \label{omegagradproof}
\rho_{,i} = \omega_{,i} = 0 \ .
\end{equation}
From this two conditions we should try to show that the vorticity is constant, and because of (\ref{omegadotlemma}) that it vanishes. In GR (\ref{omegagradproof}) corresponds to $\tilde{\nabla}_{\alpha} \rho = \tilde{\nabla}_{\alpha} \omega = 0$, which we used in combination with lemma 1 to prove the theorem. This approach doesn't work in NG: in fact in NG the condition $\omega_{,i} = 0$ cannot be combined with the Newtonian equivalent of lemma 1, as the Newtonian counterpart of lemma 1 doesn't exist. In NG indeed $\omega_{,i} = 0$ doesn't imply that the rotation is constant in time, therefore the theorem is not valid. NG is then a theory where both expanding and rotating shear-free solutions are possible. We now attempt to define what would be the Newtonian equivalent of lemma 1, and we explain why the break down of the proof takes place.
\newtheorem*{lemma4}{Newtonian attempt at lemma 1}
\begin{lemma4}
If there exist a function f such that $f_{,i} = 0$, then either $f$ is constant in time or the rotation vanishes, i.e.\ \\
hp: $\exists f \; \;  / \; \;  f_{,i} = 0$. \\
ts: $\frac{\partial f}{\partial t} = 0 \; \; \lor \; \;  \omega = 0$.
\end{lemma4}
Clearly this statement is not true in NG, and lemma 1 is a purely relativistic result. In fact, even if the gradient of $f$ is zero, $f$ may still be a function of time in NG. This is due to the fact that in NG the space is flat, and its slicing at a given time is absolute, therefore a constraint such as $\dot{f} u_{\alpha} = \nabla_{\alpha} f$ doesn't exist. The two quantities $\frac{\partial f}{\partial t}$ and $f_{,i}$ are decoupled and we cannot draw any conclusion about the time evolution of the vorticity (setting $f=\omega$) from the knowledge that its gradient is zero. In order to show how this decoupling takes place and how it is connected to the geometrical nature of the two theories we resort to the post-Newtonian approximation of GR. This is the content of the next section.
\subsection{The theorem in PNC} \label{conclusionstheorem}
We can expand the equation $\tilde{\nabla}_{\alpha} \omega = h_{\alpha}^{\phantom{\alpha} \beta} \nabla_{\beta} \omega = 0$ in PNC using the metric (\ref{epsilonphi}) and the four-velocity (\ref{PNvel}), keeping only first order terms in $\beta$. A dimensional argument immediately gives that in PNC $h_{\alpha \beta} = g_{\alpha \beta} + u_{\alpha} u_{\beta}c^{-2} = g_{\alpha \beta} + \beta^2 u_{\alpha} u_{\beta}$. In the $\alpha = i$ component of the post-Newtonian expansion of $\tilde{\nabla}_{\alpha} \omega = 0$ we get
\begin{equation} \label{limitbeta}
\omega_{,i} = - \beta^2 v_i \frac{\partial \omega}{\partial t}  \ .
\end{equation}
From this equation we deduce that the shear-free theorem holds also in post-Newtonian cosmology, as in this case the gradient of the vorticity and its time derivative are coupled.

This result is not surprising since we didn't make any assumption on the strength of the fields while proving the theorem for GR\footnote{The theorem holds also if we take the FRW metric as a background \cite{FRWshear}.}. This also illuminates the fact that PNC tends to preserve the geometrical structure of GR, and it is fundamentally different from Newtonian gravity and its geometrized version. 

In fact if we want to recover the geometric structure of NG we should take the limit of (\ref{limitbeta}) for $\beta$ going to zero, as done in the chapter dedicated to frame theory for the parameter there named $\lambda$, which corresponds here to $\beta^2$. If we do so then we get $\omega_{,i} = 0$, which is, not by accident, the second condition (after $\lambda = 0$) that we needed in section (\ref{newlim}) to restrict GNG to NG. This shows that at this point, when taking this limit, the decoupling between the gradient and the time derivative of the vorticity takes place, giving the possibility in NG to have both expanding and rotating solutions of shear-free universes \cite{shear1} \cite{shear2}. In a more geometrical sense this decoupling reflects the fact that in the limit of $\beta = 0$ the space and time dimensions themselves decouple, which was shown in the FT chapter as well.

\section{$1$PN Weyl tensor} \label{PNweyl}

In this section we give the post-Newtonian form of the magnetic and electric parts of the Weyl tensor. 

The calculation for the post-Newtonian Weyl tensor proceeds as follows: first, we take the post-Newtonian metric in the form (\ref{g00})-(\ref{gij}). We want to work in the framework of $1$PN cosmology, which is equivalent to saying that from the geodesic equation (\ref{geodesic}) we want to calculate the three-acceleration $\frac{d^{2}x^i}{dt^2}$ up to order $\beta^4$. In order to do so we need to keep perturbations up to fourth order in $g_{00}$, third order in $g_{0i}$ and second order in $g_{ij}$ \cite{weinberg}.
From this metric we calculate the Christoffel symbols and from these we derive the Riemann tensor. Using the definitions (\ref{GRelectric})-(\ref{GRmagnetic}) for the electric and magnetic parts of the Weyl tensor and the definition (\ref{upost}) for the four-velocity we write down the $1$PN form of the spatial part of $E_{\alpha \beta}$ and  $H_{\alpha \beta}$ \cite{PNhydro}:
\begin{equation} \label{PNE}
E^{(PN)}_{ij} \doteq c^2 C_{i0j0} = \phi^{,i}_{\phantom{,i} ,j} - \frac{1}{3} \phi^{,k}_{\phantom{,k} ,k} \delta_{ij} + \mathcal{O}(\beta^2) \ ,
\end{equation}
\begin{equation} \label{PNH}
\begin{split}
H^{(PN)}_{i j} \doteq \frac{1}{2} \epsilon _{i \gamma}^{\phantom{\alpha \gamma} \tau \kappa} C_{\tau \kappa j \delta} u^{\gamma} u^{\delta} = &\\ = \beta \frac{\tilde{\epsilon} _{i}^{\phantom{i} kl}}{2}   \left[ \left( \frac{1}{2}\zeta_{l,m}^{\phantom{1,m},m}  - \frac{1}{2} \zeta^{m}_{\phantom{m},ml} - \frac{1}{3} v_{l} \phi^{,m}_{\phantom{,m},m} \right) \delta_{jk}  - \zeta_{l,kj} + 2v_l (\phi)_{,kj} \right] +& \mathcal{O}(\beta^3) \ .
\end{split}
\end{equation}
Here $\tilde{\epsilon ^{ijk}} = \sqrt{-g}{\epsilon}^{0ijk}$. We notice that at leading order $E_{ij}$ coincides with its Newtonian definition (\ref{Enewton}), and it stays finite in the limit of $\beta$ going to zero. On the other hand $H_{ij}$, which we learned from FT is identically zero for NG, is defined in $1$PN cosmology as (\ref{PNH}). This shows that it's one order of perturbation smaller than the electric part and that both the gravitational potential $\phi$ and vector perturbations $\zeta^i$ contributes to it. From the $\alpha=0$ component of the harmonic gauge condition (\ref{harm2}), it follows that the gravitational potential and the vector perturbation are related by
\begin{equation}
\zeta^i_{\phantom{i},i} + 4 \phi = 0 \ ,
\end{equation}
thus (\ref{PNH}) can be given in terms of the vector perturbation $\zeta^i$ and its derivatives at leading order in the harmonic gauge.

It would appear that in the paper \cite{PNhydro} the order of magnitude in $\beta$ of $E_{ij}$ and $H_{ij}$  is wrong. In the metric expansion the authors suppose $U \sim \mathcal{O}(\beta^0) = \mathcal{O}(1)$, and in equation (40) they get $E_{ij} \cong \frac{1}{c^2} U_{,i ,j}  = \mathcal{O}(\beta^2)$, where for them $U$ equals minus the gravitational potential, which we call $\phi$. In that paper $U$ has dimension $[c^2]$, meaning velocity squared, however its perturbative order must be $\mathcal{O}(1)$, and cannot be $\mathcal{O}(\beta^{-2})$ which would make $E_{ij} = \mathcal{O}(1)$ as it should be, otherwise the expansion in the metric doesn't hold\footnote{In fact if $U=\mathcal{O}(\beta^{-2})$ then $\beta U = \mathcal{O}(\beta^{-1})$, while it should be of order $\beta$ as it's the first term in the scalar sector of metric perturbation.}. This gives an example of why it is more transparent to write the expansion parameter as $\beta = {|v_0|}c^{-1}$ and then use $|v_0|$ to fix the dimensionality, instead of using the parameter $c^{-2}$, as customarily done in the literature. This mistake is inherited from the fact that, instead of taking the zero component of the four-velocity to be of order $\mathcal{O}(\beta^{-1})$, the authors take it to be of order unity. Similar considerations seem to hold for most of the other quantities calculated in the paper which involve the four-velocity.

\section{$1$PN backreaction} \label{PNback}

The $1$PN form of the backreaction can be calculated starting from the $1$PN metric described in the previous section and from the definition of the post-Newtonian four-velocity (\ref{upost}). Using these two quantities we can derive the post-Newtonian form of the orthogonal projection tensor, expressed taking into account $c$, as
\begin{equation} \label{bla}
h_{\alpha \beta} = g_{\alpha \beta} + \beta ^2 u_{\alpha} u_{\beta} \ .
\end{equation}
We use (\ref{bla}) to calculate $\Theta$, and the spatial part of $\sigma_{\alpha \beta}$ and $\omega_{\alpha \beta}$, using their definitions in 1+3 cosmology given by (\ref{Thetadef})-(\ref{omega3+1}).
This results in:
\begin{equation}
\Theta^{(PN)} = v ^i_{\phantom{i},i} + \mathcal{O}(\beta^2)\ ,
\end{equation}
\begin{equation}
\sigma^{(PN)}_{ij} = v _{(i,j)} - \frac{1}{3} \Theta^{(PN)} \delta _{ij} + \mathcal{O}(\beta^2) \ ,
\end{equation}
\begin{equation}
\omega^{(PN)}_{ij} = v _{[i,j]} + \mathcal{O}(\beta^2) \ .
\end{equation}
At lowest order these three quantities coincide with their Newtonian analogues (\ref{thetaNG})-(\ref{omegaNG}), implying that backreaction reduces to a boundary term in PNC. This shows that backreaction proves to be small if studied in the framework of post-Newtonian hydrodynamics with periodic boundary conditions. However this is not surprising: backreaction parametrizes the non-linearity that we have in a system (its degree of inhomogeneity) and in the context of this section we are studying a system linearised a background homogeneous metric. Therefore there is no surprise in finding out that the corrections due to inhomogeneities are small.

This result leads us to make two considerations. 
\paragraph*{The linearised metric.} First, if the geometry of the universe is well described at late times by a metric linearised around a FRW background (meaning (\ref{g00})-(\ref{gij}) with the addition of a scale factor), then the backreaction is small in the real universe. This holds also at higher order in usual cosmological perturbation theory \cite{pertbackreaction}\cite{syksylast}. The success of the $\Lambda$CDM model in fitting data coming from cosmological observations is a hint in this direction. However the $\Lambda$CDM model suffers from a couple of problems, which were discussed in the introduction. First, the coincidence problem, which asks if there is any reason why the universe starts accelerating when structure formation enters in the non-linear regime. Second the homogeneity scale of Newtonian $\Lambda$CDM simulations is one order of magnitude smaller then in the real universe \cite{sylos}, which suggests that the universe is more structured than expected.

Furthermore, it is well known that metric perturbations are small as long as matter density perturbations are, but it's not clear if they keep being small even when density perturbation become non-linear \cite{pertbackreaction} \cite{syksylast} \cite{syksynewton} \cite{waldgreen}.

\paragraph*{The inhomogeneous model.}
The second consideration is the following: if we want to find out in which context the backreaction has a sizeable effect on cosmic acceleration then we should build up an inhomogeneous model of the universe. It seems however hard to write down an inhomogeneous metric. We could overcome this problem for instance by building a statistical model which bypasses the problem of which is the overall metric of the universe. This model should be fully non-perturbative and should take into account GR effects such as the tidal forces due to the Weyl tensor, and perhaps the fact that the energy-momentum tensor is not conserved globally in GR. Yet at the same time it should be numerically tractable, so in this sense close to NG.

\chapter{Conclusions} \label{summary}
In the thesis we tried to understand better the relationship between Newtonian cosmology and general relativity. It is pretty clear that in the case of homogeneous and isotropic cosmologies Newtonian gravity is close to general relativity given in terms of the FRW metric. 
Thus we focused our attention on the inhomogeneous case, which is less understood.

We reviewed some results in this field: in inhomogeneous Newtonian cosmologies the backreaction is a boundary term, and its contribution to the expansion law vanishes identically in $N$-body simulations of structure formation, which run in periodic boxes. Therefore there's no use of studying backreaction in such a way.

In principle inhomogeneous relativistic cosmologies could be studied numerically for the purpose of quantifying backreaction, as in this case its value depends on the global behaviour of the fluid. However, because of computational cost this is impossible to put in practice. 

In order to study backreaction quantitatively we need then a theory which is not NG nor GR. If we want to work out this theory we have to understand how NG and GR relate to each other. 

In the fully non-linear regime this understanding can be achieved theoretically thanks to frame theory, which includes NG and GR as degenerate cases. In the linear regime of GR it is possible to write down a system of equations which encodes both purely NG and GR degrees of freedom, and which has a well-posed initial value problem. However the effect of inhomogeneities is small by definition in this context, as  backreaction takes into account non-linear features of the system. If we apply it to a linearised system it will bring a small contribution to its dynamics. 

The problem whether the universe at late times can be well described by a linearly perturbed metric is an open issue which hasn't been addressed here. However thanks to post-Newtonian cosmology we realised that at $1$PN level the backreaction contribution to the observed cosmic acceleration is small.

In conclusion we mention that a new approach to the problem of the Newtonian limit of GR, applied in \cite{waldgreen}, appears to be promising in understanding Newtonian and relativistic inhomogeneous cosmologies. This approach is perturbative, and relies on the assumption of a background metric, but still allows a significant non-linear behaviour on small scales. It seems to be a theory encoding some of the features that we are interested in, however its structure needs further investigation.


\end{document}